\documentstyle[epsfig,epsf,subfigure,12pt,here]{article}
\def\partname{Chapter}
\def\thepart{\arabic{part}}
\def\appendixname{Appendix} 
\def\appendix{\par
 \setcounter{part}{0}
 \setcounter{equation}{0}
 \setcounter{section}{0}
 \def\@chapapp{\appendixname}
 \def\thepart{\Alph{part}}
 \def\theequation{\thepart\arabic{equation}}
 \def\partname{Appendix} 
}

\newcommand{\chap}{\section}
\newcommand{\sect}{\subsection}
\newcommand{\subsect}{\subsubsection}

\newcommand{\rmfont}[1]{{\rm #1}}
\newcommand{\stand}[1]{{  #1}}

\newcommand{\myfrac}[2]{\frac{\dsp #1}{\dsp #2}}
\newcommand{\GeV}{\ \mbox{{\rm GeV}}}
\newcommand{\MeV}{\ \mbox{{\rm MeV}}}

\newcommand{\qbf}{{\bf q}}

\newcommand{\MSbarsmall}%
{{\overline{{ \mbox{{\scriptsize  MS}}}}}}
\newcommand{\MSsmall}{{{{\scriptscriptstyle MS}}}}

\catcode`\@=11
\def\slash{\mathpalette\make@slash}
\def\make@slash#1#2{\setbox\z@\hbox{$#1#2$}%
  \hbox to 0pt{\hss$#1/$\hss\kern-\wd0}\box0}
\catcode`\@=12 

\newcommand{\FMslash}{\slash}

\def\bPsi{{\bf\Psi}}

\def\bfpsi{{\bf\psi}}

\def\bfPhi{{\bf\Phi}}

\def\mbf{{\bf m}}
\def\zrm{{\rm z}}

\newcommand{\Phibf}{{\bf \Phi}}
\newcommand{\Sbf}{{\rm\bf S}}
\newcommand{\Lc}{{\cal L}}
\newcommand{\prd}{\partial}

\newcommand{\unl}{\underline}

\newcommand{\bea}{\begin{eqnarray}}
\newcommand{\eea}{\end{eqnarray}}

\def\PLB{{ Phys. Lett.} }

\def\PR{{ Phys. Rev.} }

\newcommand{\ice}[1]{\relax}

\ice{
\setlength{\textwidth}{16cm}
\setlength{\textheight}{23cm}
\setlength{\oddsidemargin}{1.cm}
\setlength{\evensidemargin}{1.cm}
\setlength{\headheight}{0.5cm}
\setlength{\headsep}{1.0cm}
\setlength{\topmargin}{-2.0cm}
\setlength{\topskip}{0.1cm}
\setlength{\footheight}{0.5cm}

\setlength{\footskip}{1.5cm}
\frenchspacing
}

\newcommand{\api}{\frac{\alpha_s}{\pi}}
\newcommand{\apis}{\frac{\dsp\alpha_s(s)}{\dsp \pi}}
\newcommand{\apimu}{\frac{\dsp\alpha_s(\mu)}{\dsp \pi}}

\newcommand{\ba}{\begin{array}}
\newcommand{\ea}{\end{array}}
\newcommand{\ds}{\displaystyle}
\newcommand{\as}{\alpha_s}
\newcommand{\gm}{\gamma_m}
\newcommand{\G}{\Gamma}
\newcommand{\g}{\gamma}
\newcommand{\gaam}{\gamma^{\rmfont{AA}}_m}

\newcommand{\dmu}{\mu^2\frac{d}{d\mu^2}}
\newcommand{\msbar}{\overline{\mbox{MS}}}
\newcommand{\ordas}{{\cal O}(\alpha_s}

\newcommand{\dsp}{\displaystyle}

\newcommand{\re}[1]{(\ref{#1})}
\newcommand{\beq}{\begin{equation}}
\newcommand{\eeq}{\end{equation}}
\newcommand{\EQN}{\label}

\newcommand{\ovl}{\overline}
\def\bbuildrel#1_#2^#3%
{\mathrel{\mathop{\kern 0pt#1}\limits_{#2}^{#3}}}

\newcommand{\al}{\alpha}
\newcommand{\be}{\beta}
\newcommand{\dif}{{\rm d}}
\newcommand{\ex}{{\rm e}}
\newcommand{\fos}[2]{\>\>\mathop{#2}^{{}\atop{#1}}{}}

\def\L{{\cal L}}
\def\bPhi{{\bf \Phi}}

\def\ep{\epsilon}

\def\bPhi{{\bf \Phi}}

\def\pd{\partial}

\def\m{{\bf m}}

\def\bPhi{{\bf\Phi}}

\newcommand{\prev}{Phys.\ Rev.\ }

\begin{document}

\begin{titlepage}
\noindent
%
%
\mbox{}
\hfill {\small TTP94--32}
\mbox{}

\hfill  { LBL-36678}
\mbox{}

\hfill  {\small hep-ph/9503396}
\mbox{}

\hfill December   1994   \\
\protect\vspace*{.3cm}
%
%
%
\vspace{-1.0cm}
\begin{center}
  \begin{Large}
  \begin{bf}
QCD Corrections to the $e^+e^-$
Cross Section and the
$Z$ Boson Decay Rate${}^{\dagger}$
  \\
  \end{bf}
  \end{Large}
%
%
  \vspace{0.3cm}
K.G.~Chetyrkin $^{ab}$,
J.H.~K\"uhn$^{a}$,
A.~Kwiatkowski$^{c}$
\\
%

\begin{itemize}
\item[$^a$]
    Institut f\"ur Theoretische Teilchenphysik,
    Universit\"at Karlsruhe \\
    D-76128 Karlsruhe, Germany
\item[$^b$]
 Institute for Nuclear Research,
 Russian Academy of Sciences   \\
 60th October Anniversary Prospect 7a
 Moscow 117312, Russia
\item[$^c$]
              Theoretical Physics Group,
              Lawrence Berkeley Laboratory\\
              University of California,
              Berkeley, CA. 94720, USA
\end{itemize}
\vspace{0.2cm}

%
  \vspace{0.5cm}
  {\bf Abstract}
\end{center}
\begin{quotation}
\noindent
QCD corrections to the electron positron
annihilation cross section into hadrons
and to the hadronic $Z$ boson decay
rate are reviewed. Formal developments
are introduced in a form particularly
suited for practical applications.
These include
the operator product expansion,  the heavy
mass expansion, the decoupling of heavy quarks and
matching  conditions.
Exact results for the quark mass
dependence are presented whenever
available, and formulae valid in the
limit of small bottom mass
($m_{\rmfont{b}}^2\ll  s$) or
of large top mass ($m_{\rmfont{t}}^2\gg s$)
are presented. The differences between
vector and axial vector induced rates
as well the classification of  singlet
and nonsinglet rates are discussed.
Handy formulae for all contributions
are collected and their numerical
relevance is investigated.
Prescriptions for the separation of the
total rate into partial rates are
formulated.  The applicability of the
results in the low energy region,
relevant for measurements around 10 GeV
and below, is investigated and
numerical predictions are collected for
this energy region.
\end{quotation}

\vfill

\noindent
emails:
\\
chet@ttpux2.physik.uni-karlsruhe.de
\\
johann.kuehn@physik.uni-karlsruhe.de
\\
kwiat@theor2.lbl.gov  \\

\vfill

\footnoterule
\noindent
$^{\dagger}${\footnotesize
To appear  	in ``Reports of the Working Group
on Precision Calculations for the Z-resonance'' .

\noindent
The complete postscript
file of this preprint, including
figures, is available via anonymous ftp at \\
ttpux2.physik.uni-karlsruhe.de (129.13.102.139) as /ttp94-32/ttp94-32.ps
or via www at  \\
http://ttpux2.physik.uni-karlsruhe.de/cgi-bin/preprints/
}

\end{titlepage}

\tableofcontents

\renewcommand{\arraystretch}{2}
\chap{Introduction\label{intro}}
Since experiments at the $\rmfont{e}^+\rmfont{e}^-$
storage ring LEP started data-taking a few
years ago, and by the end of the 1993 run by the four
experiments,
more than seven million
hadronic events
had been collected at the $\stand{Z}$ resonance.
The accuracy of the measurements is
impressive. Numerous parameters of the
standard model can be determined with
high precision,  allowing
stringent tests of the standard
model to be performed . Among them: the mass
$M_{\stand{Z}} = (91.188\pm 0.0044)$ GeV and the
width $\Gamma_{\stand{Z}}=(2.4974\pm 0.0038)$ GeV
of the $\stand{Z}$ boson or the weak mixing angle
$\sin^2\theta_{\rmfont{eff}}^{\rmfont{lept}}
=0.2322~\pm~0.0006$ \cite{Schaile94}.
All experimental results were in remarkable
agreement with theoretical predictions and
a triumphant confirmation of the
standard model.

As well as  the electroweak sector of the
standard model, LEP provides
an ideal laboratory for the investigation
of strong interactions. Due to their purely
leptonic initial state,  events are very
clean from both the theoretical and
experimental point of view and represent
the ideal  place for testing
QCD. From cross-section measurements
$\sigma_{{\rm had}}
=(41.49\pm 0.12)$ nbarn
\cite{Schaile94} as well as from the
analysis of event topologies the strong
coupling constant can be extracted.
Other observables   measurable
with very high precision are the (partial)
$\stand{Z}$ decay rates into hadrons
$\Gamma_{\rmfont{ had}}/\Gamma_{\rmfont{e}}=
20.795\pm 0.040$
and bottom quarks
$\Gamma_{\rmfont{b}\ovl{\rmfont{b}}}/\Gamma_{\rmfont{had}}=
0.2192\pm 0.0018$.
{}From the line shape analysis of LEP
a value $\as=0.126\pm 0.005\pm 0.002$
is derived.
The program of experimentation at LEP  is
still not complete. The prospect of
an additional increase in  the number of
events by a factor of
two to four
 will further improve the level of
accuracy. This means, for example,
that the relative uncertainty of the
partial decay rate into $\rmfont{b}$ quarks
 $\Delta \Gamma_{\rmfont{b}}/\Gamma_{\rmfont{b}}$ falls below
 one percent and that an experimental
error for $\as$ of 0.002 may be achieved.

Also at lower energies significant
improvements can be expected in the
 accuracy of cross-section measurements.
The energy region of around $10$ GeV
just below the ${{\rm B}}{\rm \ovl{B}}$ threshold
will be covered with high statistics
at future {B}  meson factories. The cross
section between the charm and bottom
thresholds can be measured at the BEPC
storage ring
in Bejing. These measurements could
provide a precise value for $\alpha_s$
and --- even more important ---
a beautiful proof of the running of the
strong coupling constant.

In view of this experimental situation
 theoretical
predictions for the various observables
with comparable or even better
accuracy  become mandatory and
higher-order radiative corrections
are required.
It seems appropriate to collect all
presently available calculations and
reliably estimate their theoretical
uncertainties. The aim of this report is
to provide such a review for the QCD
sector of the standard model,
as far as cross-section
 measurements are concerned, at the $\stand{Z}$
peak as well as in the  `low energy'
region from 5 to 20 GeV.
(Related topics have been also discussed in recent reviews
\cite{reviews}.)
Higher-order QCD
corrections to the $\rmfont{e}^+\rmfont{e}^-$ annihilation
cross-section into hadrons
will be discussed as well as the
hadronic width of the $\stand{Z}$ boson.
 Further  interest lies in the partial
rates for the decay of the $\stand{Z}$ boson into specific
quark channels. Of particular importance is the
partial width
 $\Gamma(Z\rightarrow \rmfont{b}\ovl{\rmfont{b}})$, as this
quantity can be measured with high accuracy
and provides important information about
the top quark mass from the $Z\rmfont{b}\ovl{\rmfont{b}}$
vertex.
However, the decomposition
of $\Gamma_{{\rm had}}$ into partial decay rates of
different quark species is possible
in a simple, straightforward way only
up to corrections of the order of $\ordas)$. Apart from
diagrams where  `secondary quarks'
are radiated off the `primary  quarks'
one encounters flavour singlet diagrams
that  first   arise
in order $\ordas^2)$ and lead to a confusion
 of different species.  They therefore have to be
 carefully scrutinized.

For many considerations and experimental conditions
quark masses can be neglected,  compared to the
characteristic energy of the problem.
Accordingly,  higher-order
 QCD corrections to the total cross-section
were first calculated
for massless quarks. At LEP energies this is
certainly a good approximation for ${\rmfont{u,d,s}}$ and
$\rmfont{c}$ quarks. In view of the accuracy reached at LEP
much effort has been spent  in estimating  the size of
mass effects of the bottom and the top quark.
Whereas $\rmfont{b}$ quarks are present as particles in the
final state, top quarks can appear only through
virtual corrections.
A large part of this report
is devoted to these effects.
The application of these formulae and,
if necessary, their numerical
evaluation will also
 be covered.

In Part~\ref{general}  topics of a
general nature are addressed. In Section~\ref{notations}
the  notation is introduced  and
the relation between
cross-sections and decay rates on the one hand and the
corresponding current correlators
on the other  is discussed.
Furthermore, the classification of singlet versus
nonsinglet terms is introduced. The behaviour of coupling
constant, masses, operators and correlators under renormalization
group
transformations is reviewed  in Section~\ref{renorm}  and the relevant
anomalous dimensions are listed. The decoupling of heavy quarks
and the resulting matching conditions for coupling constant
masses and effective currents are treated
in Section~\ref{decoupling}.
Numerical  values of quark masses are discussed in
Section~\ref{Qmasses}.
Part~\ref{calc-tech} is
concerned  with calculational techniques
relevant to  the problems at hand. Emphasis is put on the
behaviour  of the current correlators at large momenta,
the structure of mass corrections  in the small mass limit
and the resummation of large logarithms of $m^2/s$.
And the other extreme, with  $s/m^2 \ll 1$
also  dealt with in this Part, which concludes with a discussion
of $\gamma_5$ in $D \neq 4 $ dimensions.

The analytical first-order QCD corrections
to the cross-section are recalled in Part~\ref{exact}.
Approximations in the limits of low and high
energies are given.

Nonsinglet and singlet
contributions to the QCD corrections
are presented  in Parts~\ref{nonsinglet}  and \ref{singlet},
respectively, and  the relevant  formulae
for various applications are given.
First, the calculations are reviewed
for massless quarks.
This assumption is evidently  not
justified for the  heavy top   top mass,
which appears as a virtual particle. Top mass
corrections are described in Section~\ref{top}.
The dependence on the mass of the final-state quarks
is given in Section~\ref{repnsm2}. At low energies not only do the
leading quadratic mass terms have to be taken into
account, but quartic mass terms  also  become relevant.
They are presented in Section~\ref{repnsm4}.
The influence of
secondary quark production on
determinations of the partial rate is
treated in Section~\ref{nspart}.

Flavour singlet contributions are discussed in Part~\ref{singlet}.
They arise for the first time
in second order for the axial-induced rate and in
third order for the vector current-induced rate.
${\cal O}(\as^2)$ singlet corrections would
be absent for six massless flavours, but do not
vanish due to the large mass splitting in the
$(\rmfont{b,t})$ doublet.
Massless contributions and
bottom-mass corrections
from singlet diagrams  are covered
in Sections~\ref{singl-massless} and~\ref{singl-bottom}
respectively.
The assignment of the
singlet contributions to a partial rate into
a specific quark flavour is explained in Section~\ref{singl-partial}.
and the resulting  ambiguity is discussed.
In Part~\ref{numerical} the numerical relevance of the
different contributions are studied.  Different sources
of  theoretical uncertainties are investigated and their
size  estimated.

A collection of formulae is presented
in the Appendix.  It provides  an
overview  and may serve as a quick and
convenient  reference for
later use.
\chap{General Considerations\label{general}}
\sect{Notations\label{notations}}
\subsect{Cross-Sections and Decay Rates\label{cross}}
We introduce our notations by casting the
total cross-section for longitudinally polarized
$\rmfont{e}^+\rmfont{e}^-$ into hadrons
in leading order of the electroweak coupling
as:
\begin{eqnarray}
\sigma_{{\rmfont{R} \atop \rmfont{L}}}& =&  \frac{4\pi\alpha^2}{3s} \Bigg\{
   \frac{(v_{\rmfont{e}}\mp a_{\rmfont{e}})^2}{y^2}\left|
     \frac{s}{s-M_{\stand{Z}}^2+iM_{\stand{Z}}
      \Gamma_{\stand{Z}}}\right|^2
       \frac{R^{\rmfont{V}}+R^{\rmfont{A}}}{y^2}
\nonumber\\
&&+2Q_{\rmfont{e}}
\frac{v_{\rmfont{e}}\mp a_{\rmfont{e}}}{y} {\rm Re} \left[
 \frac{s}{s-M_{\stand{Z}}^2+iM_{\stand{Z}}\Gamma_{\stand{Z}}}\right]
 \frac{R^{{\rm int}}}{y}+Q_{\rmfont{e}}^2 R^{\rmfont{em}} \Bigg\}
{}\, ,
\end{eqnarray}
with the weak couplings defined through
\beq
v_{\stand{f}}=2I_3^{\stand{f}}-4Q_{\stand{f}}\sin^2
\theta_{\rmfont{w}}\, ,\;\;\;\;
a_{\stand{f}}=2I_3^{\stand{f}}\, ,\;\;\;\;
y=4\sin\theta_{\rmfont{w}}\cos\theta_{\rmfont{w}}
{}\, .
\eeq
$R$ and $L$ denote the electron beam polarization
(positrons are  assumed to be  unpolarized).
The functions $R^{\stand{k}}$ with
$ k=\rmfont{V,A},{\rm em},{\rm int} $
are the natural generalization of the Drell ratio
$R\equiv \sigma_{{\rm had}}/\sigma_{\rm point}=
R^{{\rm em}}$, which is
familiar from purely electromagnetic interactions at lower energies.
They are induced by the vector and axial
couplings of the $\stand{Z}$ boson, the pure QED part
and an interference term.
In the massless parton model they are given by
\beq\EQN{not3}
R^{\rmfont{V}}=3\sum_{\stand{f}}
v_{\stand{f}}^2\, ,\;\;\;\;\;
R^{\rmfont{A}}=3\sum_{\stand{f}} a_{\stand{f}}^2\, ,
\;\;\;\;\;
R^{{\rm em}}=3\sum_{\stand{f}} Q_{\stand{f}}^2\, ,\;\;\;\;\;
R^{{\rm int}}=3\sum_{\stand{f}} Q_{\stand{f}} v_{\stand{f}} \, .
\eeq
Here the sum extends over all flavours $f$.

The hadronic decay rate of the $\stand{Z}$ can be
expressed in a way similar to  the
incoherent sum of its vector-  and  axial
vector-induced parts:
\beq
\ba{ll}
\dsp
\Gamma_{\stand{Z}}^{\rmfont{had}}
&\dsp
= \Gamma^{\rmfont{V}}+\Gamma^{\rmfont{A}} \\
& \dsp
= \frac{\alpha}{3}\frac{M_{\stand{Z}}}{y^2}(R^{\rmfont{V}}
+R^{\rmfont{A}})
{}\, .
\ea
\eeq
Alternatively one may express $\alpha/y^2$
 through the Fermi constant
\beq
\frac{\alpha}{y^2} = \frac{G_{\rmfont{F}} M_{\stand{Z}}^2}{8\pi\sqrt{2}}
\eeq
and absorb the large logarithms from the
running of QED.
 These formulae are  equivalent for the
 present purpose, where higher-order electroweak
 corrections are ignored.

All relevant  information needed for the
correction factors $R^{\stand{k}}$
is contained in the
 current correlation functions
\beq
\EQN{not6}
\ba{rl}
\Pi_{\mu\nu}^{ij}(q)  =
& \dsp
i \int \rmfont{d} x e^{iqx}
\langle 0| \;
T\;j_{\mu}^{\stand{i}}(x)j_{\nu}^{ j }(0)\;|0 \rangle
\\
= &
\dsp
g_{\mu\nu} \Pi_1^{ij}(-q^2) + q_{\mu}q_{\nu} \Pi_2^{ij}(-q^2)
{}\, ,
\ea
\eeq
with $(i,j)=\rmfont{(V,V),(A,A)},({\rm em},{\rm em}),
({\rm em , V})$ for $k=\rmfont{V,A},{\rm em},
{\rm int}$ respectively. The
currents under consideration are defined through
\beq
j^{\rmfont{V}}_{\mu}=\sum_{\stand{f}}
v_{\stand{f}} \ovl{\psi}_{\stand{f}} \gamma_{\mu}
\psi_{\stand{f}}\, ,\;\;\;\;\;
j^{\rmfont{A}}_{\mu}=\sum_{\stand{f}} a_{\stand{f}}
\ovl{\psi}_{\stand{f}}
\gamma_{\mu}\gamma_5 \psi_{\stand{f}}\, ,\;\;\;\;\;
j^{{\rm em}}_{\mu}=\sum_{\stand{f}} Q_{\stand{f}}
\ovl{\psi}_{\stand{f}} \gamma_{\mu} \psi_f
{}\, ,
\eeq
where the sum extends over all six flavours.

The relation between the cross-section
$\sigma_{{\rm had}}$ and the corresponding current
correlator is closely connected to the
analytic properties of
$\Pi_{\mu\nu}$.
After the Lorentz decomposition
into the functions $\Pi_1$  and $\Pi_2$,
 only $\Pi_1$ enters the
cross-section, since the contraction of
$q_{\mu}q_{\nu}\Pi_2 $ with the lepton tensor is
suppressed by the electron mass.
The threshold energies
for the production of  fermion pairs
 are branch points of the vacuum
 polarization, and
$\Pi_1(-s)$
 is analytic in the
complex plane cut along the real positive
axis. For energies above the lowest-lying
threshold ($s=4m^2$)
the function
 $\Pi_1(s)$ is discontinuous when ${s}$
 approaches the real axis from above and below.
The optical theorem
relates  the inclusive cross-section
and thus the function $R(s)$
to the discontinuity of $\Pi_1$
in the complex plane
\beq \EQN{d3}
R(s) =  \dsp
-\, \frac{12\pi}{s} \,{\rm Im}\, \Pi_1( - s -i\epsilon)
=
\frac{6\pi i}{s}
[
\Pi_1(- s - i\varepsilon)
 -
\Pi_1(- s + i\varepsilon)
]
{}\, ,
\eeq
where Schwarz's reflection
principle has been employed for the second step.
Conversely,  the vacuum polarization
is obtained through a
dispersion relation from
its absorptive part.
  Applying Cauchy's
theorem along the integration contour of
Fig.~\ref{contour}
leads to:
\beq
\EQN{d4}
\ba{rl} \dsp
\Pi_1(-s)
=&\dsp
\frac{1}{2\pi i} \oint ds' \frac{\Pi_1(-s')}{s'-s}
=
 \frac{1}{\pi} \int_0^{\infty} ds'
\frac{{\rm Im}\,
\Pi_1(-s^{\prime}-i\varepsilon)}
{s^{\prime}-s} \;\;
{\rm mod \; sub}
 \\
=&\dsp
  - \, \frac{1}{12\pi^2} \int^{\infty}_0 ds^{\prime}
\frac{s^{\prime}}{s^{\prime}-s}R(s^{\prime}) \;\;
{\rm mod \; sub}
{}\, ,
\ea
\eeq
No subtraction is needed  if
$\Pi_1(-s)$ vanishes at infinity, since the
large circle does not contribute to the integral
in this case.
If the spectral
function is only bounded by
  $s^{\rmfont{n}}$ at large
distances, one may apply the dispersion
relation to the function $\Pi_1/s^{n+1}$. This is
 achieved by $n+1$ subtractions.
For example, a  twice-subtracted
dispersion relation has to be applied
for $\Pi_1(-s)$, is given by
\[
\tilde\Pi_1(-s)
=
{\Pi}_1(-s)
-
{\Pi}_1(0)
-
(-s) {\Pi}_1'(0)
{}\, .
\]
The absorptive part is
not affected by these subtractions.
For the vector current ${\Pi}_1(0)$
vanishes as a consequence of current
conservation and the second subtraction
corresponds to charge renormalization.

Let us add an additional remark concerning the
applicability of perturbative QCD for the calculation
of radiative corrections to the cross-section
$\sigma_{{\rm had}}$.
Experimental $\rmfont{e}^+\rmfont{e}^-$ data are taken
in the physical regime of timelike
momentum transfer $q^2>0$. This region is influenced
by threshold and bound state effects which make
the use of perturbative QCD questionable.
However,
perturbative QCD is
strictly applicable for large
 spacelike momenta ($q^2=-Q^2<0$),
 since this region is far away
from non-perturbative
effects due to
hadron thresholds, bound state and
resonance effects \cite{Adl74}.
Therefore,
reliable theoretical predictions can
  be made for $\Pi_1(Q^2)$ with $Q^2>0$.
To compare theoretical predictions and
experimental results for time-like momenta,
one has to perform suitable
averaging procedures \cite{PogQuiWei76}.
For large positive ${s}$ one may appeal to the
experimentally observed smoothness of $R$
as a function of ${s}$ and to the absence
of any conceivable non-perturbative
contribution.

For later use it is convenient to
introduce the
Adler function
\beq
\EQN{d5}
 D(Q^2) = -12\pi^2 Q^2
\frac{d}{dQ^2}\left[
\frac{\Pi_1(Q^2)}{Q^2}
\right]
.\eeq
%

{
\begin{figure}[t]
\begin{center}
\hphantom{XXXXXXXXX}
\parbox{3cm}{
\epsfig{file=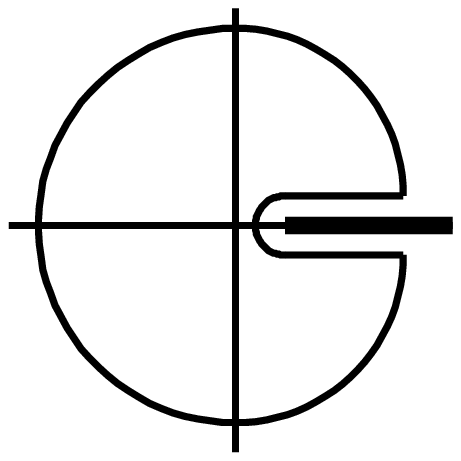,width=5.cm,height=5.cm}}
\end{center}
\caption {\label{contour}
Contour integral.}
\end{figure}
}

\noindent
It
is related to $R$ through a dispersion relation
which allows a comparison between the
 perturbatively calculated Adler function ($Q^2>0$)
and  the experiment
 if the  cross-section $R$ is known
over the full energy scale $s'>0$:
\beq D(Q^2) = Q^2 \int_0^{\infty}
 ds^{\prime}\frac{R(s^{\prime})}
                 {(s^{\prime}+Q^2)^2}
 + 12\pi^2 \frac{\Pi_1(0)}{Q^2}
{}\, .
\eeq
The relation inverse to Eq.~(\ref{d5}) finally reads
\beq   \EQN{d6}
R(s) = \frac{1}{2\pi i}
\int_{-s-i\varepsilon}^{-s+i\varepsilon}
d Q^2\frac{D(Q^2)}{Q^2}
{}\, .
\eeq

\begin{figure}[t]
\hspace*{-0.5cm}
\begin{center}
\begin{tabular}{cccccc}
Im
&
\parbox{3cm}{
\epsfig{file=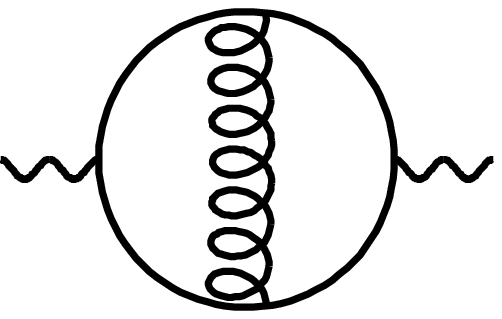,width=3.cm,height=3.cm}
            }
&
=
&
\parbox{3cm}{
\epsfig{file=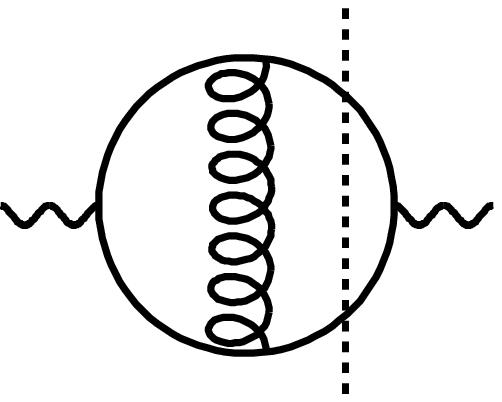,width=3.cm,height=3.cm}
            }
&
+
&
\parbox{3cm}{
\epsfig{file=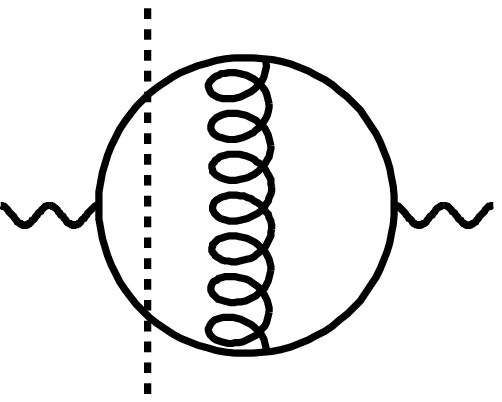,width=3.cm,height=3.cm}
            }
\\
&
&
+
&
\parbox{3cm}{
\epsfig{file=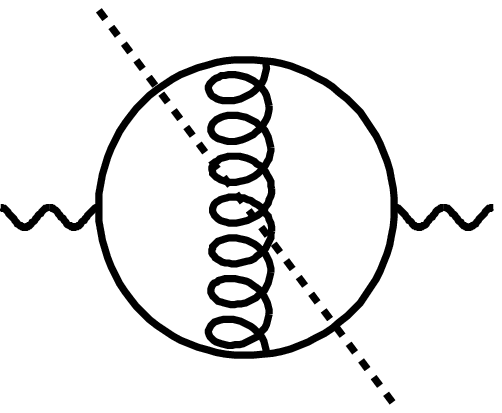,width=3.cm,height=3.cm}
            }
&
+
&
\parbox{3cm}{
\epsfig{file=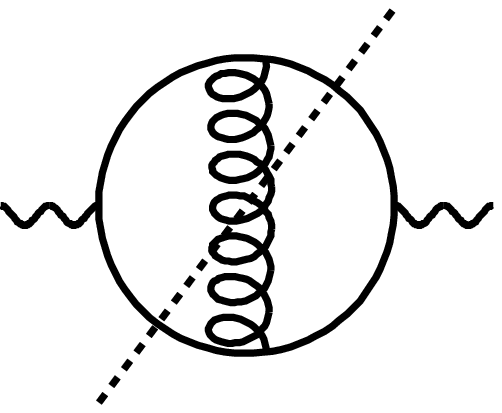,width=3.cm,height=3.cm}
            }
\end{tabular}
\end{center}
\caption[]{\label{cuts}
The absorptive part of a current correlator
is obtained by cutting the diagram in all
  possible ways.}
\end{figure}

Diagrammatically,
current correlators are depicted  as
vacuum polarization graphs. Their
 absorptive parts are obtained from the sum of
all possible cuts applied to the diagram
(see Fig.~\ref{cuts}).
This means
--- according to  Cutkosky's rule --- that
the absorptive part of a Feynman  integral is obtained,
 if the substitution
\beq
\frac{1}{p^2-m^2+i\epsilon}\rightarrow
-2\pi i \delta(p^2-m^2)\theta(p_0)
\eeq
is applied to
those propagators  associated  with to the
cut lines of the corresponding Feynman diagram.
Calculating the two-point correlator and taking
its absorptive part is  equivalent
to the evaluation of the matrix element squared
with a subsequent integration over the phase space of
 the final-state particles.
The former
method has some advantages.
Although the
vacuum polarization graph contains one
loop  more  than the amplitudes in the
direct calculation of the rate,
the problem is reduced to a propagator type
integral, for which quite elaborate techniques
have been developed and implemented in corresponding
computer packages.
Furthermore, the occurance of infrared divergences
is naturally circumvented, since
virtual and bremsstrahlung corrections
correspond only to different cuts
of the same diagram and hence are combined
 in the same amplitude. The
 cancellation  of infrared divergences  is therefore
inherent in each diagram.
Depending on the cut,
final states with a different number of particles
are represented by the same diagram, as is shown in
Fig.~\ref{cuts}.
\subsect{Classification of Diagrams
\label{classification}}
Higher-order QCD corrections
to $\rmfont{e}^+\rmfont{e}^-$ annihilation into hadrons
 were first
 calculated for the electromagnetic case in the
 approximation of massless quarks.
Considering the annihilation process through
the $\stand{Z}$ boson,
 numerous new features and subtleties
become relevant at the
 present level of precision.

The  different charge and chiral
structure of  electromagnetic and
weak currents respectively has  already been  addressed
 in the previous section:
The functions $R^{\stand{k}}$ as defined above were
classified according to the space--time
 structure of the
currents (vector versus axial vector) and their
electroweak couplings.
Another important distinction, namely  `singlet'
versus `non-singlet'  diagrams,
originates from two classes of diagrams
with intrinsically different topology and
resulting charge structure.
The first class of diagrams consists of non-singlet
contributions with one fermion loop coupled
to the external current. All these amplitudes
are proportional to the charge
structures given in Eq.~(\ref{not3}),
consisting of a sum of terms proportional
to the square of the coupling constant
or  the trivial generalization in the
interfence  term $R^{{\rm int }}$.
QCD corrections corresponding to these
diagrams contribute
a correction factor   independent of the
current under consideration
as long as masses of final  state quarks
are neglected.
Singlet contributions
arise from a second class of
 diagrams where two currents
 are coupled to two
different fermion loops and hence can be cut into two parts
by cutting gluon lines only (see Fig.~\ref{class}).
They cannot be assigned to the contribution from one
individual quark species.
In the axial vector and the vector case the first
 contribution of this type arises in order $\ordas^2)$
and $\ordas^3)$ respectively.
Each  of them has
 a charge structure different
 from the one in Eq.~(\ref{not3}).
The lowest order term is therefore ultraviolet  finite.
Furthermore, singlet contributions are
 separately invariant under renormalization
 group transformations.
 These diagrams are obviously absent
in charged-current-induced processes
like the $W$ decay.

The functions $R^{\stand{k}}$ are therefore conveniently
decomposed as follows:
\beq
R^{\rmfont{V}} = 3\left[ \sum_{\stand{f}} v_{\stand{f}}^2
r^{\rmfont{V}}_{\rmfont{NS}}(f)
            + \sum_{f,f'}v_{\stand{f}} v_{f'}
             r^{\rmfont{V}}_{\rmfont{S}}(f,f')
      \right]
{}\, .
\eeq
It will be shown below that $r^{\rmfont{V}}_{\rmfont{S}}(f,f')$
 is independent
of $f$ and $f'$ (meaning the respective
quark masses) up to terms of order
$\alpha_s^4 m_{\rmfont{q}}^2/s$,
where $\rmfont{q}$ stands for one of the five light quarks.
Hence
\beq
R^{\rmfont{V}}
\approx 3\left[\sum_{\stand{f}} v_{\stand{f}}^2
r^{\rmfont{V}}_{\rmfont{NS}}(f) +
      (\sum_{\stand{f}} v_{\stand{f}})^2
         r^{\rmfont{V}}_{\rmfont{S}}\right]
{}\, .
\eeq
The functions
$r^{\rmfont{V}}_{\rmfont{NS}}$
and $r^{\rmfont{V}}_{\rmfont{S}}$ are independent
of the quark charges and arise identically in the
decompositions of $R^{{\rm int}}$ and
 $R^{{\rm em}}$:
\beq
\ba{ll}
\dsp
R^{{\rm int}}
& \dsp
= 3 \left[\sum_{\stand{f}} v_{\stand{f}}
Q_{\stand{f}} r^{\rmfont{V}}_{\rmfont{NS}}(f)
+ \sum_{f,f'}v_{\stand{f}} Q_{f'}
r^{\rmfont{V}}_{\rmfont{S}}(f,f')\right]
\\
& \dsp
\approx 3 \left[\sum_{\stand{f}} v_{\stand{f}} Q_{\stand{f}}
r^{\rmfont{V}}_{\rmfont{NS}}(f)
+ (\sum_{\stand{f}} v_{\stand{f}})(\sum_{f'} Q_{f'})
r^{\rmfont{V}}_{\rmfont{S}}\right]
\\
\dsp
R^{{\rm em}}
& \dsp
= 3 \left[\sum_{\stand{f}} Q_{\stand{f}}^2
r^{\rmfont{V}}_{\rmfont{NS}}(f)
+ \sum_{f,f'}Q_{\stand{f}} Q_{f'}
r^{\rmfont{V}}_{\rmfont{S}}(f,f')\right]
\\
& \dsp
\approx 3 \left[\sum_{\stand{f}}
Q_{\stand{f}}^2 r^{\rmfont{V}}_{\rmfont{NS}}(f)
    + (\sum_{\stand{f}} Q_{\stand{f}})^2
       r^{\rmfont{V}}_{\rmfont{S}}\right]
{}\, .
\ea
\eeq
A similar decomposition can be derived for $R^{\rmfont{A}}$:
\beq
R^{\rmfont{A}} =
3\left[ \sum_{\stand{f}} a_{\stand{f}}^2 r^{\rmfont{A}}_{\rmfont{NS}}(f)
            + \underbrace{\sum_{f,f'}a_{\stand{f}} a_{f'}
     r^{\rmfont{A}}_{\rmfont{S}}(f,f')}_{r^{\rmfont{A}}_{\rmfont{S}}}
      \right]
{}\, .
\eeq
%
{
\begin{figure}[t]
\begin{center}
\begin{tabular}{lll}
\parbox{3cm}{
\epsfig{file=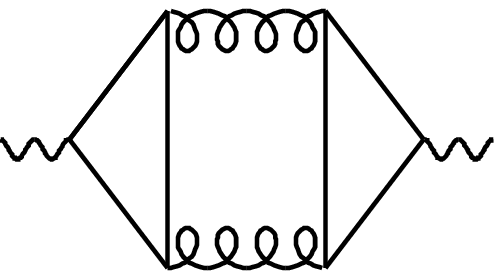,width=5.cm,height=5.cm}
            }
&
\hphantom{XXXXXXX}
&
\parbox{3cm}{
\epsfig{file=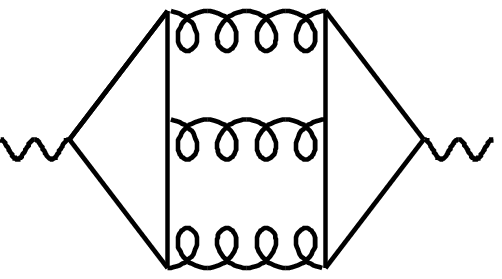,width=5.cm,height=5.cm}
            }
\end {tabular}
\end{center}
\caption[]{\label{class}
Singlet contribution of order $\ordas^2)$
and $\ordas^3)$.}
\end{figure}
}
\noindent
\hspace*{-0.3cm}In the limit of massless $\rmfont{u,d,s}$ and $\rmfont{c}$
  quarks the second term receives contributions from
$f,f'=b$ or $\rmfont{t}$ only, or --- more precisely
--- the light
quarks compensate mutually.
The advantage of this decomposition becomes
even more manifest in the limit $m_{\rmfont{q}}^2/s\rightarrow 0$.
Then the nonsinglet functions
 $r^{\rmfont{V}}_{\rmfont{NS}}$ and $r^{\rmfont{A}}_{\rmfont{NS}}$
are identical and the corrections for non-vanishing, but
small, masses are easily calculated.
\sect{\protect $\beta$ Function and Anomalous Dimensions
\label{renorm} }
In this section several aspects of the
renormalization procedure in QCD are
recalled, which will be of importance
for the subsequent calculations.  The
renormalization of various currents and
the corresponding current correlators
will be considered.  Green functions
with the insertion of two external
currents require subtractive
renormalization.  The corresponding
renormalization constants lead to
anomalous dimensions for the
correlators.   The  presentation will be
rather short, and for more detail  the
reader should consult, for example,
Refs.~[5--8].

\subsect{Renormalization in QCD
\label{QCD}}
The  QCD Lagrangian is given by
\begin{eqnarray}
\EQN{r1}
{\cal L }\{\bPhi;\unl{g},\mu\}&= &
{\cal L}
\{A_{\mu}^{\stand{a}},\Psi,\ovl{\Psi},
C^{\stand{a}},\ovl{C}^{\stand{a}};g,{\bf m},\xi,\mu\}
\nonumber\\
\dsp
& =&  \dsp
-\frac{1}{4}G^{\stand{a}}_{\mu\nu}G_{\stand{a}}^{\mu\nu}
+\overline{\Psi}(i\not\!\!D -{\bf m})\Psi
+  {\cal L}^{\rm GF}
+  {\cal L}^{\rm FP}
{}\, ,
\end{eqnarray}
\begin{eqnarray}
G^{\stand{a}}_{\mu\nu}&=&\prd_\mu A^{\stand{a}}_\nu -
\prd_\nu A^{\stand{a}}_\mu
+
g\mu^\ep f^{abc} A^{\rmfont{b}}_\mu A^{\rmfont{c}}_\nu \, ,
\nonumber\\
\dsp
D_\mu&=&\prd_\mu -ig\mu^\ep
           A^{\stand{a}}_\mu  \frac{\lambda^{\stand{a}}}{2}\, ,
\nonumber\\
\nabla^{ab}_\mu&=& \delta^{ab}\prd_\mu
                - g\mu^\ep f^{abc} A^{\rmfont{c}}
{}\, .
\EQN{r1b}
\end{eqnarray}
Here
$f^{abc}$ are the structure constants of the colour
$SU_{\rmfont{C}}(3)$ group, $\xi$ is  the gauge parameter,
$G^{\stand{a}}_{\mu\nu}$
is the
gluon field strength tensor
and $C^{\stand{a}}$ is the ghost fields, with
\beq
{\cal L}^{\rm GF}
=
-\frac{1}{2\xi} (\prd_\nu A_\nu)^2
\ \ \
\rmfont{and}
\ \ \
{\cal L}^{\rm FP}
=
(\prd_\mu \ovl{C})
\nabla_\mu{C}
{}\, .
\EQN{GFandFP}
\eeq
The quark masses are denoted as
${\bf m} = \{m_{\rmfont{q}}\}$;
$\Psi=\{\Psi_{\rmfont{q}}| q = \rmfont{u,d,s,c,b,t}\}$
represents the quark fields;  while
$\bPhi$ stands  for the collection of all
fields, and $\unl{g} = \{g,{\bf m},\xi\}$ for
the `coupling constants'.  Anticipating
the use of  dimensional  regularization,
the unit mass $\mu$  has been introduced in
\re{r1}  to keep
the coupling constant $g$ dimensionless even if the
Lagrangian is considered in
$D = 4 -2       \ep$ dimensions.

A convenient representation of all
(connected) Green functions is provided
by the   generating functional
\beq
Z_{\rmfont{c}}(I;\Sbf) =
\left\{
\int [\dif\bPhi]
\exp \,(
     i I + \bPhi\cdot\Sbf
    )
\right\}_{\rmfont{c}} \,\, ,
\EQN{gen.func.def}
\eeq
with  the normalization condition
$Z_{\rmfont{c}} (I,0) \equiv 1 $.
Here the action
\beq
I(\bPhi,\unl{g},\mu) = \int {\cal L}(\bPhi,\unl{g}) {\rm d} x
\EQN{action}
\eeq
and
the functional integration
is to be understood  in the standard
manner within the perturbation theory
framework.

Finite Green functions
can be constructed from the Lagrangian
Eq.~(\ref{r1}) in three equivalent ways:
The first method is based on the
renormalized Lagrangian,
obtained from the original
one by a rescaling of fields and parameters,
expressing them in terms of renormalized
quantities:
\beq
{\cal L }_{R}\{\bPhi;\unl{g},\mu\} =
{\cal L } \{Z_3^{\frac{1}{2}}
A_{\mu}^{\stand{a}},Z_2^{\frac{1}{2}}\Psi,
Z_2^{\frac{1}{2}}\ovl{\Psi},
\tilde{Z}_3^{\frac{1}{2}}
C^{\stand{a}},\tilde{Z}_3^{\frac{1}{2}}\ovl{C}^{\stand{a}},
Z_\xi \xi,
Z_{\stand{g}} g ,{ Z}_{\stand{m}} m\}
\EQN{r4}
{}\, .
\eeq
The explicit form of the
 renormalization constants depends on the
renormalization scheme adopted.
The most  powerful method, which is
particularly suitable for applications in
 QCD, the procedure of dimensional
regularization \cite{dim.reg-a,dim.reg-b}
 and minimal subtraction
\cite{ms}
is nowadays widely used.
After  continuation of the Feynman integrals
 to $D=4-2\epsilon$ space--time
dimensions  divergences reappear as
 poles in $\epsilon$.
The renormalization constants
may then be expanded
in the coupling constant
\beq
\EQN{r4b}
 Z=1+\sum_{i,j}^{0<j\leq i}Z_{ij}
\left(\api\right)^i
\frac{1}{\ep^{\stand{j}}}
{}\, .
\eeq
In the minimal subtraction scheme
(we will use the $\msbar$-scheme \cite{MSbar}
throughout in this work) the coefficients
$Z_{ij}$ are just dimensionless constants.
There exists a  choice of the renormalization
constants such that
every Green function of elementary
fields computed with the help of $\L_{\rmfont{R}}$ is
finite in the limit $\ep\to0$ in
every order of perturbation theory.
Hence, too,   the generating functional
\beq
Z_{\rmfont{c}}(I_{\rmfont{R}};\Sbf) \ \ \rmfont{\rm      with}
 \ \ I_{\rmfont{R}}
 =
 \int \L_{\rmfont{R}} {\rm d} x
\label{r4c}
\eeq
is finite  in every order of perturbation theory.

An example of a renormalized
finite Green function obtained from $\L_{\rmfont{R}}$ is the
two-point function of two quark fields,
namely the renormalized quark propagator
\beq  \EQN{r5}
S(p,g ,m,\mu) \equiv i\int  {\rm d}
e^{ipx}\langle 0|\;
T\; [q(x)\overline q (0)]
\; |0\rangle
{}\, ,
\eeq
whose contribution  will serve to define the quark field
renormalization constant $Z_2$.

A second calculational method is based on
the bare Lagrangian
\beq
\L_{\rmfont{B}}\{\bfPhi_{\rmfont{B}};\unl{g}_{\rmfont{B}} \}
\equiv
\L\{\bfPhi_{\rmfont{B}};\unl{g}_{\rmfont{B}}, 1 \}
\EQN{r6}
\eeq
\ice{
\beq \EQN{r6}
\L_{\rmfont{B}}\{A_{B,\mu}^{\stand{a}},
\Psi_{\rmfont{B}},\ovl{\Psi}_{\rmfont{B}},
C_{\rmfont{B}}^{\stand{a}},\ovl{C}_{\rmfont{B}}^{\stand{a}},
g_{\rmfont{B}},{\bf m_{\rmfont{B}}}\},
\eeq
}
and the resulting generating functional of bare Green
functions
\beq
Z_{\rmfont{c}}(I_{\rmfont{B}};\Sbf_{\rmfont{B}}) =
\left\{
\int [\dif \bPhi_{\rmfont{B}} ]
\exp \, (
     i I_{\rmfont{B}} + \bPhi_{\rmfont{B}}\cdot\Sbf_{\rmfont{B}}
     )
\right\}_{\rmfont{c}} \,\, ,
\EQN{gen.func.bare}
\eeq
with  the
bare action
\beq
I_{\rmfont{B}}(\bPhi_{\rmfont{B}},\unl{g}_{\rmfont{B}})
 =
\int \L_{\rmfont{B}}\{\bPhi_{\rmfont{B}},\unl{g}_{\rmfont{B}},1\}
{\rm d} x
{}\, .
\EQN{bare_action}
\eeq
The functional change of variables
\beq
 A_{B,\mu}^{\stand{a}}  = (Z_3)^{1/2}A^{\stand{a}}_{\mu},\;\;
 \Psi_{\rmfont{B}}  =(Z_2)^{1/2}\bfpsi, \;\;
 \ovl{\Psi}_{\rmfont{B}}  =(Z_2)^{1/2}\ovl{\Psi}, \;\;
 C_{\rmfont{B}}^{\stand{a}}  =(\tilde{Z}_3)^{1/2}C^a
\EQN{r7a}
\eeq
leads to the immediate  conclusion
that
\beq
Z_{\rmfont{c}}(I_{\rmfont{B}};\Sbf_{\rmfont{B}})
\equiv
Z_{\rmfont{c}}(I_{\rmfont{R}};\Sbf)
\EQN{r7b}
{}\, ,
\eeq
{\em provided}
bare and renormalized sources
and parameters are related
through
\beq
\ba{c}
\dsp
 (S_{\rmfont{A}}^{\mu a})_{\rmfont{B}}
=  (Z_3)^{-1/2}S_{\rmfont{A}}^{\mu a}\,  ,
\ \ \
 (S_{\rmfont{C}}^{\stand{a}})_{\rmfont{B}}
= (\tilde{Z}_3)^{-1/2}S_{\rmfont{C}}^{\stand{a}} \, \, ,
\ \ \
 (S_{\ovl{C}}^{\stand{a}})_{\rmfont{B}}
= (\tilde{Z}_3)^{-1/2}S_{\ovl{C}}^{\stand{a}} \, \,  ,
\\
\dsp
 (S_{\ovl{\Psi}})_{\rmfont{B}} =  (Z_2)^{-1/2}S_{\ovl{\Psi}}\, \,  ,
\ \ \ \  \ \
 (S_{\Psi})_{\rmfont{B}} =(Z_2)^{-1/2}S_{\Psi}
\ea
\EQN{r7c}
\eeq
and
\beq
g_{\rmfont{B}}   = \mu^{\ep}Z_gZ_3^{-1/2}Z_2^{-1}g\, ,
\ \ \ \
\;\;
\m_{\rmfont{B}}   = Z_{\stand{m}} Z_2^{-1}\m\, ,
\;\;
\ \ \ \
\xi_{\rmfont{B}}  = (Z_3)^{-1}Z_\xi \xi\, .
\EQN{r7d}
\eeq

The  bare  Green function
corresponding to Eq.~(\ref{r5}) is given by
\beq
S_{\rmfont{B}}(p,g_{\rmfont{B}},m_{\rmfont{B}}) = i\int
 \rmfont{d} x e^{ipx}\langle 0|\;
T [ {{q}}_{\rmfont{B}}(x) \ovl{{q}}_{\rmfont{B}}(0) ]
\; |0\rangle
{}\, .
\EQN{r8}
\eeq
Equations (\ref{r5}) and (\ref{r8}) show that
after having introduced a
renormalized coupling constant, masses and
gauge-fixing parameters,  all remaining
divergences of the Green function
can
be eliminated by wave function renormalization:
\beq
 S(p,g,m,\mu)=Z_2^{-1}S_{\rmfont{B}}(p,g_{\rmfont{B}},m_{\rmfont{B}})
{}\, .
\EQN{r9}
\eeq

A third way of obtaining  finite Green functions
is  based on the so-called $R$-operation
 --- a recursive subtraction scheme --- to
remove  ultraviolet (UV) divergences
from a given (arbitrary)  Feynman integral in
a way compatible with adding local counterterms to the
Lagrangian\footnote{A good pedagogical introduction  to the
apparatus of the  $R$-operation in the MS-scheme and its\break\hfill\indent
$\,\,\,\,\,$applications may be found in Refs.~\cite{JCC84,Kennedy82}.}.
Using the $R$-operation the renormalized generating functional
\re{r4c} can  be conveniently presented in  the form:
\beq
Z_{\rmfont{c}}(I_{\rmfont{R}};\Sbf) =
Z^{\rmfont{R}}_{\rmfont{c}}(I;\Sbf)
\equiv
R_{\overline{\rmfont{MS}}}
\left\{
\int [\dif\bPhi]
\exp \, (
     i I + \bPhi\cdot\Sbf
    )
\right\}_{\rmfont{c}} \,\, .
\EQN{r10}
\eeq\subsect{Running Coupling Constant and  Masses
\label{running}}
In comparison with the classical  Lagrangian \re{r1}
(now  considered  in the physical $D=4$ number of
space--time dimensions)
the renormalized one  \re{r4}  depends on an additional parameter
--- the  {}'t Hooft unit mass $\mu$.
This     naturally leads to the well-known
renormalization group (RG) constraint: any
physical  prediction  (that is
measurable at least in principle) obtained with the help
of \re{r4} must not depend on the value of  $\mu$ {\em provided}
bare fields and parameters are kept fixed. If $P(\as,\mbf,\mu)$
denotes a physical quantity computed with the Lagrangian \re{r4}
then it must meet the RG equation
\beq
\dmu P(\as,\mbf,\mu) \equiv 0
\, ,
\EQN{rg-eq}
\eeq
where
\beq
   \mu^2\frac{d}{d \mu^2} \equiv \dmu
|{{}_{\displaystyle
   g_{\scriptscriptstyle \rmfont{B}},
   m_{\scriptscriptstyle \rmfont{B}} }}
\label{rgop}
\eeq or, equivalently,
\beq
   \mu^2\frac{d}{d\mu^2} = \mu^2\frac{\partial}{\partial\mu^2} +
   \pi\beta(\alpha_s) \frac{\partial}{\partial \alpha_s}
 +  2\ovl{m}^2
\gm(\alpha_s)\frac{\partial}{\partial \ovl{m}^2}
\, \, .
\label{rgop2}
\eeq
Note that we  follow the  common convention by denoting
\[
\alpha_s  \equiv \frac{g^2}{4\pi}
\, \, .
\]
In addition, in complicated formulae
we shall for brevity use the {\em couplant } $a$
defined as
\[
a
\equiv  \frac{\alpha_s}{\pi}
= \frac{g^2}{4\pi^2}
\, .
\]
The  $\beta$-function
and the quark mass anomalous dimension $\gm$
are
\beq
\dmu \left[ \frac{\as(\mu)}{\pi}
\right]|{{}_{\displaystyle g_{\scriptscriptstyle \rmfont{B}},
                           m_{\scriptscriptstyle \rmfont{B}}
        }   }
= \beta(\as) \equiv
-\sum_{i\geq 0}\beta_{\stand{i}}\left(\api\right)^{i+2},
\EQN{beta-def}
\eeq
\beq
\dmu \ovl{m}(\mu)|{{}_{\displaystyle g_{\rmfont{B}}, m_{\rmfont{B}} }}
 =  \ovl{m}(\mu)\gm(\as) \equiv
-\ovl{m}\sum_{i\geq0}\gm^{\stand{i}}\left(\api\right)^{i+1}.
\EQN{anom-mass-def}
\eeq
Their expansion coefficients  to three loops
are well known
\cite{TarVlaZha80,Tar82} and
read ($n_{\stand{f}} $ is the number of quark flavours; note
that the results \re{beta3} and \re{anom.mass3}
have been recently confirmed in
Refs.~\cite{Larin:betaQCD,Larin:massQCD} respectively)
\beq
\ba{c}\dsp
\beta_0=\left(11-\frac{2}{3}~n_{\stand{f}}\right)/4\, ,  \  \
\beta_1=\left(102-\frac{38}{3}~n_{\stand{f}} \right)/16\, , \\ \dsp
\beta_2=\left(\frac{2857}{2}-\frac{5033}{18}~n_{\stand{f}} +
\frac{325}{54~}~n_{\stand{f}} ^2\right)/64\, ,
\ea
\EQN{beta3}
\eeq
\beq
\ba{c}\dsp
\g^0_{\stand{m}}=1\, , \  \ \
\g^1_{\stand{m}}=
\left(\frac{202}{3} - \frac{20}{9}~n_{\stand{f}} \right)/16\, ,
\\ \dsp \g^2_{\stand{m}}=\left\{1249 -
\left[\frac{2216}{27}+\frac{160}{3}~\zeta(3)\right]
n_{\stand{f}} -\frac{140}{81}~n_{\stand{f}}^2\right\}/64\, .
\ea
\EQN{anom.mass3}
\eeq

According to  \re{beta-def} and  \re{anom-mass-def}  both the minimally
renormalized coupling constant $g$ and
quark mass $m_{\rmfont{q}}$ {\em run}
(that is depend on) with $\mu$. This demonstrates clearly  that  $g$
and $m$ are just parameters entering the QCD Lagrangian and that
their connection to measurable physical quantities is not direct.
%
%
In this sense the $\msbar$ renormalization scheme is not unique
or distinguished by  physical considerations.
However,  it allows one to employ the RG equation \re{rg-eq}  in
order to very efficiently and conveniently  `improve' the
perturbation expansion by neatly summing up potentially dangerous
logarithms of momenta and masses appearing in higher-orders.

Indeed, in any  necessarily  finite order of
perturbation theory,  the master RG
Eq. \re{rg-eq} is met only partially: that is,
its r.h.s.  does not
vanish but rather is a polynomial in the coupling constant
that includes
only terms of  order  higher than the one taken into account in
the calculation of $P$. If  additionally   the characteristic scale $Q$
on   which the physical quantity $P$ depends
is  taken large,  then,  as is
well known, the general structure of $P$ may be visualized as
follows --- it is understood that the $P$ starts from  an $\alpha_s$
independent constant, with  all  power (suppressed) mass effects are
neglected for the moment  ---
\beq
P =
\sum_{n_1 < n_2}
c_{n_1 n_2}
\left(
\ln\,\frac{\mu^2}{Q^2}
\right)^{n_1}
\left(
\frac{\alpha}{\pi}\right)^{n_2}
{}\, .
\EQN{large-Q}
\eeq
Even if  a
renormalization prescription has  already been
specified,
there remain  two problems:
the residual $\mu$
dependence and  the invalidation of the
perturbation expansion by large logarithms of $Q$
irrespective of the smallness of the
initial value of the coupling constant.
The well-known solution of both
problems is to fix the value of $\mu$
to be of order $Q$ or, in many cases,
just equal to $Q$.  Such a prescription
clearly eliminates the dangerous momentum
logs and, as a side effect, helps to
specify the value of  $\mu$ .

Of course, the above consideration cannot fix the $\mu$ value
exactly. In other words, in the limit of asymptotically large $Q$ the
choices of $\mu =Q$ or $\mu = \sqrt{2}Q $ are {\em mathematically}
equivalent but still   generally lead to slightly different predictions.
A considerable amount of literature discussing this
ambiguity exists and
various recipes  for  overcoming  it
have been suggested
(we cite only a few of them
[18--23]).
Below,
a  commonly accepted and pragmatic approach will be adopted:
the $\msbar$ scheme will be chosen with $\mu$
set to a characteristic momentum of the problem at hand and, finally,
$\mu$ will be varied somewhere around the scale to test the
sensitivity of the result with respect
to not-yet-computed corrections  of higher
order.


In what follows we will always identify the $\msbar$  quark mass
with the running one and occasionally denote the latter  with a bar.
If not stated otherwise a  running mass without an argument will be
understood as taken at scale $\mu$; so that
\[
m(\mu) \equiv
\ovl{m}(\mu) =
\ovl{m}
{}\, .
\]

We finish this subsection by writing out  the
explicit solution for the
running $\alpha_s$  and  quark mass.
%
The solution of Eq.~(\ref{beta-def}) reads
($L\equiv \ln\, \mu^2/\Lambda^2_{\overline{MS}}$)
\beq
\frac{\as(\mu)}{\pi} = \frac{1}{\beta_0 L} \left\{
1 - \frac{1}{\beta_0L}~\frac{\beta_1\ln\, L}{\beta_0}
 + \frac{1}{\beta_0^2L^2} \left[\frac{\beta_1^2}{\beta_0^2}
(\ln^2L -
\ln\, L - 1) + \frac{\beta_2}{\beta_0} \right] \right\}
\eeq
while Eq.~(\ref{anom-mass-def}) is solved by
\renewcommand{\arraystretch}{3}
\beq \EQN{a2}
\ba{rl}
 \dsp
\ovl{m}(\mu ) =
& \dsp
\ovl{m}(\mu_0 )\left[
\frac{\as(\mu )}{\as(\mu_0 )}\right]
^{{\g_m^0}/{\beta_0}}
\left\{ 1+  \left(\frac{\g_m^1}{\beta_0}
-\frac{\beta_1\g_m^0}{\beta_0^2}\right)
\left[\frac{\as(\mu )}{\pi}
-\frac{\as(\mu_0 )}{\pi}\right]
\right.
\\ & \dsp
  +~\frac{1}{2}
  \left(\frac{\g_m^1}{\beta_0}
-\frac{\beta_1\g_m^0}{\beta_0^2}\right)^2
\left[\frac{\as(\mu )}{\pi}
-\frac{\as(\mu_0 )}{\pi}\right]^2
\\ & \dsp
 \left.
  +~\frac{1}{2}
\left( \frac{\g_m^2}{\beta_0}
-\frac{\beta_1\g_m^1}{\beta_0^2}
   -\frac{\beta_2\g_m^0}{\beta_0^2}
+\frac{\beta_1^2\g_m^0}{\beta_0^3}\right)
\left(\left[\frac{\as(\mu )}{\pi}\right]^2
-\left[\frac{\as(\mu_0 )}{\pi}\right]^2\right)
\right\}
{}\ .
\ea
\EQN{run-mass}
\eeq
\renewcommand{\arraystretch}{2}
\ice{
\beq \ba{rl}\dsp
\ovl{m}(\mu)
=& \dsp
\hat{m}\left(\frac{\beta_0}{2}
\frac{\as(\mu )}{\pi}\right)^{\g_m^0/\beta_0}
\left\{ 1+\apimu
\left[\frac{\g_m^1}{\beta_0}-\frac{\beta_1\g_m^0}{\beta_0^2}\right]\right.
\\[3ex] &\dsp
  +\frac{1}{2} \left(\apimu\right)^2\left[
  \left[\frac{\g_m^1}{\beta_0}-\frac{\beta_1\g_m^0}{\beta_0^2}\right]^2
 \left.
+  \frac{\g_m^2}{\beta_0}-\frac{\beta_1\g_m^1}{\beta_0^2}
       -\frac{\beta_2\g_m^0}{\beta_0^2}+\frac{\beta_1^2\g_m^0}{\beta_0^3}
\right] \right\}
{}\, .
\ea
\EQN{run-mass}
\eeq
}
\subsect{$\protect\msbar$  Mass Versus Pole  Mass
\label{versus}}
There are situations  where it is convenient to  deal with quark mass
definitions different from that  given by
the $\protect\msbar$ scheme.  For
instance,  for very heavy quarks a non-relativistic description is
believed to be relevant. In this case   the pole mass seemingly should
be used.  The pole mass
$M_{\rmfont{q}}$
presents a gauge-invariant, infrared-finite,
scheme-independent object  which is defined  as the position of pole
of a  renormalized quark propagator.  It should be heavily emphasized
that by definition the renormalized quark propagator  is to be
understood in   a strictly perturbative framework.

There is the  firm belief that non-perturbative effects should change
significantly the pole structure of the propagator of even a quite
heavy quark.  This belief has been   supported by
the  recent
observation \cite{Shifman94} that
there are some  non-perturbative effects in
the heavy quark propagator which defy their description  in terms of
familiar vacuum condensate contributions. However, with the
qualification above the pole, mass remains a valuable characteristic
of heavy quark masses.

The explicit relation between both masses was  obtained in
Refs. [25--29].
The most advanced calculation presented  in Ref.~\cite{GraBro90}
leads to the following result:
\begin{eqnarray}
\dsp
\ovl{m}_{\rmfont{q}}(\mu)
&=&
{M_{\rmfont{q}}}
\left\{
\rule{0mm}{5mm}
\right.
1 -~\apimu\left[
\frac{4}{3}
+
\ln\, \frac{\mu^2}{M_{\rmfont{q}}^2}
\right]
\nonumber
\\
&&-
\dsp
\left[
\apimu
\right]^2
\left[
K_{\rmfont{q}}({\bf M})
- \frac{16}{9}
+
\left(
\frac{157}{24} - n_{\stand{f}} \frac{13}{36}
\right)
\ln\, \frac{\mu^2}{M_{\rmfont{q}}^2}
+
\left(
\frac{7}{8}
-\frac{n_{\stand{f}}}{12}
\right)
 \ln^2 \frac{\mu^2}{M_{\rmfont{q}}^2}
\right]
\nonumber
\\
\dsp
&&+
\left.
{\cal O}(\as^3)
\right\}
{}\, ,
\EQN{m-from-M}
\eea
with
${\bf M} = \{ M_{\stand{f}}\}$ and
\begin{equation}
K_{\rmfont{q}}({\bf M}) =
\frac{3817}{288}
+\frac{2}{3}(2
+\ln\,2)\zeta(2)
-\frac{1}{6}\zeta(3)
-\frac{n_{\stand{f}}}{3}
\left[
\zeta(2)
+\frac{71}{48}
\right]
+\frac{4}{3}
\sum_{f} \Delta\biggl(\frac{M_{\stand{f}}}{M_{\rmfont{q}}}\biggr)
{}\, ,
\label{eq:K}
\end{equation}
and $\Delta(r)$ being a complicated function of $r$.
For our aims it is enough to know that
it has the limiting behaviours \cite{BroadGraySch91}:
\bea
\Delta(r) &\bbuildrel{=\!=\!=}_{r \to \infty}^{} &
\frac{1}{4} \ln^2 r
+\frac{13}{24} \ln\, r
+\frac{1}{4} \zeta(2)
+\frac{151}{288}
+O(r^{-2} \ln\, r)
\label{asymp1}
{}\, ,
\\
\Delta(r ) &\bbuildrel{=\!=\!=}_{r \to 0}^{} &
\frac{3}{4}~r \zeta(2)
+O(r^{2})~,
\label{asymp2}
\eea
with $\Delta(1) = \frac{3}{4}\zeta(2) - \frac{3}{8}$.
Numerically Eq.~~\re{eq:K} reads
\beq
K_{\rmfont{q}}=16.00650 - 1.04137~n_{\stand{f}}
+ \frac{4}{3}\sum_{\stand{f}}
\Delta\left(\frac{M_{\stand{f}}}{M_{\rmfont{q}}}\right)
\EQN{eq:K:num}
\eeq
or, equivalently,
\beq
K_{\rmfont{q}}=17.1514 - 1.04137~n_{\stand{f}}
+ \frac{4}{3}\sum_{f\not= q}
\Delta\left(\frac{M_{\stand{f}}}{M_{\rmfont{q}}}\right)
{} .
\EQN{eq:K:num2}
\eeq
If $0 \le r \le 1$  then the function
$\Delta(r)$ may be conveniently approximated as follows
\beq
\Delta(r) = \frac{\pi^2}{8}~r - 0.597~r^2 + 0.230~r^3
\EQN{K-appr}
\eeq
which is accurate to $1$\%.
\subsect{External Currents
\label{external}}
So far we have addressed the properties
of Green functions of elementary
fields. The discussion may be extended
to Green functions with
insertions of composite operators,
 for which we want to
consider at first  various external currents
coupled to quark fields. Let
\beq  \EQN{j1}
\dsp j(x)=\ovl{{q'}}(x)\G {q}(x)
\eeq
denote a general current
where $\G$ stands for an arbitrary combination
 of $\g$-matrices.
Green functions with the insertion
of one current $j$  remain in general
divergent even after wave function
 renormalization of all
elementary fields,   coupling
constants and masses.
The remaining divergence is removed by
multiplicative renormalization:
\beq
 j_{\rmfont{R}}=Z_{\stand{j}} \ovl{{q}}_{\rmfont{R}}'
\G {{q}}_{\rmfont{R}}
{}\, ,
\eeq
where
\beq
Z_{\stand{j}}=1+ \sum_{i,k}^{0<k\leq i}(Z_{\stand{j}})_{ik}
\left(\frac{\as}{\pi}\right)^{\stand{i}}\frac{1}{\ep^{\stand{k}}}
{}\, \, .
\EQN{j2}
\eeq

The renormalized Green function
$G^{(n)}_{\stand{j}} $ with  insertion of the
external current $j(0)$ is given by
\bea
\dsp
G^{(n)}_{\stand{j}}({\bf p},{\bf m},g,\mu,\ep)
&=&
\int  \left(\prod_{i=1}^n
 d y_ie^{i p_{\stand{i}} y_{\stand{i}}}\right)
\Biggl\langle 0|\; T \;j_{\rmfont{R}}(0) \prod_{i=1}^{\rmfont{n}}
\Phi_{\stand{i}}(y_{\stand{i}})\;|0\Biggl\rangle^{\rmfont{c}}
\EQN{j4}
\\
\nonumber
\dsp
&
\equiv
&
\int  \left(\prod_{i=1}^{\rmfont{n}}
d y_{\stand{i}} e^{i p_{\stand{i}} y_{\stand{i}}}\right)
\left\{
\int
[\dif \bfPhi]
\exp \,(i\int \rmfont{d} x \L_{\rmfont{R}}\{\Phi(x)\})
j_{\rmfont{R}}(0)
\prod_{i=1}^{\rmfont{n}}\Phi_{\stand{i}}(y_{\stand{i}})
\right\}_c
{}\, .
\eea
The bare Green function
 is defined   through the  insertion
 of the bare current
\beq  \EQN{j5a}
j_{\rmfont{B}}(x)=Z_{\stand{j}}^{-1}Z_2 j_{\rmfont{R}}
{}\, ,
\eeq
and in an
analogous manner
\bea
\dsp
G^{(n)}_{j,\rmfont{B}}
({\bf p},{\bf m_{\rmfont{B}}},g_{\rmfont{B}},\ep)
&=&
\int  \left(\prod_{i=1}^{\rmfont{n}} d y_i
e^{i p_{\stand{i}} y_{\stand{i}}}\right)
\Biggl\langle 0|\;T\;j_{\rmfont{B}}(0) \prod_{i=1}^n
 \Phi_{B,i}(y_{\stand{i}})\;|0\Biggl\rangle
\EQN{j5}
\\
\nonumber
\dsp
& \equiv &
\int  \left(\prod_{i=1}^{\rmfont{n}} d y_i
e^{ip_{\stand{i}} y_{\stand{i}}}\right)
\int  [\dif \bfPhi]_{\rmfont{B}}
 \exp \,(i\int \rmfont{d} x
\L_{\rmfont{B}}\{\Phi_{\rmfont{B}}(x)\})
j_{\rmfont{B}}(0)\prod_{i=1}^{\rmfont{n}} \Phi_{B,i}(y_{\stand{i}})
{}\,\, .
\eea

Comparing eqs.(\ref{j4}) and (\ref{j5}) one
may relate bare and renormalized Green
functions
\beq
\EQN{j6}
G^{(n)}_{\stand{j}}({\bf p},{\bf m},g,\mu,\ep)=(Z_{\stand{j}}/Z_2)
\prod_{i=1}^{\rmfont{n}}(Z_{\stand{i}})^{-1/2} \, G^{(n)}_{j,\rmfont{B}}
({\bf p},{\bf m}_{\rmfont{B}},g_{\rmfont{B}},\ep)
{}\, .
\eeq
The connection between the renormalized and the bare
Green functions allows the  derivation
of the current renormalization constant
$Z_{\stand{j}}$
and its  anomalous dimension
\beq
\g_{\stand{j}}=\mu^2\frac{d}{d\mu^2}
\ln\,\left(\frac{Z_{\stand{j}}}{Z_2}\right)
{}.
\EQN{j3}
\eeq
The relation between bare and renormalized
quark propagator
 Eq.~(\ref{r9}) can be combined with
 Eq.~(\ref{j6}) employed for the special case of
the vertex function
\beq   \EQN{j7}
G^{(2)}_{\stand{j}}(p_1,p_2)
=
i^2
\int dy_1dy_2
 e^{ip_1y_1-ip_2y_2}
\langle 0|\; T \;q(y_1)
 j(0)\ovl{q}(y_2)\;|0\rangle
\, ,
\eeq
to discuss the renormalization
  of  scalar,
pseudoscalar, vector and axial vector
currents respectively:
\beq  \EQN{j8}
j_{\rmfont{S}}=  \ovl{q}'q\, \, ,\qquad
j_{\rmfont{P}} =  \ovl{q}'i\g_5 q\, \, ,\qquad
j^{\rmfont{V}}_{\mu} =  \ovl{q}'\g_{\mu} q\, \, ,\qquad
j^{\rmfont{A}}_{\mu} =  \ovl{q}'\g_{\mu}\g_5 q
\, .
\eeq
For the diagonal {\it vector current}
$j^{\rmfont{V}}_{\mu} = \ovl{q}\g_{\mu} q$
the Ward--Takahashi identity
\beq   \EQN{j9}
 (p_1-p_2)^\alpha
G^{(2)}_{\rmfont{V}\alpha}(p_1,p_2,g,m,\mu)=
S(p_1,g,m,\mu)-S(p_2,g,m,\mu)
\eeq
can be employed.
A corresponding identity relates the
bare vertex function
with the bare quark propagator
in the same manner,
which implies in view of Eqs.~(\ref{r9})
and (\ref{j6}) for the renormalization
constant and the vector current anomalous dimension
\beq\EQN{j10}
Z_{\rmfont{V}} = Z_2\,\, ,\;\;\;
\g_{\rmfont{V}} =0
{}\, .
\eeq
These identites are
 valid also for a
nondiagonal vector current $j^{\rmfont{V}}_{\mu} =
 \ovl{q}'\g_{\mu} q$
composed of two different quark fields,
 because $Z_{\rmfont{V}}$ does not depend on the
quark mass.

In the case of the {\it axial vector current}
one may, in a first step, ignore the
 axial anomaly and
assume a Hermitian,
anticommuting $\g_5$
($\g_5^2=1, \{\g_5,\g_{\mu}\}=0$).
 As long as diagrams are considered
which involve either nondiagonal
 axial currents or
diagrams without traces of an
 odd number of $\g_5$ matrices,
these assumptions are justified and lead to
\beq \EQN{j11}
Z_{\rmfont{A}} = Z_2~,\;\;\;
\g_{\rmfont{A}} = 0
{}\, .
\eeq
The more involved discussion of the
correct treatment of
$\g_5$ in $D$ dimensions
is given
below  in
Section~\ref{gamma5}.

The quark propagator
and the two-point Green function
with a {\it scalar current} insertion
are related by
\beq   \EQN{j12}
G^{(2)}_{\rmfont{S}}(p_1,p_1,g,m,\mu)=-~\frac{\pd}{\pd m}
S(p_1,g,m,\mu)
{}\, .
\eeq
{}From the comparison of  this identity with the
analogous identity for the
bare Green functions one then obtains
\beq  \EQN{j13}
Z^{\rmfont{S}}= Z_{\stand{m}} Z_2
{}\, .
\eeq
The scalar current therefore has
a non-vanishing anomalous dimension.
This result holds true also
for nondiagonal currents.

With the same qualifications as
discussed above for the axial vector current
and with similar arguments
one obtains for the  {\it pseudoscalar current}
\beq  \EQN{j14}
Z^{\rmfont{P}}=Z_{\stand{m}} Z_2
{}\, .
\eeq
Whereas neither scalar
nor  pseudoscalar
 currents are  RG invariant, this holds
true for the combinations
$m_{\rmfont{B}} j^{\rmfont{S/P}}_{\rmfont{B}}
= m_{\rmfont{R}} j^{\rmfont{S/P}}_{\rmfont{R}}
$.
\subsect{Current Correlators
\label{correlators}}
The renormalization
properties and anomalous dimensions
of the two-point current correlators,
defined  as the vacuum
expectation value of the time-ordered
product of the respective currents,
will be discussed in this section. In
coordinate space the renormalized correlator
of a nondiagonal current $j=\ovl{q}'\Gamma q$
is given by
\[
 \tilde{\Pi}_{\stand{j}}(x,g ,m,m',\mu)=
\langle 0|\; T\;j(x) j^{\dagger}(0)\; |0\rangle
{}\, .
\]
The masses of the quark fields $\rmfont{q}$ and $\rmfont{q}'$
are denoted by
$m$ and $m'$
and all other quarks are assumed to be
massless. The correlator
 contains two renormalized composite
operators and hence is finite in the limit
$\ep\rightarrow 0$ as long as $x\neq 0$.
 However, in the vicinity
of the point $x=0$ this
function does exhibit singularities which
 are not   removed  by
 renormalization
of the coupling constant, the quark masses,
 the quark fields and the current
as discussed above.
In momentum space the renormalized
polarization function
$\Pi_{\stand{j}}(q,g ,m,m',\mu)  $
is thus    obtained
from its bare counterpart by adding
new
renormalization constants.
For the vector
 current correlator two independent
constants appear:
\bea
\dsp
\!\!\!\!\!\!\Pi^{\rmfont{V}}_{\mu\nu}(q,g,m,m',\mu) \;
&=&
\;
i\int \rmfont{d} x e^{iqx}  \langle 0|\; T\;j_{\mu}(x)
 j^{\dagger}_{\nu}(0)\; |0\rangle
=
\dsp\mu^{2\ep}\Pi^{\rmfont{V}}_{B,\mu\nu}
(q,g_{\rmfont{B}},m_{\rmfont{B}},m_{\rmfont{B}}')
\nonumber
\\
\dsp
&&+~(q_{\mu} q_{\nu}
 -g_{\mu\nu}q^2)Z_{{q}}^\rmfont{VV}\frac{1}{16\pi^2}
+ g_{\mu\nu}({m-m'})^2
Z^\rmfont{VV}_{\stand{m}}\frac{1}{16\pi^2}
{}\, .
\EQN{cc2}
\eea
The transversality  of the
polarization operator for $m=m'$
is explicitly taken into account.
The factor $\mu^{2\ep}$ is
introduced  in order
to make the dimension of the function
$\Pi^{\rmfont{V}}_{\mu\nu}(q,g ,m,m',\ep)$ independent
of $\ep$ and the
factors $1/(16\pi^2)$ have been
 introduced for convenience.

The subtractive renormalization
constants $Z_{\stand{q}}^\rmfont{VV}, Z_{\stand{m}}^\rmfont{VV}$ can be
expanded in the coupling constant
 and have the following form in the
minimal subtraction scheme:
\beq
\ba{ll}\EQN{cc3}
\dsp
 Z_{\stand{q}}^\rmfont{VV} & \dsp =\sum_{0\leq
j-1\leq i}(Z_{\stand{q}}^\rmfont{VV})_{ij}\left(\frac{\alpha_s}
{\pi}\right)^{i}\frac{1}{\ep^{\stand{j}}}
{}\, ,
\\
\dsp
Z_{\stand{m}}^\rmfont{VV}& \dsp =\sum_{0\leq j-1\leq i}
(Z_{\stand{m}}^\rmfont{VV})_{ij}
\left(\frac{\alpha_s}{\pi}\right)^{i}\frac{1}{\ep^{\stand{j}}}
{}\, .
\ea
\eeq
The dimensionless expansion coefficients
$(Z_{\stand{q}}^\rmfont{VV})_{ij}$ and
$(Z_{\stand{m}}^\rmfont{VV})_{ij}$ are pure numbers.
The quadratic dependence of the subtractions in
Eq.~({\ref{cc2}) on the
momentum $q$ and the quark masses
is a trivial consequence of
the mass dimension of the
function $\Pi^{\rmfont{V}}_{\mu\nu}$ and the fact that
renormalization constants
are polynomial in masses and momenta
\cite{Collins75}.

Applying
$\dsp \mu^2 \frac{d}{d \mu^2}$
to both sides of Eq.~(\ref{cc2}),
one obtains the  RG equation
\beq
\EQN{cc4}
\mu^2 \frac{d}{d \mu^2}\Pi^{\rmfont{V}}_{\mu\nu}
(q,g,m,m',\mu)=
(q_{\mu} q_{\nu}  -g_{\mu\nu}q^2)
\g_{\stand{q}}^\rmfont{VV}\frac{1}{16\pi^2} +
g_{\mu\nu}({m-m'})^2 \g^\rmfont{VV}_{\stand{m}}\frac{1}{16\pi^2}
{}\, ,
\eeq
with the anomalous dimensions
\beq
\ba{ll}
\g^\rmfont{VV}_{\stand{q}} & \dsp =\mu^2
 \frac{d}{d \mu^2}(Z^\rmfont{VV}_{\stand{q}})-\ep
Z^\rmfont{VV}_{\stand{q}}
{}\, ,
\\
\dsp \g^\rmfont{VV}_{\stand{m}} & \dsp =
\mu^2
\frac{d}{d \mu^2}(Z^\rmfont{VV}_{\stand{m}})
-\ep Z^\rmfont{VV}_{\stand{m}}
+
2\g_mZ^\rmfont{VV}_m
{}\, \, .
\ea
\EQN{cc5}\dsp
\eeq
After insertion
of Eq.~(\ref{cc3}) one observes
that $\g_{\stand{q}}^\rmfont{VV}$ and $\g_{\stand{m}}^\rmfont{VV}$ are
already completely
determined by the coefficients of the
simple poles $1/\ep$ in the expansion
of the renormalization constants:
\beq
\ba{ll}
\EQN{cc6}
\dsp
\g^\rmfont{VV}_{\stand{q}} & \dsp =-\sum_{i\geq 0}(i+1)
(Z^\rmfont{VV}_{\stand{q}})_{i1}
\left(\frac{\alpha_s}{\pi}\right)^{i}
{}\,  ,
\\
\dsp
 \g^\rmfont{VV}_{\stand{m}} & \dsp =-\sum_{i\geq 0}(i+1)
(Z^\rmfont{VV}_{\stand{m}})_{i1}
\left(\frac{\alpha_s}{\pi}\right)^{i}
{}\, .
\ea
\eeq
For the axial current correlator
\beq   \EQN{cc7}
\Pi_{\mu\nu}^{A}=i\int \rmfont{d} x
e^{iqx}\langle 0|\; T \; j_{\mu}^{\rmfont{A}}(x)
j_{\nu}^{A\dagger}(0)  \;|0\rangle
{} \, ,
\eeq
the renormalization properties  correspond
to those of the vector
case, since
\beq
\Pi^{\rmfont{A}}_{\mu\nu}(q,g,m,m',\mu)=
\Pi^{\rmfont{V}}_{\mu\nu}(q,g,\pm m,\mp m',\mu)
{}\, ,
\EQN{cc8}
\eeq
which leads to the equivalent RG equation
\beq
{\dsp\mu^2 \frac{d}{d \mu^2}
\Pi^{\rmfont{A}}_{\mu\nu}(q,g,m,m',\mu)= \dsp
(q_{\mu} q_{\nu}  -g_{\mu\nu}q^2)
\g_{\stand{q}}^\rmfont{AA}\frac{1}{16\pi^2} +
g_{\mu\nu}({m+m'})^2 \g^\rmfont{AA}_{\stand{m}}\frac{1}{16\pi^2}}
{}\, ,
\EQN{cc9}
\eeq
with
\beq
\g^\rmfont{AA}_{\stand{q}}=\g^\rmfont{VV}_{\stand{q}}, \;\;\;
\g^\rmfont{AA}_{\stand{m}}=\g^\rmfont{VV}_m
{}\, .
\EQN{cc9b}
\eeq
Similar considerations apply to the
two-point correlation function
of  pseudoscalar currents
\beq   \EQN{cc10}
\Pi^{P}(-q^2,g,m,m',\mu) =i\int \rmfont{d} x
e^{iqx}\langle0|\; T\; j_{\rmfont{P}}(x) j^{\dagger}_{\rmfont{P}}(0)
\;|0\rangle
{}\, .
\eeq
The subtractive renormalization of the bare correlator
(we limit ourselves below to the massless case)
\beq
\Pi^{\rmfont{P}}(Q^2,g,\mu) =
(Z_{\stand{m}})^2 \mu^{2\ep}
\Pi_{\rmfont{B}}^{\rmfont{P}}(Q^2,g_{\rmfont{B}})
+Q^2 Z_{\stand{q}}^{\rmfont{PP}}\frac{1}{16\pi^2}
\EQN{cc11}
\eeq
leads to the   RG equation
\beq
\EQN{cc12}
\mu^2 \frac{d}{d \mu^2}\Pi^{\rmfont{P}}(Q^2,g,\mu)=
Q^2\g_{\stand{q}}^{\rmfont{PP}}\frac{1}{16\pi^2}
+2\g_{\stand{m}} \Pi^{\rmfont{P}}(Q^2,g,\mu)
{}\, ,
\eeq
with the anomalous dimension
\beq
\ba{rl}\dsp
\g^{\rmfont{PP}}_{\stand{q}}= & \dsp \mu^2 \frac{d}
{d \mu^2}(Z^{\rmfont{PP}}_{\stand{q}})-\ep Z^{\rmfont{PP}}_q
\dsp
  =
-\sum_{i\geq 0}(i+1)(Z^{\rmfont{PP}}_{\stand{q}})_{i1}
\left(\frac{\alpha_s}{\pi}\right)^{i}
{}.
\ea
\EQN{cc13}
\eeq

\noindent
The scalar current correlator
\beq     \EQN{cc14}
\Pi^{\rmfont{S}}=i\int   \rmfont{d} x
e^{iqx}\langle 0|\; T\;j_{\rmfont{S}}(x)j^{\dagger}_{\rmfont{S}}(0)
\;|0\rangle
 \eeq
and the pseudoscalar
current correlator are related in a simple
manner:
\beq
\Pi^{\rmfont{S}}(Q^2,m,m',\mu) = \Pi^{\rmfont{P}}(Q^2,\mp m,\pm m',\mu)
{}~.
\EQN{cc15}
\eeq
For vanishing quark masses
scalar and  pseudoscalar current
correlators  are therefore identical:
$\Pi^{\rmfont{S}}= \Pi^{\rmfont{P}}$.

The axial vector and pseudoscalar correlators are connected
through  the axial Ward identity
\beq
q_\mu q_\nu
\Pi^{\rmfont{A}}_{\mu\nu} = (m +  m')
\langle
\ovl{\psi}_{\rmfont{q}} \psi_{\rmfont{q}}
\ovl{\psi}_{\rmfont{q'}}      \psi_{\rmfont{q'}}
\rangle
{}\, ,
\EQN{axial-ward}
\eeq
where the vacuum expectation values on the r.h.s.  are
understood within the perturbation theory framework.
Equation~\re{axial-ward} leads to the following relation between the
corresponding anomalous dimensions \cite{CheKueKwi92}:
\beq
\gamma_{\stand{m}}^\rmfont{AA}
\equiv
-
\gamma_{\stand{q}}^{\rmfont{PP}}
{}\, .
\EQN{AA-PP}
\eeq
This relation was used in Ref.~\cite{CheKueKwi92}
in order to find  the anomalous dimension
$\gamma_{\stand{m}}^\rmfont{AA}$ at the $\alpha_s^2$
order starting from
the results of Ref.~\cite{GorKatLarSur90}.
\sect{Decoupling of Heavy Quarks\label{decoupling}}
This section deals with the issue of heavy quarks decoupling
when MS-like renormalization schemes are employed.  The matching
conditions relating the parameters of minimally renormalized theories
describing the physics well below and above a heavy quark threshold
are formulated and a short discussion of power-suppressed effects is
given.

\subsect{Decoupling Theorem in MS-like Schemes\label{decouplingl2}}

Masses of known quark species  differ vastly in their magnitude.  As
a result, in many QCD  applications  the mass of
a heavy quark $h$ is much larger than the characteristic momentum
scale $\sqrt{s}$ intrinsic to the physical process. In
such a situation there appear two interrelated problems when using
an MS-like scheme.
\begin{itemize}
\item
First, one has {\em two} large but in
general quite {\em different}  mass
scales, $\sqrt{s}$ and $m_{\rmfont{h}}$,  and thus
two different types of potentially
dangerously large  logarithms.  The
standard trick of a clever choice of
the renormalization scale $\mu$ is no
longer effective; one can not set
{\em  one } parameter $\mu$ equal to two
different mass scales
simultaneously.
\item
Second, according to the
 Appelquist--Carrazone theorem \cite{AppelquistCarazzone75} heavy
particles should  be   eventually `decoupled' from  low-energy
physics\footnote{{{\footnotesize It should be  stressed
that the statement is  literally
valid only if  power-suppressed corrections of order\break\hfill\indent
$\,\,\,\,$$(s/m_{\rmfont{h}}^2)^{\stand{n}}$ with $n > 0$ are neglected.
It is also understood that the effective
Lagrangian without\break\hfill\indent$\,\,\,\,$the  heavy quark field remains
renormalizable.  Fortunately, the QCD Lagrangian \re{r1} meets
this\break\hfill\indent$\,\,\,\,$demand. The standard model
however,  does not fulfil this requirement.
This leads to the\break\hfill\indent$\,\,\,\,$well-known deviations from the
theorem such as the $m_{\rmfont{t}}^2$ effects in
$\G(Z \to \ovl{\rmfont{b}} \rmfont{b}) $
or the $\rho$ parameter.}}}.
However, a peculiarity of  mass-independent renormalization schemes
is  that the decoupling  theorem does not hold
in its na\"{\i}ve form  for theories
renormalized in such schemes:  the effective  QCD
action   to appear will not be canonically normalized.
Even worse, large mass logarithms in
general appear when one calculates  a
physical observable!  (See the  example
below.)
\end{itemize}

Fortunately, both problems are  controlled  once a proper
choice of the expansion parameter is made
and the renormalization group
improvement is performed [35--37].

In order to be specific,
consider QCD with $n'_{f} = n_{\stand{f}} -1$ light quarks
$\psi =
\{
\psi_{\stand{l}}| l= 1 \mbox{-} n'_{f}
\}
$
with masses
$m =
\{
m_{\stand{l}}| l= 1 \mbox{-} n'_{f}
\}
$
and one  heavy quark $h$ with mass $m_{\rmfont{h}}$.

The respective action
$I(\bPsi,h,\ovl{h},g,m,m_{\rmfont{h}},\mu)$
is determined by
integrating  over the space--time  the Lagrangian density
\beq
\ba{rl}
\dsp
\Lc
= & -\frac{1}{4} (G^{\stand{a}}_{\mu\nu})^2
+
\sum_{l}\ovl\psi_{\stand{l}}
(i\!\not{\!\! D} - m_{\stand{l}})
{\psi}_{\stand{l}} + \ovl h (i\!\not{\!\! D} - m_{\rmfont{h}}) h
\\
\dsp
         &      + \ \mbox{\rm terms with ghost fields and the gauge-fixing
term}.
\ea
\EQN{Lagr}
\eeq
In  the  condensed notation      of  Section~\ref{renorm}
the collection of  fields $\bPhi$ is now decomposed as
$\bPhi = \{\bPsi,h,\ovl{h}\}$
with
$\bPsi=\{\psi,\ovl{\psi},A^{\stand{a}}_\mu\}$.
The (renormalized)  generating functional
of (connected) Green functions of light fields may now be
written as
\beq
Z^{\rmfont{R}}_{\rmfont{c}}(I;\Sbf,s) =
R_{\overline{\rmfont{MS}}}
\left\{
\int [\dif\bPsi\dif h \dif \ovl{h}] \exp \,(i I + \bPsi\cdot\Sbf
+ J \cdot s)
\right\}_{\rmfont{c}} \ .
\EQN{gen.func}
\eeq
Here
$R_{\overline{\rmfont{MS}}}$
is the ultraviolet $R$-operation in
$\msbar$-scheme.
$\Sbf$ and ${s}$ are sources for the (light) elementary fields
from $\bPsi$ and  for an  external quark
current $J=\ovl{\psi}\Gamma\psi$
respectively.
For the sake of notational simplicity  we
shall proceed  in the Landau gauge
and ignore the ghost field variables.

Integrating out the heavy quark should transform the generating
functional \re{gen.func} to that  corresponding to the effective QCD
with  $n'_{\stand{f}}$  remaining quark
flavours plus additional higher dimension
interaction terms  suppressed by powers of the
inverse heavy mass.
The current $J$ as well as
any other composite operator will generically develop a
non-trivial coefficient  function even if  one neglects
all power-suppressed terms.

In a  more formal language  the result
of integrating out  the heavy  quark may be summarized in the
following master expansion for the generating functional
\re{gen.func}:
\beq
\dsp
Z^{\rmfont{R}}_{\rmfont{c}}(I;\Sbf,s)
\bbuildrel{=\!=\!=}_{\scriptstyle{m_{\rmfont{h}}\to\infty}}^{}
R_{\overline{\rmfont{MS}}}
\left\{
\int [\dif\bPsi] \exp \,
\left[ i I^{\rmfont{eff}}(\bPsi , g)
+
\bPsi'\cdot\Sbf'
+
\Biggl( J' \zrm_{\stand{j}}
+ \sum_{n}\frac{ \zrm_{n}}{m_{\rmfont{h}}^{\delta_{\stand{n}} -3}}
J_{\stand{n}}\Biggl)
\cdot s_\mu\right]
\right\}_c
{}\, ,
\EQN{master.exp}
\eeq
where  the sum is performed over   operators
constructed from the light fields,
with the quantum numbers of
those of the initial   current $J$ and
of mass dimension $\delta_{\stand{n}}$.
The effective action   $I^{\rmfont{eff}}(\Phibf,g)$  can be written as
\beq
I^{\rmfont{eff}}(\bPsi,g) =
I(\bPsi',g')|{{}_{\scriptstyle h=0}}
+
\sum_{\stand{n}}
\int \frac{\zrm_{\stand{n}}
\dsp O_{\stand{n}}(x)}{\dsp m_{\rmfont{h}}^{d_{\stand{n}} -4}} \dif x,
\EQN{effect.action}
\eeq
with
\beq
g'   = z_{\stand{g}} g,  \ \ \
m_{\rmfont{q}}' = z_{\stand{m}} m_{\rmfont{q}}
\EQN{match1}
\eeq
and
$\bPsi' =
\{ \psi'= z_2^{1/2}\psi,(A^{\stand{a}}_\mu)'
= z_3^{1/2} A^{\stand{a}}_\mu\},
\ J' = \ovl{\psi'}\G \psi'$.
Here
$\{O_{\stand{n}} \}$ are Lorentz scalars of
dimension
$d_{\stand{n}} > 4$,
again constructed from the (primed)  light fields only. At last,
$$
{\Sbf}' = \{ S_\psi/\sqrt{\zrm_2}, S_{\ovl{\psi}}/\sqrt{\zrm_2},
S_{\rmfont{A}}/\sqrt{z_3}\}\, ,
$$
and
all `normalization constants'  $z'$s with various subscripts
are series of the generic  form
\beq
z_?  \equiv
          \left\{
          \ba{ll}
          \dsp
1 + \sum_{i>0} a^?_{\stand{i}} g^{2 i}
\ \ \rmfont{if} \ \ ?=\psi,\ovl\psi,2,3,g \ \ \rmfont{or } \ \ m\, ,
          \\
          \dsp
\sum_{i>0} a^?_{\stand{i}} g^{2 i}  \ \ \rmfont{if}
         \ \      ? = J,n \, .
          \ea
          \right.
\EQN{z}
\eeq
with {\em finite} coefficients $a_{\stand{n}}$,  which are
polynomials in $\ln\, (\mu^2/m_{\rmfont{h}}^2)$.

The master equation   \re{master.exp}  requires some comments.
\begin{itemize}
\item
The expansion  \re{master.exp} should be  understood in the strictly
perturbative sense; once it is performed
it is necessary to utilize the usual
renormalization group methods in order to resum all large logs
of the heavy quark mass (see below).
\item
The master equation \re{master.exp}
governs  the $ m_{\rmfont{h}} \to \infty $ asymptotic behaviour
of {\em all} light Green functions: if one neglects
power-suppressed terms and does not consider extra
current insertions,
then  their asymptotic behaviour is  completely  determined by
a few finite normalization constants.  Even more: in the
calculation    of physical quantities,
which  do not  depend on  the
normalization of quantum fields,
only two constants  remain,
viz.  $z_1$ and $z_{\stand{m}}$.
\item
There exist  several methods of computing
the finite renormalization constants.
The most advanced  approach is  based
on the so-called heavy mass expansion
algorithm and will be discussed in
Section \ref{repmtexp}.
\end{itemize}
\subsect{Matching Conditions
for $\alpha_s$ and  Masses\label{mathcing}}
In this subsection we review   the so-called matching conditions
which allow   the relating of
the parameters of effective low-energy
theory without a  heavy quark to  those of the full theory.

The master equation \re{master.exp}
states that the effective coupling constant $\alpha_s^{\prime} $ and
the (light) quark masses
$m_{\rmfont{q}}^{\prime}$ are expressed in terms of
those of the full theory, viz.
$\alpha_s $ and $m_{\rmfont{q}},m_{\rmfont{h}}$,  via Eq.
 \cite{Bernreuther82b} --- see (\ref{match1}):
\bea
\alpha_s^{\prime}(\mu) &=&  \alpha_s(\mu) \, C(\alpha_s(\mu),x)
\EQN{match2a}
{}\, ,
\\
m_{\rmfont{q}}^{\prime}(\mu) &=&
m_{\rmfont{q}}(\mu) \, H(\alpha_s(\mu),x)
{}\, .
\EQN{match2b}
\eea
Here $x= \ln\,(\ovl{m}_{\rmfont{h}}^2/\mu^2)$ and
the functions $C$ and  $H$ exhibit the following
structure:
\bea
C(\alpha_s,x) &=& 1
+
\sum_{k \geq 1}
C_{k}
\left(
\frac{\alpha_s}{\pi}
\right)^k
\label{alpha2},
\  \
C_{\stand{k}}(x)
=
\sum_{0 \leq i \leq k }
 C_{ik} x^{\stand{i}}\, ,
\\
H(\alpha_s(\mu),x)
 &=&
1
+
\sum_{k \geq 1}
H_{k}
\left(
\frac{\alpha_s}{\pi}
\right)^{\stand{k}} \, ,
\  \
H_{\stand{k}}(x)
=
\sum_{0 \leq i \leq k }
H_{ik} x^i
\, ,
\label{mass2}
\eea
with $C_{ik}$ and $H_{ik}$ being {\em pure}
numbers.
At present, the functions
$C$ and $H$ are known at two-loop level
[37--39]
and read\footnote{The  constant term in $C_2$
is cited according to Ref.~\cite{timo2}, where it has been
recalculated  using two\break\hfill\indent$\,\,\,$ different approaches.}
\bea
C_1 &=& \frac{1}{6}~x\, ,\quad\,
C_2 =
\frac{11}{72} + \frac{11}{24}~x  + \frac{1}{36}~x^2\, ,
\EQN{match3a}
\\
H_1 &=& 0\, , \qquad\,\,
H_2 =
\frac{89}{432} + \frac{5}{36}~x + \frac{1}{12}~x^2
\, \, .
\EQN{match3b}
\eea
Another useful form of \re{match2a}
is obtained after expressing its r.h.s. in terms of
the pole mass $M_{\rmfont{h}}$:
\begin{equation}
\alpha'_s(\mu) = \alpha_s(\mu)
\left\{
1
+  \frac{X}{6} \frac{\alpha_s(\mu)}{\pi}
+  \left(-\frac{7}{24} + \frac{19 X}{24} + \frac{X^2}{36}\right)
\left[
\frac{\alpha_s(\mu)}{\pi}
\right]^2
\right\}
{}\,  ,
\EQN{match2aPole}
\end{equation}
with $X = \ln\, (M_{\rmfont{h}}^2/\mu^2)$.

The effective $\alpha'_s$ and  the
light quark masses evolve with $\mu$ according to
their own effective RG equations \cite{Ovrut80}.
It is  important to
stress that the master equation and hence \re{match2a}
were derived under the requirement that the normalization scale
$\mu$ is much less than $m_{\rmfont{h}}$. However, once obtained,
Eqs.~(\ref{match2a},~\ref{match2b}),  present  {\em universal }
relations,  valid order
by order in  perturbation theory.

This implies that on formal grounds one is free
to choose the {\em  matching } value of $\mu= \mu_0$  to
determine  the value of, say,  $\alpha'_s$ in terms of
the parameters  of the full theory.  The final result should
not depend on $\mu_0$.
However, in practice, some dependence remains from the
truncation of higher-orders.
Thus the problem is completely similar  to that
discussed in Section~\ref{running}. The correct prescription,
hence, is to solve the matching conditions
(\ref{match2a},~\ref{match2b})
with $\mu$  fixed somewhere in the vicinity of
$m_{\rmfont{h}}$  to suppress all mass logarithms.
A popular particular choice is to set $\mu = \ovl{m}_{\rmfont{h}}(\mu)$
and thus   nullify all mass logarithms.
The  mass $m_{\rmfont{h}} =
\ovl{m}_{\rmfont{h}}(m_{\rmfont{h}})$ is sometimes
referred to as {\em scale invariant mass}
of the quark $h$. Finally, one should run the effective coupling
constant and quark masses to a lower normalization scale with
the effective renormalization group equations.
\subsect{Matching Equations for
Effective Currents \label{mathcing2}}
{}From  a fundamental  point of view the treatment
of effective currents
does not differ significantly  from the one    discussed
for the effective
coupling constant and masses. Moreover, for the customary  case of
bilinear quark currents it is even easier: in many instances there
exist some extra constraints like Ward identities which help to fix
the constant $z_{\rmfont{J}}$. Two cases are of particular interest.

\vspace{2mm}

\noindent
{\em Vector current:} This is the most simple
and well-known case.  For
$ J = \ovl{\psi_{\rmfont{q}}}\g_\mu \psi_{\rmfont{q}} $
one derives from  the
vector Ward identity\footnote{An explicit derivation
may be found e.g. in Ref.~\cite{me93}.}
that
\beq
z_{\rmfont{V}}  \equiv
          \left\{
          \ba{ll}
1 \ \ \mbox{if \ \   q \ \ \  is a light quark}
          \\
0 \ \ \mbox{if      \ \  q = h }\, .
          \ea
          \right.
\EQN{zV}
\eeq
Thus the functional form of  a (light) vector
quark current is unchanged
after integrating out a heavy quark and rewriting it in terms of the
effective (that is properly normalized) light quark fields.
\vspace{2mm}

\noindent
{\em Axial vector current:}
Here  the situation is more complicated
due to the famous axial vector
anomaly. A statement similar to  \re{zV}  may be proved only
for non-singlet axial vector current constructed  from light quark
fields [41--43]. Explicitly, if
$J_{\rmfont{A}} =
\sum_{l,l'} a_{ll'}\ovl{\psi}_{\stand{l}}
\gamma_5 \g_\mu \psi_{l'}
$ with
a traceless matrix
$\{ a_{ll'} \}$
then  the corresponding effective current
reads
\beq
J'_{\rmfont{A}} =
\sum_{l,l'} a_{ll'}\ovl{\psi}'_{\stand{l}}
\gamma_5 \g_\mu \psi'_{l'}\, .
\EQN{zA}
\eeq
It is understood in \re{zA}
that   $\gamma_5$ is treated in a way which does not violate the
(non-anomalous) chiral Ward identity.  In fact this requirement is
{\em unmet} if the axial vector currents are minimally renormalized
with the {}'t Hooft--Veltman
definition of $\gamma_5$.  The necessary
modifications are discussed in Section~\ref{gamma5}.

If, however, one has a non-singlet combination of light
{\em and } heavy diagonal axial vector currents
then there are no simple
formulas like \re{zV} and \re{zA}: the
resulting effective current is
in general
not
a non-singlet
combination of some light axial  vector currents. This  case
is discussed in Refs.~\cite{Collins78,CK3}.
\subsect{Power Suppressed Corrections\label{power}}
The apparatus of the  effective theory also  allows
the taking into account of power suppressed corrections.
These can in turn be separated into the corrections
to the effective Lagrangian and an effective
current.  Below we list for illustrative purpose
some  well-known results.
\vspace{2mm}

\noindent
{\em QCD Lagrangian}

\noindent
The least  power suppressed contribution to  the
sum in \re{effect.action}   is given by  a four-quark
operator of dimension 6 (see Ref.~\cite{NSVZ84}), viz.
\beq
-\frac{\as'^2}{15m_{\rmfont{h}}^2}
\sum_{ll'}
(\bar\psi'_{\stand{l}} \g_\al t^{\stand{a}}\psi'_{l})
\,
(\bar\psi'_{l'} \g_\al t^{\stand{a}}\psi'_{l'})
{}\, .
\EQN{4quark}
\eeq
Here the colour group generators $t^{\stand{a}}$ are
normalized in the standard way
$
Tr(t^{\stand{a}} t^{\rmfont{b}}) = \delta^{ab}/2
{}.
$
\vspace{2mm}

\noindent
{\em Vector and axial vector currents}

\noindent
The formulae look  almost identical
for  vector  and axial vector (non-singlet) currents
(if, of course, the ``correct" treatment of $\gamma_5$ is
employed, see above and Section~\ref{gamma5}).
For the case  most useful in
practice, namely that  of a massless light quark (axial)
vector current,   one obtains \cite{me93}
\beq
\ba{c}
\dsp
\ovl\psi_{\stand{l}}\g_\mu(\g_\mu\gamma_5)\psi_{\stand{l}}
\bbuildrel{=\!=\!=}_{{\scriptstyle{m_{\rmfont{h}}\to\infty}}}^{}
\ovl{\psi'}_{\stand{l}}\g_\mu(\g_\mu\gamma_5)\psi _{\stand{l}}'
\\
\dsp
+
\left\{
\frac{1}{135}\ln\,
\left(
\frac{\mu^2}{m_{\rmfont{h}}^2}
\right)
-\frac{56}{2025}
\right\}
\left(
\frac{\as'}{\pi}
\right)^2
\frac{\prd^2}{m_{\rmfont{h}}^2}
[\ovl{\psi'_{\stand{l}}}\g_\mu(\g_\mu\gamma_5)\psi'_{\stand{l}}]
+ O(\as^3)
+ O(1/m_{\rmfont{h}}^4)
{}\, .
\ea
\EQN{current.exp}
\eeq
\subsect{Example\label{example}}
To provide an example of a peculiar realization of
the decoupling theorem in  MS-like schemes we  now
discuss the  evaluation of  a `physical'
quantity  --- the pole mass $M_{\stand{l}}$ of a light
quark $q_{\stand{l}}$ in the full and the effective theories.

First of all  we recall that the pole mass is defined as the position
of the pole of the quark propagator  computed in
perturbation theory. It is a renormalization scheme and gauge
invariant object \cite{Tarrach81,Nar87}
whose numerical value should
obviously not  depend on the   theory
 ---   full or   effective  ---
in which it is evaluated in.

The result of the evaluation of $M_{\stand{l}}$ in the
$\msbar$-scheme at two-loop
level in the QCD with a heavy quark
reads --- the formula below is  in fact just an inversion of
\re{m-from-M}:
\begin{eqnarray}
\!\!\!\!\!M_l
&=&
{m_{\stand{l}}(\mu)}\left\{
\rule{0mm}{5mm}
\right.
\!\!1 +
\apimu
\left(
\frac{4}{3} + \ln\, \frac{\mu^2}{m_{\stand{l}}^2}
\right)
\\
\dsp
\rule{-8mm}{3mm}
&&+
\left[
\apimu
\right]^2
\left[
K_{\stand{l}}(\m)
-\frac{8}{3}
+
\left.
\left(
\frac{173}{24} -  \frac{13}{36}~n_f
\right)
\ln\, \frac{\mu^2}{m_{\stand{l}}^2}
+
\left(
\frac{15}{8} - \frac{1}{12}~n_f
\right)
\ln^2 \frac{\mu^2}{m_{\stand{l}}^2}
\right]
\right\}\nonumber        .
\EQN{M-from-m}
\end{eqnarray}
If  $m_{\rmfont{h}} \to  \infty$
then, according to \re{asymp1},
the function  $K_{\stand{l}}(\m)$  behaves as
\nonumber
$\ln^2(m_{\rmfont{h}}/m_{\stand{l}})/3$ and thus the r.h.s. of
Eq.~\re{M-from-m}   is {\em not} well defined!
This is,   of course,  a manifestation of the fact that
in this limit  the initial parameters of the full theory
are not adequate  to construct the  perturbative theory
expansion for a low energy quantity.

However, using the relations
\re{match2a}  and \re{match2b}  and expressing
the r.h.s of
\re{m-from-M} in terms of the {\em effective}
$\alpha'_s$ and $\m'$, the resulting expression becomes
well-defined at the $m_{\rmfont{h}} \to \infty$ limit and
reads
\begin{eqnarray}
\dsp M_l
&=&
m_{\stand{l}}(\mu)
\left\{
\rule{0mm}{5mm}
\right.
\!\!1
+
\frac{\alpha'_s}{\pi}
\left[\frac{4}{3}
+
\ln\, \frac{\mu^2}{(m'_{\stand{l}})^2}
\right]
\nonumber
\\
&&+
\left(
\frac{\alpha'_s}{\pi}
\right)^2
\left[
\frac{3817}{288}
+ \frac{2}{3}(2 + \ln\, 2)\zeta(2)
-
\frac{1}{6} \zeta(3)
- \frac{n_{\stand{f}}'}{3}
\left(
\zeta(2)  + \frac{71}{48}
\right)
\right.
\EQN{M-from-m-prime}
\\
\nonumber
&&+~
\frac{4}{3} \sum_{1 \leq f \leq n_{\stand{f}}'}
\Delta\left(\frac{m_{\stand{f}}}{m'_{\stand{l}}}\right)
+
\Biggl(\frac{173}{24} -  \frac{13}{36}~n'_{\stand{f}}\Biggl)\ln\,
\frac{\mu^2}{(m'_{\stand{l}})^2}
+\Biggl(\frac{15}{8} - \frac{1}{12}~n'_{\stand{f}}\Biggl)\ln^2
\frac{\mu^2}{(m'_{\stand{l}})^2}
\Biggl]
\Biggl\}
{}\,    .
\end{eqnarray}
Now it can be easily seen that  \re{M-from-m-prime}
is nothing but \re{m-from-M} written in the
effective theory with the decoupled heavy quark!

\sect{\label{Qmasses}Quark Masses}
In this section we briefly discuss
the presently available numeric values of
pole and running quark masses at different
scales. The exposition below serves to explain and
motivate the choice of the input quark masses in the numerical
discussion of Part~\ref{numerical}.
It is not intended to
provide a comprehensive  review  of this involved
issue (for some recent reviews see, for example, Refs.
\cite{Narison94a,Narison94b}).

\subsect{\label{light} Light $\rmfont{u},\rmfont{d}$
and $\rmfont{s}$ Quarks}
For a  light quark   $\rmfont{q}= \rmfont{u,d,s}$
the concept of the pole mass
$M_{\rmfont{q}}$ is
clearly meaningless, at least in the framework of
the perturbative definition given above.
In contrast, the running
mass $\ovl{m}_{\rmfont{q}}$ is well defined, provided the scale parameter
$\mu$ is not too small.  Traditionally, the reference scale
$\mu$ is taken to be  1 GeV.
The latest available values for these masses are
\bea
\ovl{m}_{\rmfont{u}}(1 \GeV) +  \ovl{m}_{\rmfont{d}}
(1 \GeV) &=& (12 \pm 2.5) \MeV, \qquad
\frac{\dsp
\ovl{m}_{\rmfont{u}}}{\dsp \ovl{m}_{\rmfont{d}}} = 0.4 \pm 0.22~,
\label{ud:quarks}
\\
\ovl{m}_s(1 \GeV) &=&  189 \pm 32 \MeV~.
\label{s:quark}
\eea
(These values  \re{ud:quarks} and  \re{s:quark}
are  cited according to Refs.
[47--49];
some earlier
determinations can be found in
Refs.~[50--53]).

In all of the applications considered  in  the present work, it is clearly
more than legible to consider $\rmfont{u}$ and $\rmfont{d}$
quarks as massless.
Also, the strange quark mass
will be neglected
everywhere except for
small corrections induced by  $m_{\rmfont{s}}$ in  the relation
between pole and running masses of
$\rmfont{c}$ and $\rmfont{b}$ quarks,  respectively,
as  discussed below.

\subsect{\label{CandB} Charm and Bottom}
Within the  effective four-quark  theory
the relation between the pole and the running
masses of the charmed quark reads
(it is, in fact, Eq.~\re{M-from-m}
with   $n_{\stand{f}}=4 $)
\begin{eqnarray}
\dsp
M_c
&=&
{\ovl{m}_{\rmfont{c}}(\mu)}\left\{
\rule{0mm}{5mm}
\right.
\!\!1~+
\apimu\Biggl(\frac{4}{3} + \ln\, \frac{\mu^2}{\ovl{m}_{\rmfont{c}}^2}\Biggl)
\EQN{Mc-from-mc}
\\
&&+
\dsp
\left[
\apimu
\right]^2
\left[
10.319 + \frac{4}{3}
\Delta\left(\frac{\ovl{m}_s}{\ovl{m}_{\rmfont{c}}}
\right)
\left.
+
\frac{415}{72}\ln\, \frac{\mu^2}{\ovl{m}_{\rmfont{c}}^2}
+
\frac{37}{24}\ln^2 \frac{\mu^2}{\ovl{m}_{\rmfont{c}}^2}
\right]
\right\}
\, .
\nonumber
\end{eqnarray}
This equation  may be used in two ways. First,
if one is given a  value of $\ovl{m}_{\rmfont{c}}(\mu)$
then  \re{Mc-from-mc} may be used to construct
$M_{\rmfont{c}}$ in the following way. One runs
$\ovl{m}_{\rmfont{c}}(\mu)$
(using RG equations in the $n_{\stand{f}}=4$ theory)
to find the scale invariant mass
$m_{\rmfont{c}} =
\ovl{m}_{\rmfont{c}}(m_{\rmfont{c}})$, and then  evaluates the
r.h.s. of \re{Mc-from-mc} with $\mu = m_{\rmfont{c}}$.
Second, let us suppose that $M_{\rmfont{c}}$ is  known and we would like
to find the running mass $\ovl{m}_{\rmfont{c}}(\mu)$ at some reference
point $\mu$. Even in this case the  use of \re{Mc-from-mc}
is  preferable  to that of \re{m-from-M}, as the latter
would contain a contribution proportional to the
ill-defined  pole mass of the strange quark.

In the case of the  $\rmfont{b}$  quark   the relation
\re{m-from-M} assumes  the following form
(all running masses
and the coupling constant are now defined
in the $n_{\stand{f}}=5$ effective
QCD)
\begin{eqnarray}
\dsp
\!\!\!\!\!M_b
&=&
{\ovl{m}_{\rmfont{b}}(\mu)}\left\{
\rule{0mm}{5mm}
\right.
\!\!1~
+
\apimu\Biggl(\frac{4}{3} + \ln\, \frac{\mu^2}{\ovl{m}_{\rmfont{b}}^2}\Biggl)
\EQN{Mb-from-mb}
\\
\dsp
\rule{-8mm}{3mm}
&&+
\left[
\apimu
\right]^2
\left[
9.278
+ \frac{4}{3}
\Delta\left(\frac{\ovl{m}_s}{\ovl{m}_{\rmfont{b}}}\right)
+ \frac{4}{3}
\Delta\left(\frac{\ovl{m}_{\rmfont{c}}}{\ovl{m}_{\rmfont{b}}}
   \right)
\left.
+
\frac{389}{72}\ln\, \frac{\mu^2}{\ovl{m}_{\rmfont{b}}^2}
+
\frac{35}{24} \ln^2 \frac{\mu^2}{\ovl{m}_{\rmfont{b}}^2}
\right]
\right\}
\, .
\nonumber
\eea
This equation is to   be used in  the same way as \re{Mc-from-mc}.

In the literature there is a variety of somewhat different
results for the masses of  $\rmfont{b}$ and $\rmfont{c}$  quarks.
\ice{
For instance, the central value of $M_{\rmfont{b}}$
varies from $4.87 \GeV$ in Ref.~\cite{Titard94} till
$4.806 \GeV$ in Ref.~\cite{Narison94b}.
Similar situation occurs with  the charmed quark.
}
Also, there exist strong indications that
the very concept of the pole mass is plagued
with severe non-perturbative ambiguities \cite{Shifman94}.
It may well  happen that
eventually  the most accurate and unambiguous
mass parameter related to a quark will be its running
mass taken at some convenient reference point.
However, for illustrative purposes we will  use
the following {\em ansatz} for the pole masses
$M_{\rmfont{c}}$  and $M_{\rmfont{b}}$:
\beq
M_{\rmfont{c}} = 1.6 \pm 0.10 \GeV \qquad \rmfont{{\rm and }}
\qquad M_{\rmfont{b}} = 4.7 \pm 0.2 \GeV
{}\, .
\EQN{c-b:mass}
\eeq
The central values and uncertainty bars
in \re{c-b:mass}  are in broad agreement
with Refs. \cite{Narison94b,Titard94} and also
with those used by the
Electroweak Precision Calculation
Working Group  \cite{DimaBardin}.
Table~1 shows the running masses
 obtained from (\ref{Mc-from-mc},\ref{Mb-from-mb}),
and RG equations at various relevant scales in dependence on
$\alpha_s(M_{\stand{Z}})$ with $ M_{\stand{Z}} = 91.188 $ GeV.
\begin{table}
\begin{center}
{\bf Table 1}
\end{center}
\vskip0.3cm
{\small{Values of
$\protect \Lambda_{\MSbarsmall}^{(5)} $,
$\protect \Lambda_{\MSbarsmall}^{(4)} $,
$\protect \ovl{m}_{\rmfont{c}}^{(4)}(M_{\rmfont{c}})$,
$\protect \ovl{m}_{\rmfont{c}}^{(5)}(M_{\rmfont{b}})$,
$\protect \ovl{m}_{\rmfont{b}}^{(5)}(M_{\rmfont{b}})$,
$\protect \ovl{m}_{\rmfont{b}}^{(5)}(M_{\stand{Z}})$
and
$\protect {m}_{\rmfont{b}}(m_{\rmfont{b}})$
(in GeVs)
for different values of
$\protect \alpha_s^{(5)}(M_{\stand{Z}})$,
and the default values of
$M_{\rmfont{c}}$  and $M_{\rmfont{b}}$ as in (110).}}
(\protect\ref{c-b:mass})
\begin{center}
\vskip0.2cm
\begin{tabular}{|c|c|c|c|c|c|c|c|}
\hline
$\as^{(5)}(M_{\stand{Z}})$
&
$\Lambda_{\MSbarsmall}^{(5)}$
&
$\Lambda_{\MSbarsmall}^{(4)}$
&
$\ovl{m}^{(4)}_{\rmfont{c}}(M_{\rmfont{c}})$
&
$\ovl{m}^{(5)}_{\rmfont{c}}(M_{\rmfont{b}})$
&
$\ovl{m}^{(5)}_{\rmfont{b}}(M_{\rmfont{b}})$
&
$\ovl{m}^{(5)}_{\rmfont{b}}(M_{\stand{Z}})$
&
${m}_{\rmfont{b}}(m_{\rmfont{b}})$
\\ \hline\hline
0.11$\phantom{0}$  &  0.129  &  0.188  &  1.27$\phantom{0}$  &
1.03$\phantom{0}$  &  4.13  &  3.01  &  4.20\\  \hline
0.115  &  0.175  &  0.248  &  1.21$\phantom{0}$  &
0.953  &  4.07  &  2.89  &  4.15\\  \hline
0.12$\phantom{0}$  &  0.233  &  0.32$\phantom{0}$  &  1.12$\phantom{0}$  &
0.855  &  3.99  &  2.77  &  4.10\\  \hline
0.125  &  0.302  &  0.403  &  1.01$\phantom{0}$  &
0.734  &  3.91  &  2.64  &  4.04\\  \hline
0.13$\phantom{0}$  &  0.383  &  0.499  &  0.853  &
0.583  &  3.82  &  2.5$\phantom{0}$  &  3.97\\  \hline
\end{tabular}
\end{center}

\end{table}

\subsect{\label{topmass} Top}

The top quark mass value as reported by the CDF collaboration
\cite{TOP} is
\beq
M_{\rmfont{t}}  =  174~\pm 10^{+13}_{-12} \GeV
{}\, .
\EQN{top:mass}
\eeq
In our numerical discussions we shall use a conservative
input value of  $M_{\rmfont{t}} = 174~\pm$\break\hfill 20  GeV.

In order to  find the corresponding
running mass we use
the equation below obtained from \re{m-from-M}
(we deal now  with the  fully-fledged
$n_{\stand{f}} =6 $ theory, and
discard completely negligible terms caused by
the masses of $\rmfont{s}$ and $\rmfont{c}$ quarks)
\begin{eqnarray}
\dsp
\ovl{m}_{\rmfont{t}}(\mu)
&=&
M_{\rmfont{t}}\left\{
\rule{0mm}{5mm}
\right.
\!\!1 - \apimu
\left(
\frac{4}{3} + \ln\, \frac{\mu^2}{M_{\rmfont{t}}^2}
\right)
\EQN{}
\\
&&-
\dsp
\left[
\apimu
\right]^2
\left[
9.125
+
\frac{4}{3} \Delta\left(\frac{M_{\rmfont{b}}}{M_{\rmfont{t}}}\right)
\left.
+
\frac{35}{8}\ln\, \frac{\mu^2}{M_{\rmfont{t}}^2}
+
\frac{3}{8} \ln^2 \frac{\mu^2}{M_{\rmfont{t}}^2}
\right]
\right\}
\,      .
\nonumber
\eea
After setting $\mu = M_{\rmfont{t}}$ and
evaluating $\alpha_s^{(6)}(M_{\rmfont{t}})$
one finds
\[
\alpha_s^{(6)} (M_{\rmfont{t}}) = 0.109    , \  \
\ovl{m}_{\rmfont{t}} (M_{\rmfont{t}})   =  164 \GeV~,
\]
for our default value
$\alpha_s^{(5)}(M_{\stand{Z}}) = 0.120$ corresponding
to $\Lambda^{(5)}_{\MSbarsmall} = 233 \MeV $.

\chap{Calculational Techniques \label{calc-tech}}
In this  Part we  discuss available calculational
techniques  to perform small and heavy mass expansions
of two-point correlators as well  as the problem
of $\gamma_5$ in dimensional regularization.
\sect{Current Correlators at Large  Momentum
\label{large}}
A lot of  results on higher-order radiative corrections were
derived after  neglecting  quark masses,   originating
from massless diagrams  and  resulting in a drastic
simplification of calculations.  However, problems arise when quark
masses are  taken into account,  at least in the form of power
corrections.   In the simplest cases
the evaluation of, say,   a quadratic quark mass correction
may be reduced to  the computation of
massless diagrams which are obtained by na\"{\i}vely
expanding the massive propagators in the quark mass.
However, this strategy  fails in the general case starting
from  quartic mass terms.
The so-called
logarithmic mass singularities appear and render
the  simple  Taylor expansion meaningless.
In this section  the general structure of
non-leading mass corrections will be discussed as well as
some approaches for  their evaluation.  The
presentation  is mainly based  on
 Refs.~[57--59].

In  investigating the  asymptotic behaviour of various correlators at
large momentum  transfer, it proves to be very useful to employ
the Wilson expansion  in the framework of the MS scheme.
Consider  vector  current correlator
\begin{equation}
i \int \langle 0|
T J_\mu(x) J_\nu(0)
               |0\rangle
{\rm e}^{iqx} {\rm d} x
= (g_{\mu\nu} - q_\mu q_\nu/q^2) \ \Pi(Q^2)
\,  ,
\label{hm1}
\end{equation}
with
$J_\mu=\overline q \g_\mu q $.
Here ${\rmfont{q}}$ is a quark with
mass $m_{\rmfont{q}}\equiv m$. To simplify the
following discussion
we will consider the second
derivative $\Pi''(Q^2) \equiv d^2\Pi(Q^2)/d(Q^2)^2$, which can be
seen from \re{cc4} to satisfy a homogeneous RG equation,
\begin{equation} \mu^2 \frac{d}{d\mu^2}\Pi''(Q^2) = 0\,.
\label{hm2}
\end{equation}
The high energy behaviour of $\Pi''(Q^2)$ in the deep Euclidean
region may be reliably evaluated in QCD by employing the operator
product expansion:
\begin{eqnarray}
& &Q^2\Pi''(Q^2,\alpha_s,m ,\mu)
\bbuildrel{=\!=\!\Longrightarrow}_{\scriptstyle{Q^2\to\infty}}^{}
K_0(Q^2,\alpha_s,m ,\mu)\, \rmfont{{\bf \large 1}}\nonumber\\
& & + \sum_{n}\frac{1}{(Q^2)^{n/2}}\sum_{dim\, O_{\stand{i}} = n}
K_{\stand{i}}( Q^2,\alpha_s,m ,\mu)
\langle 0| O_{\stand{i}} (\mu)|0\rangle\,.
\label{hm9}
\end{eqnarray}
We have explicitly  separated the
contribution  of the unit operator from
that of the operators with  non-trivial
dependence on the field variables.  The
coefficient functions $K_0$ and $K_{\stand{i}}$
depend upon the details of the
renormalization prescription for the
composite operators $O_{\stand{i}}$.  The usual
procedure of normal ordering for the
composite operators appearing on the
r.h.s. of Eq.~\re{hm9} becomes
physically inconvenient  if  quark mass
corrections are to be included
\cite{CheSpi88}.  From the
calculational view-point it also does
not lead to any insight in computing
power-suppressed mass corrections
involving mass logarithms.

Indeed, the coefficient function in
front of the unit operator in \re{hm9}
represents the usual perturbative
contributions and, if normal ordering
is used, it contains in general  mass
and momentum logarithms of the form
\begin{equation}
\left(
\frac{m ^2}{Q^2}
\right)^{\stand{n}}
\left(\ln\,\frac{\mu^2}{Q^2}
\right)^{n_1}
\left(
\ln\,\frac{\mu^2}{m ^2}
\right)^{n_2},
\label{hm10}
\end{equation}
with $n, \, n_1$ and $n_2$  being non-negative integers.
More specifically,  one can write
\begin{equation}
K^{NO}_0(Q^2,\alpha_s,m ,\mu)
\bbuildrel{=\!=\!\Longrightarrow}_{\scriptstyle{Q^2\to\infty}}^{}
\sum_{n \ge 0,\, l > 0}
\left(\frac{m ^2}{Q^2}
\right)^{n}
\left(\frac{\alpha_s}{\pi}
\right)^{l - 1}
F_{nl}(L,M)\,,
\label{hm11}
\end{equation}
where
$L = \ln\,(\mu^2/Q^2)$,  $M = \ln\,(\mu^2/m ^2)$, and
the superscript NO is a reminder of the normal ordering prescription
being used.
The function $F_{nl}(L,M)$ corresponds to the contribution of the
$l$-loop diagrams, and is a polynomial of degree not higher than $l$,
in both $L$ and $M$. The contributions  due to
non-trivial operators --- that is containing some dependence on
field variables  ---  are completely decoupled from those of the unit
operator if the normal ordering is
employed,  since the vacuum expectation value vanishes
for every non-trivial operator $O$:
\[
\langle 0 |O|  0\rangle \equiv 0 \, .
\label{hm3}
\]

The situation improves  drastically if one abandons  the normal
ordering  prescription. It was realized some time ago
\cite{BroadGen84,CheSpi88} that  all logarithms of quark masses may be
completely shifted  to the vacuum expectation values (VEV) of
non-trivial composite operators appearing on the r.h.s.  of \re{hm9}
if the latter are minimally subtracted.

To give a simple example, let us consider the correlator \re{hm1} in
the lowest order one-loop approximation.  First, we use the normal
ordering prescription for the composite operators which appear
in the  OPE of the time ordered product in \re{hm1}. To determine the
coefficients of the various operators, one possible method is to
sandwich both sides of the OPE between appropriate external states.
By choosing them  to be the vacuum, only the unit operator {\bf 1}
will contribute on the r.h.s., {\em if}  the normal ordering
prescription is used. This means that the
bare loop of Fig.~\ref{high-mom}a
contributes entirely to the coefficient $K_0$ in \re{hm9}. A simple
calculation gives  (in the sequel we
neglect all  terms of order $1/Q^6$ and higher):
\begin{figure}
 \begin{center}
  \begin{tabular}{cc}
  \subfigure[]{\epsfig{file=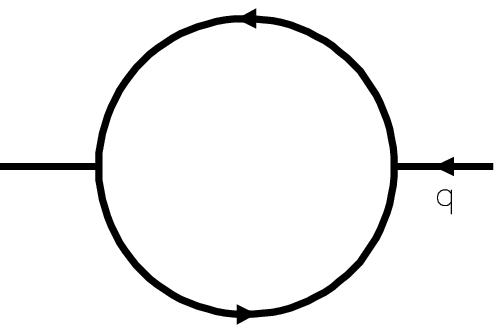,width=5cm,height=5cm}}
  &
\subfigure[$\langle \ovl{\Psi}\Psi \rangle $]%
{\epsfig{file=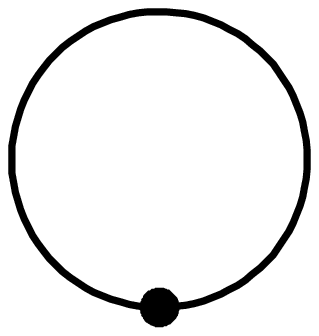,width=5cm,height=5cm}}
  \end{tabular}
 \end{center}
\caption {
          \label{high-mom}
(a) Lowest order contribution to the correlator $\Pi''(Q^2)$.
(b) Vacuum diagram contributing to the perturbative VEV of the
operator $\bar q q $.
         }
\end{figure}
\beq
Q^2\Pi''(Q^2)
\bbuildrel{=\!=\!\Longrightarrow}_{\scriptstyle{Q^2\to\infty}}^{}
K_0(Q^2)\, \rmfont{{\bf \large 1}}
+ \frac{4}{Q^4}
\langle 0| m\ovl{q} q |0\rangle\, ,
\label{hm11a}
\eeq
\begin{equation}
K_0^{NO}(Q^2,m ,\mu) = Q^2\Pi''(Q^2,m ,\alpha_s,\mu)|
{{}_{{}_{\scriptstyle \alpha_s =0}}}
=-\frac{1}{16\pi^2}
\left[1 + \frac{48m^4}{Q^4}(1 + L - M)\right].
\label{hm12}
\end{equation}
The coefficient function $K_0^{NO}$
contains mass singularities (the M-term).
On the other hand, if one does
not follow the normal ordering prescription,
then the operator $m \ovl{q} q $
develops a non-trivial vacuum expectation
value even if the quark gluon interaction is turned off by
setting $\alpha_s = 0$.
Indeed, after minimally removing its pole singularity,
the one loop diagram of Fig.~\ref{high-mom}b
leads to the following result \cite{Broad81}:
\begin{equation}
\langle 0|\bar q q|0\rangle^{\rmfont{PT}}=\frac{3m^3}{4\pi^2}
\left(\ln\, \frac{\mu^2}{m^2}+1\right)\,.
\label{hm13}
\end{equation}
By inserting this into \re{hm9}, the new coefficient
function $K_0$ can be
extracted, with the result:
\begin{equation}
 K_0 = -\frac{1}{16\pi^2}
\left[1 + \frac{48m^4}{Q^4} (2 + L)\right].
\label{hm14}
\end{equation}
The mass logarithms are now completely transferred from the CF $K_0$ to
the VEV of
the quark operator \re{hm13}! The same phenomenon continues to hold
even after the $\alpha_s$ corrections are taken into account for
(pseudo)scalar and pseudovector correlators, independently of their
flavour structure \cite{BroadGen87,Gen89}.

The underlying reason for this was first
found in Ref.~\cite{me82b}.
There it was shown that
no coefficient function can depend on  mass logarithms in every
order of perturbation theory if the minimal subtraction procedure is
scrupulously observed\footnote{In particular it also excludes the
normal ordering, as the latter amounts to a specific
non-minimal\break\hfill\hspace*{0.5cm} subtraction of diagrams contributing to
VEV's of composite operators.}.
This is true
irrespective of the specific model
and  correlator under discussion.
Three important observations
may be made in
 this context:
\begin{itemize}
\item
     Prior knowledge  of the fact that  any conceivable correlator
     can be expanded in   a series  of the form (\ref{hm10}) makes
     it possible to obtain without  calculation important
     information  on  the structure of  mass logs as they
     appear in various correlators. For example, in QCD  any
     correlator should contain no mass logarithms in the quadratic
     in  mass terms \cite{BroadGen84,CheSpi88}.
     This holds true  because there
     does not exist a gauge-invariant non-trivial operator of
     (mass) dimension two in QCD.
\item
      From the purely calculational point of view the problem
      of computing non-leading mass  corrections to  current
      correlators becomes much simpler. This is due to two facts.
      First, all coefficient functions are
      expressed in terms of {\em massless }  Feynman integrals
      while VEV's of composite operators are by definition
      represented in terms of some  massive integrals without
      external momenta (tadpole diagrams).
      Second,   methods  have been elaborated  for
      computing  analytically both types of Feynman integrals.
\item
      The abandonment of  the normal ordering slightly   complicates
      the  renormalization properties of composite operators.  An
      instructive  example is provided by the `quark mass operator'
      $O_2 = m \ovl q q$.
      The textbook statement (see, for example, Ref.~\cite{JCC84})
      that this operator is  RG invariant is no longer valid.
      Indeed, the  vacuum diagram of Fig.~\ref{high-mom}b
      has a divergent part which
      has to be removed by a new counterterm proportional to the
      operator $m^4\rmfont{{\bf \large 1}}$. In other words,
      $m\bar qq$ begins to mix with the `operator'
      $m^4\rmfont{{\bf      \large 1}}$
      \cite{CheSpi88}.
\end{itemize}
To lowest order, the corresponding anomalous dimension matrix reads:
\begin{equation}
\mu^2\frac{d}{d\mu^2}
\left(
\begin{array}{c}
m\bar q q\\
m ^4
\end{array}\right)=
\left(\begin{array}{cc}
0 & \myfrac{3}{4\pi^2} \\
0 & -4\myfrac{\alpha_s}{\pi}
\end{array}\right)
\left(\begin{array}{c}
 m\bar q q  \\
 m^4
\end{array}\right)\,.
\label{hm16}
\end{equation}
The non-vanishing, off-diagonal matrix element describes the mixing of
the two operators under renormalization and was obtained from the
divergent part of the vacuum diagram in Fig.~\ref{high-mom}.
The diagonal matrix
elements are just the anomalous dimensions of the respective
operators in the usual normal-ordering scheme. The lower one is equal
to $4\gamma_{\stand{m}}(\alpha_s)$.  Note that the  general structure of
the anomalous dimension matrix of all gauge-invariant operators of
dimension four  has been established in Refs.~\cite{Spi84,CheSpi88}.
This information was used recently
\cite{TTP94-08} to evaluate the corrections
of order $m_{\rmfont{q}}^4 \alpha_s^2$ to the
vector current correlator (see   Section~\ref{repnsm4}).
\sect{Top Mass Expansion in $s/m_{\rmfont{t}}^2$\label{repmtexp}}
Our discussion of the dependence of cross-sections and decay rates on
the quark masses has up to now dealt with  five flavours
light enough to be produced in $\rmfont{e}^+\rmfont{e}^-$ collisions.
The top quark, on
the other hand,  is too heavy to be present in the final state, even at
LEP energies.  Nevertheless it constitutes  a  virtual particle.
Virtual top loops appear
for the first time in second order $\as^2$.  Massive multi-loop
integrals may conveniently be simplified considering the heavy top
limit $m_{\rmfont{t}} \rightarrow \infty$.
In this effective field theory
approach the top is integrated out from the theory. Then the
Lagrangian of the effective theory contains only light particles.
The effects of the top quark are accounted for  through the
introduction of additional operators in the effective Lagrangian. For
the vector current correlator their contributions are suppressed by
inverse powers of the heavy quark mass $s/m_{\rmfont{t}}^2$.
As we will also
explicitly  see,
no decoupling is operative in the case of the
axial vector correlator. A logarithmic top mass dependence signals
the breakdown of anomaly cancellation if the top quark is removed
from the theory.

The heavy mass expansion is constructed as follows
(see Refs.~[65--68];
a rigorous mathematical formulation can be
found in Ref.~\cite{Smi91}):
Let the Feynman integral $\langle
\Gamma\rangle$ of a Feynman graph $\Gamma$
depend on a heavy mass $M$ and some other  `light' masses
and  external momenta which we will generically denote as
$m$ and $q$  respectively.  In the limit  $M \to \infty$
with  $q$ and $m$ fixed $\langle \G \rangle$
may    be represented by
the asymptotic expansion:
\beq
\langle \Gamma \rangle
\bbuildrel{=}_{{\scriptstyle{m_{\rmfont{t}} \to\infty}}}^{}
\sum_{\gamma}
C^{(t)}_{\gamma}
\star \langle \Gamma /\gamma\rangle^{\rmfont{eff}}~.
\EQN{t1}
\eeq
The diagrams $\langle \Gamma   /\gamma\rangle^{\rmfont{eff}}$ of the
effective theory consist of light particles only, whereas the top
mass is only present  in the  `coefficient functions'
$C^{(t)}_{\gamma}$. The notation $\langle \Gamma
/\gamma\rangle^{\rmfont{eff}}$ means that
the hard subgraph $\gamma$ of the
original diagram $\Gamma$ is contracted to a blob. By definition a
hard subgraph contains at least all heavy quark lines and becomes one
particle irreducible if each  top quark propagator is  contracted to a
point.  The Feynman integral of the hard subgraph is expanded in a
formal (multidimensional) Taylor expansion with respect to the small
parameters, namely the light masses and the external momenta of $\g$.
It should be noted that the set of external momenta for a subgraph
$\g$   is defined {\em with respect to} $\g$ and thus in general
consists   of some genuine external momenta (that is, those  shared by
$\g $  and the very diagram $ \G  $)  as
well as momenta flowing through {\em internal} lines of
$ \G  $,
which are {\em external ones } of $ \g $ (see the
example below).  This Taylor series $C^{(t)}_{\gamma}$ is inserted in
the effective blob and the resulting Feynman integral has to be
calculated.  All possible hard subgraphs have to be identified and
the corresponding results must be added.

The   prescription for the construction of the coefficent function
$C_\g^{(t)}$  for a hard subgraph $\g$ can be formulated as
follows: Suppose  the Feynman integral
$\langle \g\rangle (M,\qbf^\g,\mbf^\g,\mu)$ corresponds
to a hard subgraph $\g$ and
depends on  external momenta $\qbf^\g$
and  light masses $\mbf^\g$ in addition
to the  heavy mass $M$. Then
\beq
C_\g^{(t)} =
t_{\{\qbf^\g, \mbf^\g\}}
\langle \g\rangle
(M,\qbf^\g,\mbf^\g,\mu)
\, ,
\EQN{t2}
\eeq
where the operator $t_{\{x_1,x_2\dots\}}$ performs the formal  Taylor
expansion according to the rule:
\bea
t_{\{x_1,x_2\dots\}} &=&
\sum_{n \ge 0}^\infty  t_{\{x_1,x_2\dots\}}^{(n)} \, ,
\EQN{t3a}
\\
t_{\{x_1,x_2\dots\}}^{(n)}
F(x_1,x_2\dots)
&\equiv&
\frac{1}{n!}
\left( \frac{\rmfont{d}}{\rmfont{d} \xi} \right)^n
F(\xi x_1,\xi x_2\dots) |_{{}_{\scriptstyle \xi = 0}}
{}\,\, .
\EQN{t3b}
\eea
Here several  comments are in order.
\begin{itemize}
\item
The differentiation
with respect to $\xi$ in \re{t3b} may be carried out in
two ways.  One could simply differentiate
the  Feynman integral, which is a
smooth function of $\xi$ at $\xi\not=0$.
A more practical  way is to differentiate the corresponding
{\em integrand.}
\item  The operation of setting $\xi$ zero
is to  act on the
differentiated integrand.
\item
It may be immediately seen  that the
 expression
\[
t^{(n)}_{\{\qbf^\g , \mbf^\g\}}
\langle \g\rangle (M,\qbf^\g,\mbf^\g,\mu)
\]
scales with $M$ as $M^{\omega(\g) - n}$ where
$\omega(\g)$  is the (mass) dimension of the
Feynman integral $\langle \g\rangle $
determined  without counting any dimensionful
coupling constant  as well
the {}'t Hooft  mass $\mu$.
Therefore,  in every application  of the hard mass expansion
the terms with  too high value of $n$ in \re{t3a}
may be dropped.
\item  By construction the coefficient function $C_\g^{(t)}$
is  a polynomial with respect to its external momenta
$\qbf^\g$  and the light masses $\mbf^\g$.
\end{itemize}
\begin{figure}
 \begin{center}
  \begin{tabular}{ccccc}
  \parbox{3cm}{\epsfig{file=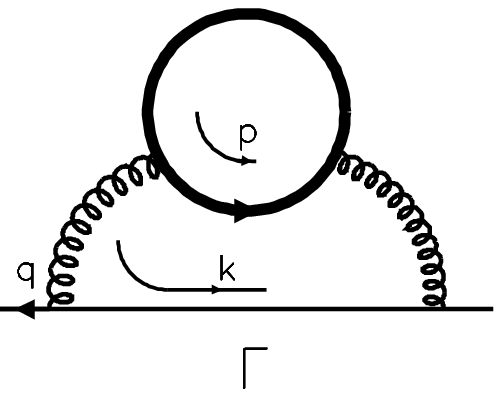,width=4cm,height=4cm}}
  & \hspace{.5cm} $\bf \longrightarrow$ &
  \parbox{3cm}{\epsfig{file=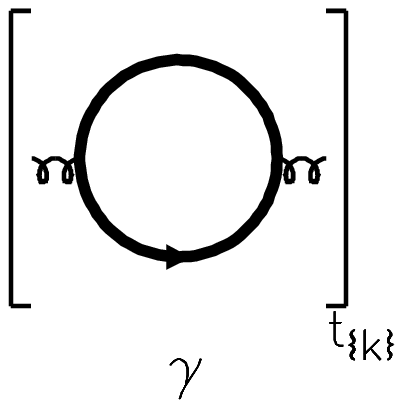,width=4cm,height=4cm}}
  & \hspace{.5cm} $\bf *$ &
  \parbox{3cm}{\epsfig{file=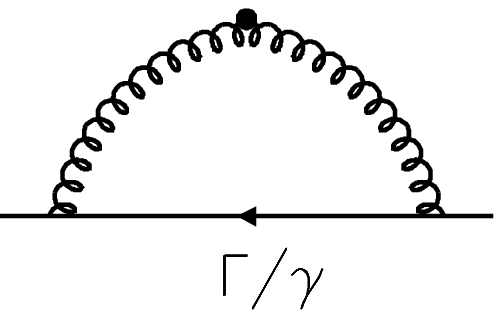,width=4cm,height=4cm}} \\
  & \hspace{.5cm} $\bf +$ &
  \parbox{3cm}{\epsfig{file=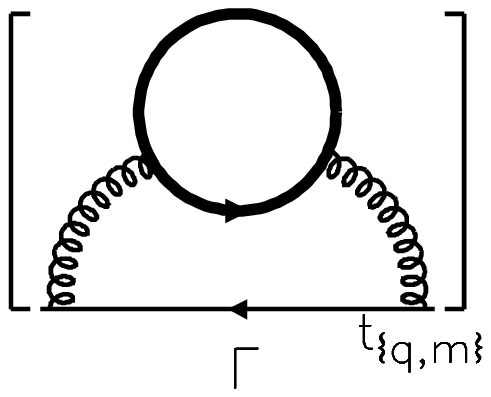,width=4cm,height=4cm}}
  \end{tabular}
 \end{center}
 \caption{
         \label{sdia}
{{Hard Mass Procedure.}}
         }
\end{figure}
As an example we consider
the  two-loop diagram $\Gamma$ depicted in Fig.~\ref{sdia} which
contributes to the fermion propagator in QED.
The heavy fermion of  mass
$M$ is contained in the  virtual fermion loop, whereas  the open
fermion line  corresponds to  a propagating   light fermion
with mass $m$.
The  integral reads (in  Feynman gauge):
\beq
\mu^{4\ep}
i^2
\int\frac{d^Dp}{(2\pi)^{\rmfont{D}}}
\int\frac{d^Dk}{(2\pi)^{\rmfont{D}}}
\frac{
\rmfont{{\rm Sp}}\left[
     \g_\al( \FMslash{p} + M)\g_\be
     (\FMslash{p} -\FMslash{k} + M)
                    \right]
               [
\g_\al(\FMslash{q}-\FMslash{k} + m)\g_\be
               ]
    }{
(-k^2)^2 (M^2 - p^2)[M^2 - (p-k)^2] [m^2 - (q-k)^2]
     }
{}\, .
\EQN{t4}
\eeq
The integration momenta are denoted as $k$ and $p$ for the outer and
the inner loops respectively.
Two different  integration regions can be identified.
In the first region  is
$k \ll M, p \simeq M$.  The corresponding hard subgraph
$\gamma_1$
is shown in Fig.~\ref{sdia} and  $\langle \g_1 \rangle $
has to be expanded
 with respect to its only
external momentum, $k$.  The second region is characterized by
$k,p \approx M$.  The hard subgraph $\gamma_2$  coincides with
$ \G $ and the Feynman integral
$\langle \g_2 /\gamma_2\rangle $
reduces  to  unity.  In this case the hard subgraph $\langle
\G\rangle $ must be expanded with respect to the external momentum
$q$ and,  in case of a non-vanishing light mass $m$,  also with
respect to $m$. The sum of all contributions results in a power series
in the inverse top mass.

Working up to the power-suppressed terms of order $q^2/M^2$, one has
\beq
C_{\g_1}^{(t)} = ( t^{(0)}_{\{k\}} + t^{(1)}_{\{k\}} + t^{(2)}_{\{k\}})
   \langle \g_1\rangle (M,k,\mu)
\ \
\rmfont{and}
\ \
C_{\g_2}^{(t)} =   ( t^{(0)}_{\{q,m\}} +t^{(1)}_{\{q,m\}} )
\langle \g_2\rangle (M,q,m,\mu)
{}\, .
\label{t5}
\nonumber
\eeq
In  explicit form these coefficient functions
are given by  the following  Feynman integrals
\beq
C_{\g_1}^{(t)} =
\mu^{2\ep}
i
\int\frac{d^{\rmfont{D}} p}{(2\pi)^{\rmfont{D}}}
\frac{
\rmfont{{\rm Sp}}\left[
     \g_\al(\FMslash{p} + M)\g_\be
     (\FMslash{p} -\FMslash{k} + M)
   \right]
     }{
      (M^2 - p^2)^2
      }
(\,  1 +  r + r^2 \, )
\EQN{t6}
\, ,
\eeq
with
$
\dsp
\ \
r = \frac{k^2 - 2pk}{M^2 - p^2}
\ \
$
and
\begin{equation}
C_{\g_2}^{(t)}  = {}
\mu^{4\ep}
i^2
\int\frac{d^Dp}{(2\pi)^{\rmfont{D}}}
\int\frac{d^Dk}{(2\pi)^{\rmfont{D}}}
\times
{}
\left[
\frac{
\rmfont{{\rm Sp}}\left[
     \g_\al(\FMslash{p} + M)\g_\be
     (\FMslash{p} -\FMslash{k} + M)
  \right]
\g_\al (\FMslash{q} - \FMslash{k} + m) \g_\be
    }{
(-k^2)^3 (M^2 - p^2)[M^2 - (p-k)^2]
     }
\right]
\, .
\label{t7}
\end{equation}
It is  of course understood that the terms of higher than second
order in the expansion parameters are  discarded in the integrand of
\re{t6}.
\sect{$\gamma_5$ in $D$ Dimensions
\label{gamma5}
        }
Multi-loop calculations with dimensional
regularization often encounter the
question of how to treat $\gamma_5$ in $D$
dimensions.
Occasionally the problem can be circumvented
by exploiting chiral symmetry which allows,
for example,  the relating of the non-singlet axial
correlator in the massless limit to the
corresponding vector correlator.
In general, however, a consistent definition
must be formulated. A rigorous choice
is based
on the original definition by 't Hooft and
Veltman \cite{dim.reg-a}, and formalized by
Breitenlohner and   Maison \cite{BreMai77}
with a  modification introduced in
\cite{{AkyDel73}}.
In this  self-consistent approach
$\gamma_5$ is defined as
 \cite{dim.reg-a}
 \beq\EQN{gam1}
\gamma_5 =
\frac{i}{4!}
\epsilon_{\mu\nu\rho\sigma}
\gamma^{\mu}\gamma^{\nu}\gamma^{\rho}
\gamma^{\sigma}\, ,
\eeq
with
$\epsilon_{0123} \equiv 1$.
For our discussion we consider
the cases for both the non-singlet
axial current $j_{5\mu}^{(\rmfont{NS})a}$ and
the singlet one  $j_{5\mu}^{(\rmfont{S})}$,
which are defined with the help of
the antisymmetrized combination
$\gamma^{[\nu\rho\sigma]}=
(\gamma^{\nu}\gamma^{\rho}\gamma^{\sigma}-
\gamma^{\sigma}\gamma^{\rho}\gamma^{\nu})/2$
in order to guarantee  Hermiticity
 for noncommuting $\gamma_5$
\beq\EQN{gam2}
\ba{ll}\dsp
j_{5\mu}^{(\rmfont{NS})a}
& \dsp
= \frac{1}{2}\ovl{\Psi}
(\gamma_{\mu}\gamma_5-\gamma_5\gamma_{\mu})
t^{\stand{a}}\Psi
\\ & \dsp
=
\frac{i}{3!}
\epsilon_{\mu\nu\rho\sigma}
\ovl{\Psi}
\gamma^{[\nu\rho\sigma]}
t^{\stand{a}}\Psi
=
\frac{i}{3!}
\epsilon_{\mu\nu\rho\sigma}
A_{\rmfont{NS}}^{[\nu\rho\sigma]a}
{}\, \,  ,
\\
\dsp
j_{5\mu}^{(\rmfont{S})}
& \dsp
= \frac{1}{2}\ovl{\Psi}
(\gamma_{\mu}\gamma_5-\gamma_5\gamma_{\mu})
\Psi
\\ & \dsp
=
\frac{i}{3!}
\epsilon_{\mu\nu\rho\sigma}
\ovl{\Psi}
\gamma^{[\nu\rho\sigma]}
\Psi
=
\frac{i}{3!}
\epsilon_{\mu\nu\rho\sigma}
A_{S}^{[\nu\rho\sigma]}
{}\, \, .
\ea
\eeq
Here $t^{\stand{a}}$ are the generators of the
$ SU(n_{\stand{f}}) $  flavour group.

The four-dimensional Levi-Civita tensor
$\epsilon_{\mu\nu\rho\sigma}$ is kept
outside the renormalization procedure
where all indices can be considered as
four dimensional whereas the calculation
is performed with the generalized
currents $A_{\rmfont{NS}}^{[\nu\rho\sigma]a},
A_{S}^{[\nu\rho\sigma]}$ in D dimensions.
As a consequence of the lost anticommutativity
of $\gamma_5$,  standard properties of the
axial current as well as the Ward identities are
violated. In particular,  it turns out that
the renormalization constant
$Z^{\rmfont{NS}}$
of the non-singlet
current is not one any more. To restore the
correctly renormalized non-singlet axial current an
extra  finite renormalization is introduced
with corresponding
finite renormalization constant
$z^{\rmfont{NS}}$
\cite{JCC84,Trueman79}.
One thus has for the renormalized non-singlet
axial current the following
expression:
\beq
\left(j_{5\mu}^{(\rmfont{NS})a}\right)_{\rmfont{R}}
\dsp
= z^{\rmfont{NS}} Z^{\rmfont{NS}}
\left(j_{5\mu}^{(\rmfont{NS})a}\right)_{\rmfont{B}}
\EQN{gam3-ns}
\eeq
\ice{
\beq
\EQN{gam3}
\ba{ll}\dsp
\left(j_{5\mu}^{(\rmfont{NS})a}\right)_{\rmfont{R}}
& \dsp
= Z_5^{\rmfont{NS}}Z^{\rmfont{NS}}_{\MSsmall}
\left(j_{5\mu}^{(NS)a}\right)_{\rmfont{B}}
\\ \dsp
\left(j_{5\mu}^{(\rmfont{S})}\right)_{\rmfont{R}}
& \dsp
= Z_5^{(\rmfont{S})}Z^{(\rmfont{S})}_{\MSsmall}
\left(j_{5\mu}^{(\rmfont{S})}\right)_{\rmfont{B}}
\ea
\eeq
}
with \cite{Larin91,Larin93}
\begin{eqnarray}
Z^{\rmfont{NS}}
&=& 1+
a^2
\frac{1}{\ep}
\left[
   \frac{11}{6}
 - \frac{1}{9}~n_{\stand{f}}
\right]
\nonumber\\
\dsp
&&+
{}~a^3
\frac{1}{\ep^2}
\left[
 - \frac{121}{36}
 + \frac{11}{27}~n_{\stand{f}}
 - \frac{1}{81}~n_{\stand{f}}^2
+ \ep
      \left(
  \frac{391}{72}
- \frac{44}{81}~n_f
+ \frac{1}{486}~n_{\stand{f}}^2
     \right)
\right]
\EQN{gam4}
\end{eqnarray}
and
\beq
z^{\rmfont{NS}} =
1
-
\frac{4}{3} a
+
a^2
\left(- \frac{19}{36} + \frac{1}{54}~n_f
\right)
{} \, .
\EQN{z-ns}
\eeq
The prescription described above and the use
of the non-singlet axial current defined according to
Eq.~(\ref{gam3-ns})
lead to the same characteristics for nonanomalous
amplitudes as would be
obtained within a na\"{\i}ve approach featuring completely
anticommutating $\gamma_5$.
First, the Ward identity  is
recovered.  Second, the anomalous
dimension of the non-singlet axial
current vanishes.  For diagrams with an
even number of $\gamma_5$ connected to
the external current it has been
checked that the treatment based on an
anticommutativity of $\gamma_5$ leads
to the same answer \cite{BroadKataev93}.

Similar considerations may be carried
out for the singlet axial vector
current. However, in this case there is
some freedom in defining the
renormalized current.  This is due to
the fact that in any physical
application the current never appears
as it is but only in a (virtually
non-singlet) combination with another
axial vector current. A physically
motivated definition has been
suggested in Ref.~\cite{CK3},   where the
singlet axial vector current has been
defined with the help of the following
limiting procedure:
\beq
(j^{(\rmfont{S})}_{5\mu})_{\rmfont{R}}
\bbuildrel{=\!=\!=}_{m_{\rmfont{T}} \to \infty}^{}
z^{\rmfont{NS}}Z^{\rmfont{NS}}
\left[
j^{(\rmfont{S})}_{5\mu}
-
n_{\stand{f}}\frac{i}{3!}
\epsilon_{\mu\nu\rho\sigma}
\ovl{\Psi}_{\rmfont{T}}
\gamma^{[\nu\rho\sigma]}
\Psi_{\rmfont{T}}
\right]_{\rmfont{B}}
{}\, .
\EQN{lim-proc}
\eeq
Here,  $\Psi_{\rmfont{T}}$ is the field of an   auxiliary quark
$\rmfont{T}$
and thus the combination in the squared brackets is
a non-singlet one (in the extended QCD with $n_{\stand{f}} +1$
flavours!). Due to the asymptotic freedom,  the
large $m_{\rmfont{T}}$ limit of \re{lim-proc}
does exist and is naturally identified with the
renormalized singlet axial current.
Explicitly, the r.h.s. of \re{lim-proc}  can
be written without any auxiliary fields in the
form (note that the renormalization constant
$Z^{\rmfont{S}}$ was  first found in Ref.~\cite{Larin93})
\beq
\left(j_{5\mu}^{(\rmfont{S})}\right)_{\rmfont{R}}
\dsp
= z^{\rmfont{S}} Z^{\rmfont{S}}
\left(j_{5\mu}^{(\rmfont{S})}\right)_{\rmfont{B}}
\EQN{gam3-s}
{}\, ,
\eeq
with
\begin{eqnarray}
Z^{\rmfont{S}}
 \dsp
&=& 1+
a^2
\frac{1}{\ep}
\left[
  \frac{11}{6}
 + \frac{5}{36}~n_f
\right]
\nonumber\\
\dsp
&&+~a^3
\frac{1}{\ep^2}
\left[
- \frac{121}{36}
- \frac{11}{216}~n_f
+ \frac{5}{324}~n_{\stand{f}}^2
+\ep
   \left(
   \frac{391}{72}
  + \frac{61}{1296}~n_f
  + \frac{13}{1944}~n_{\stand{f}}^2
   \right)
\right]
\EQN{zs}
\end{eqnarray}
and
\begin{eqnarray}
\dsp
\!\!\!z^{\rmfont{S}}& =&
1
+
a \frac{( - \frac{5}{18}~n_{\stand{f}} - \frac{11}{3})}{\beta_0}
\\
\dsp
&&+
{}~a^2 \left[
 \frac{1}{\beta_0^2}
 \left(
   - \frac{185}{2592}~n_{\stand{f}}^2
   + \frac{391}{864}~n_f
   + \frac{2651}{144}
 \right)
-\frac{1}{\beta_0}
\left(\frac{13}{1296}~n_{\stand{f}}^2
 + \frac{61}{864}~n_f
 + \frac{391}{48}
\right)
\right],\nonumber
\EQN{z-s}
\end{eqnarray}
where $\beta_0 = (11 - \frac{2}{3}~n_{\stand{f}})/4$.
It should be noted that an equivalent
definition of the singlet axial  vector current
is obtained by demanding that it  have
a vanishing anomalous dimension.
\chap{Exact Result of Order $\ordas)$
\label{exact}}
The exact QCD corrections for arbitrary
quark masses
are
known in order $O(\alpha_s)$. The
result is different for vector and axial current
correlators. Whereas the
former can be taken directly
from QED \cite{Schw73}
the latter have been obtained in Ref.~\cite{Zerwas80}.
(For the non-diagonal current and arbitrary,
different masses the result can be found in
\cite{chang-gaemers}.)
With $v^2=1-4m^2/s$ they read:
\beq \ba{ll}\EQN{ap1}
r^{\rmfont{V}}_{\rmfont{NS}} = & \dsp
v\frac{3-v^2}{2}\left[1+\frac{4}{3}\apis K_{\rmfont{V}}\right]
{}\, ,
\\ \dsp
r^{\rmfont{A}}_{\rmfont{NS}} = & \dsp
v^3\left[1+\frac{4}{3}\apis K_{\rmfont{A}}\right]
{}\, .
\ea
\eeq
$K_{\rmfont{V}}$ and $K_{\rmfont{A}}$ have been calculated
in Refs.~[76--78].
A compact form for the correction can be found
in Ref.~\cite{KniKue90b}:
\beq
 \ba{ll}\EQN{ap2}
K_{\rmfont{V}} = & \dsp
\frac{1}{v}\left[ A(v) +
 \frac{P_{\rmfont{V}}(v)}{(1-v^2/3)}\ln\,\frac{1+v}{1-v}
     + \frac{Q_{\rmfont{V}}(v)}{(1-v^2/3)}\right],
\\ \dsp
K_{\rmfont{A}} = & \dsp
\frac{1}{v}\left[ A(v)
+ \frac{P_{\rmfont{A}}(v)}{v^2}\ln\,\frac{1+v}{1-v}
                       + \frac{Q_{\rmfont{A}}(v)}{v^2}\right]
{},
\ea
\eeq
with
\beq
\ba{ll}\EQN{ap3}   \dsp
A(v) =& \dsp (1+v^2)
\left[
{\rm Li}_2\left(
      \left[\frac{1-v}{1+v}\right]^2
          \right)
+2{\rm Li}_2\left(\frac{1-v}{1+v}\right)
+\ln\,\frac{1+v}{1-v}\ln\,\frac{(1+v)^3}{8v^2}
 \right]
\\ & \dsp
          +~3v\ln\,\frac{1-v^2}{4v}-v\ln\, v
{}\, ,
\ea
\eeq
\beq
\ba{ll}
\EQN{ap4}
\dsp
P_{\rmfont{V}}(v) = \frac{33}{24}+\frac{22}{24}v^2
-\frac{7}{24}v^4 \, ,
& \dsp
Q_{\rmfont{V}}(v) = \frac{5}{4}v-\frac{3}{4}v^3  \, ,
\\ \dsp
P_{\rmfont{A}}(v) = \frac{21}{32}+\frac{59}{32}v^2
-\frac{19}{32}v^4-\frac{3}{32}v^6 \, ,
\;\;\;\;& \dsp
Q_{\rmfont{A}}(v) = -\frac{21}{16}v+\frac{30}{16}v^3
+\frac{3}{16}v^5
{}\, .
\ea
\eeq
Convenient parametrizations are
\cite{Kuhn85}:
\beq
\ba{ll}
\EQN{n18}
K_{\rmfont{V}} = & \dsp
\frac{\pi^2}{2v} - \frac{3+v}{4}
\left(\frac{\pi^2}{2}-\frac{3}{4}\right)
{},
\\
\dsp
K_{\rmfont{A}} = & \dsp
\frac{\pi^2}{2v} - \left[\frac{19}{10}
-\frac{22}{5}v+\frac{7}{2}v^2\right]
\left(\frac{\pi^2}{2}-\frac{3}{4}\right)
{}.
\ea
\eeq
Let us consider this result  in the limit
where ${s}$ approaches the threshold region
($v\rightarrow 0$) as well as the
high energy regime ($v\rightarrow 1$).

For $v\rightarrow 0$ the correction factors simplify
to:
\beq
\ba{ll} \dsp
1+\frac{4}{3}\as K_{\rmfont{V}} \stackrel{v\rightarrow 0}
{\longrightarrow}
& \dsp
\frac{2\pi\as}{3v}+\left(1-\frac{16}{3}~\api\right)
{},
\\ \dsp
1+\frac{4}{3}\as K_{\rmfont{A}} \stackrel{v\rightarrow 0}
{\longrightarrow}
& \dsp
\frac{2\pi\as}{3v}+\left(1-\frac{8}{3}~\api\right)
{}.
\ea
\eeq
For very small $v$ higher-order contributions
must be taken into consideration.
In QED these can be summed to yield the
Sommerfeld rescattering factor:
\beq
R_{\rmfont{QED}} = \frac{\pi\alpha/v}
          {1-e^{-\pi\alpha/v}}
\eeq
In QCD the coupling constant $\alpha$ would be
replaced in this formula by $4\as /3$.

However, the scale of $\as(Q^2)$ cannot be fixed
 with certainty, since subleading logarithms have
not yet been evaluated. It has been argued in
Ref.~\cite{Kuhn85} that the choice $\as(|P_{\rmfont{t}}|)$,
combined with
Eq.~(\ref{n18}) allows for an adequate
 description of $R$ in
the threshold region
and provides a smooth connection
 between resonances and continuum.
For top quarks a new element enters
through their large decay rate. Resonance and open
$t\ovl{t}$ production merge. An account of the
resulting phenomena is beyond the scope
of this paper and can be found
in Refs.~[81--86].
\begin{figure}
\begin{center}
\epsfxsize=12.0cm
\leavevmode
\epsffile[130 300 460 525]{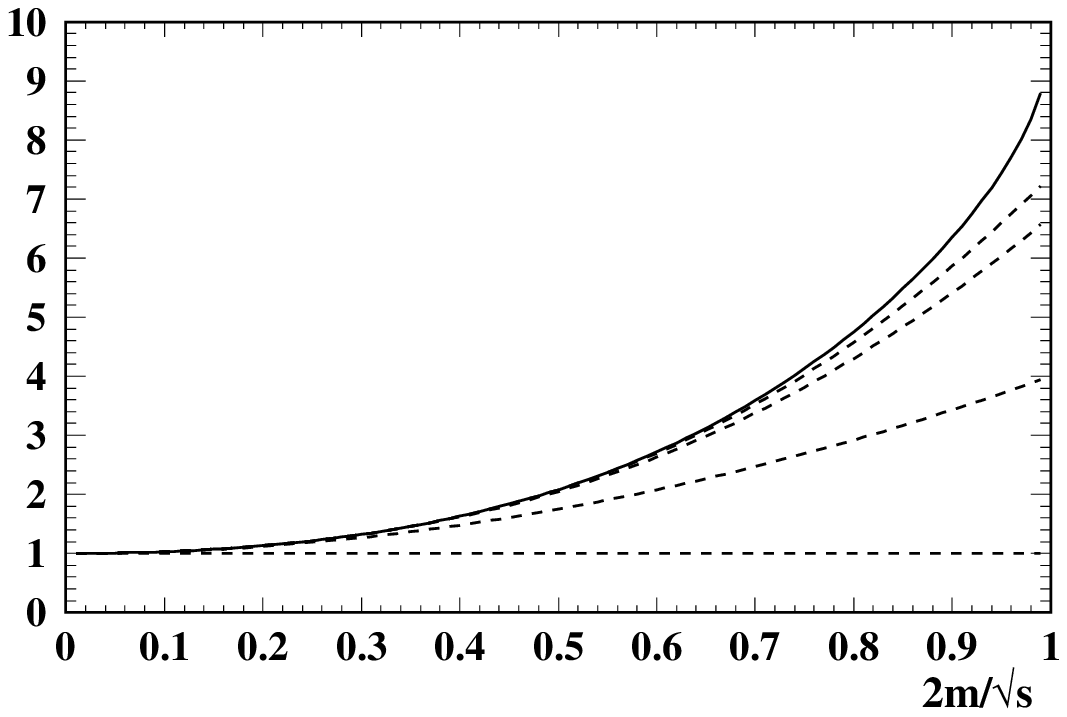}\\
\epsfxsize=12.0cm
\leavevmode
\epsffile[130 300 460 525]{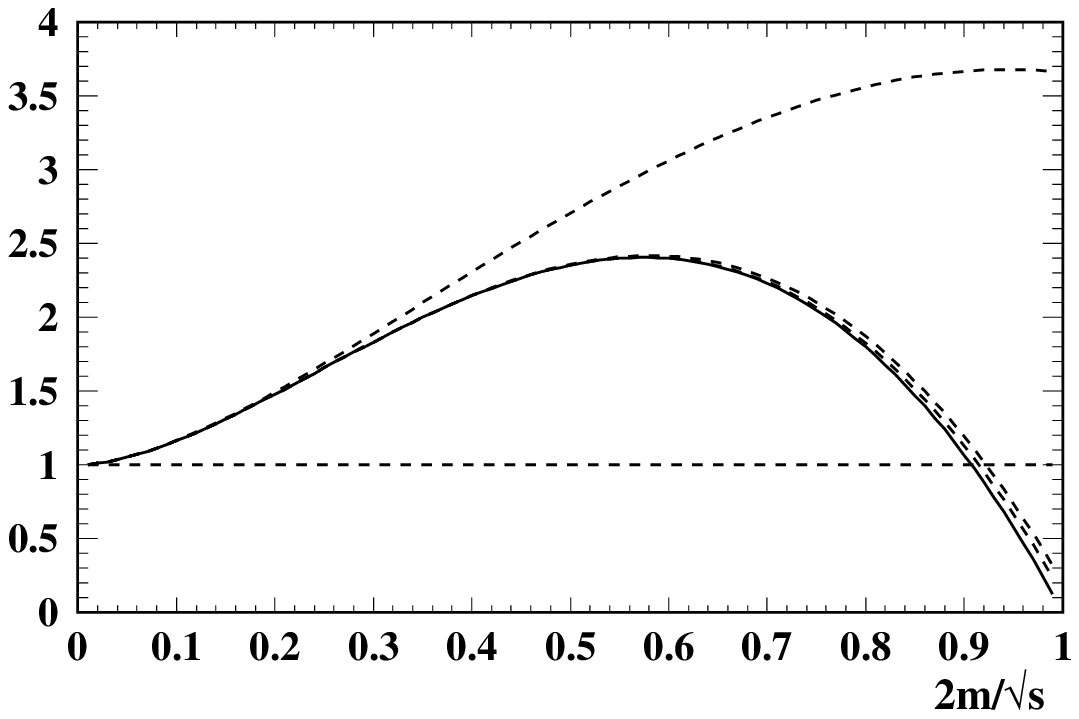}
\caption{\label{kvexp}{{Comparison between the
complete ${\cal O}(\alpha_s)$ correction
function (solid line) and approximations of increasing order
(dashed lines) in $m^2$ for vector (upper graph)
and axial vector current (lower graph) induced rates.}}}
\end{center}
\end{figure}

The behaviour of the result for large $s$
can easily be extracted from the analytic formulae
\cite{KniKue90b,Kuhn90b,CheKue90}.
In Born approximation the leading term
of the  vector and the axial vector
 correlators are of order
$m^4/s^2$ and $m^2/s$ respectively:
\beq
\ba{rl} \dsp
v\frac{3-v^3}{2}
& \dsp
\longrightarrow 1-6\frac{m^4}{s^2}
+{\cal O}(m^6/s^3)
{}\, ,
\\ \dsp
v^3
& \dsp
\longrightarrow 1-6\frac{m^2}{s}+6\frac{m^4}{s^2}
+{\cal O}(m^6/s^3)
{}\, .
\ea
\eeq
Including first-order QCD corrections leads to:
\beq\EQN{f1}
\ba{ll}\dsp
r^{\rmfont{V}}_{\rmfont{NS}}
\stackrel{v\rightarrow 1}{\longrightarrow}
& \dsp
1-6\frac{m^4}{s^2}
\\ & \dsp
+~\api
\left[
1+12\, \frac{m^2}{s}+\frac{m^4}{s^2}
\left(
10-24 \, \ln\,\frac{m^2}{s}
\right)
\right]
{}\, ,
\\ \dsp
r^{\rmfont{A}}_{\rmfont{NS}}
\stackrel{v\rightarrow 1}{\longrightarrow}
& \dsp
1-6\frac{m^2}{s}+6\frac{m^4}{s^2}
\\ & \dsp
+~\api
\left[
1-\frac{m^2}{s}
\left(
6+12\, \ln\,\frac{m^2}{s}
\right)
+\frac{m^4}{s^2}
\left(
-22+24\, \ln\,\frac{m^2}{s}
\right)
\right]
{}\, .
\ea
\eeq
The approximations to the correction functions
for the vector and the axial vector current
correlators
(including sucessively higher-orders without
the factor $\as/\pi$)
 are compared to the full result
in Fig.~\ref{kvexp}.
As can be seen in this figure,
for high energies --- say for $2 m_{\rmfont{b}}/\sqrt{s}$
 below 0.3 ---  an
excellent approximation is provided by the
constant plus the $m^2$ term.
In the region of $2m/\sqrt{s}$ above 0.3
the $m^4$ term becomes
increasingly important. The inclusion of
 this term improves the
agreement significantly and leads to an
excellent approximation,  even up
to $2m/\sqrt{s}\approx 0.7$ or 0.8. For the
narrow region between 0.6
and 0.8 the agreement is further improved
 through the $m^6$ term.

The mass $m$ in this formula is understood
as the physical mass, defined through the
location of the pole of the quark propagator in
complete analogy with the treatment of the electron
mass in QED.
However, if one tried to control fully the
 $m^2/s$ and $m^4/s^2$ terms
one might worry about the logarithmically enhanced
coefficient which could  invalidate perturbation
theory.
These leading logarithmic terms
may be  summed through renormalization group
techniques.

In order $\as$ this can be trivially achieved by
substituting
\[
m^2=\ovl{m}^2 \Biggl[1+\api(8/3-2)\ln\,(\ovl{m}^2/s)\Biggl]
\]
which implies (for completeness also $m^6/s^2$
 terms from Ref.~\cite{TTP94-08} are included)
\beq\EQN{f2}
\ba{ll}\dsp
r^{\rmfont{V}}_{\rmfont{NS}}
\stackrel{v\rightarrow 1}{\longrightarrow}
& \dsp
1-6\frac{\ovl{m}^4}{s^2}-8\frac{\ovl{m}^6}{s^3}
\\ & \dsp
+~\api
\left[
1+12\frac{\ovl{m}^2}{s}-22\frac{\ovl{m}^4}{s^2}
-\frac{16}{27}
\left(
6\ln\,\frac{\ovl{m}^2}{s}+155
\right)\frac{\ovl{m}^6}{s^3}
\right]
{}\, ,
\\ \dsp
r^{\rmfont{A}}_{\rmfont{NS}}
\stackrel{v\rightarrow 1}{\longrightarrow}
& \dsp
1-6\frac{\ovl{m}^2}{s}+6\frac{\ovl{m}^4}{s^2}
+4\frac{\ovl{m}^6}{s^3}
\\ & \dsp
+~\api
\left[
1-22\frac{\ovl{m}^2}{s}+10\frac{\ovl{m}^4}{s^2}
+\frac{8}{27}
\left(
-39\ln\,\frac{\ovl{m}^2}{s}+149
\right)
\frac{\ovl{m}^6}{s^3}
\right]
{}\, .
\ea
\eeq
A systematic discussion of higher-order terms
will be given in the subsequent sections.

\chap{Non-singlet Contributions
\label{nonsinglet}}
\sect{Massless Limit\label{repnsm0}}
This section will cover those results
obtained in the limit of massless
quarks. As discussed in the previous part,
non-singlet contributions
exhibit a universal charge
 factor which is given by the Born result
and can be trivially factored.
The  first-order correction was  derived in the
context of QED some time ago \cite{Schw73}.
The second order coefficient has been calculated
by several groups \cite{CheKatTka79}.
The initial calculation of the
$O(\alpha_s^3)$ term described  in
\cite{Gorishny88} was later corrected
by two  groups \cite{Gorishny91,Surguladze91}.
An implicit test of the results has recently  been
performed \cite{BroadKataev93}.

The full result for the
non-singlet current reads as follows:
\bea
\dsp
r^{(0)}_{\rmfont{NS}}(s)& =&
1
+
 \frac{\alpha_s(s)}{\pi}
+
\left[\frac{\alpha_s(s)}{4\pi}\right]^2
\left\{
  \frac{730}{3} - 176~\zeta(3) +
  \left[-\frac{44}{3} + \frac{32}{3} \zeta(3)\right] n_f
\right\}
\nonumber
\\
\dsp
&&+
\left[\frac{\alpha_s(s)}{4\pi}\right]^3
\left\{
   \frac{174058}{9}  - 17648~\zeta(3) + \frac{8800}{3} \zeta(5)
\right.
\EQN{ns1}
\\
&&+
\left[-\frac{62776}{27} + \frac{16768}{9}
       \zeta(3) - \frac{1600}{9} \zeta(5)
     \right] n_f
\nonumber
\\
\dsp
&&+
\left.
\left[  \frac{4832}{81} - \frac{1216}{27} \zeta(3)
     \right] n_{\stand{f}}^2
          - \left[
       \frac{484}{3} - \frac{176}{9}~n_{\stand{f}}
         + \frac{16}{27}~n_{\stand{f}}^2
     \right] \pi^2
\right\}
{}\, .
\nonumber
\eea
This leads to the following
numerical result:
\bea
\dsp
r^{(0)}_{\rmfont{NS}}(s) &=&
1
+
\apis
     +\left[\apis\right]^2
  (
1.9857 - 0.1153\,~n_f
  )
\nonumber
\\
\dsp
&&+
\left[\apis\right]^3
  (
         - 6.6369
         - 1.2001 \, n_f
         - 0.0052 \,~n_{\stand{f}}^2
  )
{}\, .
\EQN{ns2}
\eea
Those terms which depend on the number
of quark flavours $n_{\stand{f}}$ are due to virtual
fermion loops with light quarks. They appear
for the first time at second order $\as^2$.

Mixed QED and QCD corrections can be deduced from
the QCD results in a straightforward manner
\cite{Kat92}.
One obtains
\beq\EQN{ns3}
r^{(0)}_{\rmfont{QED}} = Q_{\stand{f}}^2~\frac{3}{4}
{}~\frac{\alpha(s)}{\pi}
\left[
  1-\frac{1}{3}~\apis
\right]
{}\, .
\eeq
Corrections of order $\alpha^2$ are
also given in Ref.~\cite{Kat92}.  They are
small and will not be considered
here.
\ice{
\beq \EQN{ns1}
\ba{l} \dsp
r^{(0)}_{\rmfont{NS}}(s) = 1 +
 \frac{\alpha_s}{4\pi}(3C_{\rmfont{F}})
\\ \dsp
+\left(\frac{\alpha_s}{4\pi}\right)^2
\left[C_{\rmfont{F}}^2\left( -\frac{3}{2}\right)
      + C_{\rmfont{F}} C_{\rmfont{A}}
\left(\frac{123}{2}-44\zeta(3)\right)
      +n_fTC_{\rmfont{F}}(-22+16\zeta(3))\right]
\\  \dsp
+\left(\frac{\alpha_s}{4\pi}\right)^3
\left[C_{\rmfont{F}}^3\left(-\frac{69}{2}\right)
      +C_{\rmfont{F}}^2C_{\rmfont{A}}(-127-572\zeta(3)
    +880\zeta(5))\right.
\\  \dsp
\hphantom{(\frac{\alpha_s}{4\pi})^3}
+ C_{\rmfont{F}} C_{\rmfont{A}}^2 \left(\frac{90445}{54}
          -\frac{10948}{9}\zeta(3)
                  -\frac{440}{3}\zeta(5)\right)
\\  \dsp
\hphantom{(\frac{\alpha_s}{4\pi})^3}
+ n_{\stand{f}} T C_{\rmfont{F}}^2 (-29+304\zeta(3)-320\zeta(5))
\\  \dsp
\hphantom{(\frac{\alpha_s}{4\pi})^3}
+ n_fT C_{\rmfont{F}} C_{\rmfont{A}} \left(-\frac{31040}{27}
               +\frac{7168}{9}\zeta(3)
               + \frac{160}{3}\zeta(5)\right)
\\  \dsp \left.
\hphantom{(\frac{\alpha_s}{4\pi})^3}
+ n_{\stand{f}}^2T^2 C_{\rmfont{F}} \left(\frac{4832}{27}
               -\frac{1216}{9}\zeta(3)\right)
-\pi^2C_{\rmfont{F}}\left(\frac{11}{3}-\frac{4}{3}n_fT\right)^2
\right]
\ea
\eeq
}
\sect{Top Mass
Corrections\label{top}}
The top quark is also
present at second order through a virtual quark loop.
The corrections in the
corresponding double bubble diagram
(Fig.~\ref{doubub}) are known in analytical
form, if the masses of the quarks in the
external loop are neglected
\cite{Kniehl90,Hoang94}.
The absorptive part from the cut through
the two (massless) quark lines contributes for
$s>0$ and is calculated in Ref.~\cite{Kniehl90}.
The one from the cut through all four
quark lines contributes for $s>4m_{\rmfont{t}}^2$ and
can be found in Ref.~\cite{Hoang94}. Only
the former is of relevance for the present
 discussion.
Its contribution to
$r_{\rmfont{NS}}^{(0)}$ reads:
\begin{eqnarray}
r_{\rmfont{NS}}^{(0)} &=&
 \left[\apis\right]^2
 \dsp
\left\{
\frac{4}{9}\left(1-6x^2\right)
\left[
{\rm Li}_3(A^2) -\zeta(3)-2\zeta(2)\ln\, A
   + \frac{2}{3}\ln^3 A
\right]
\right.
\nonumber\\  \dsp
&&+~\frac{2}{27}\left(19+46x\right)
\sqrt{1+4x}
\left[
{\rm Li}_2(A^2) -\zeta(2)+\ln^2 A
\right]
\\
\dsp
&&+~\frac{5}{54}\left(\frac{53}{3}+44x\Biggl)\ln\, x
+\frac{3355}{648}+\frac{119}{9} x
\right\}\, ,\nonumber
\end{eqnarray}
where
$A=(\sqrt{1+4x}-1)/\sqrt{4x}$ with $x=m_{\rmfont{t}}^2/s$.

\begin{figure}[t]
\begin{center}
\parbox{3cm}{
\mbox{\epsfig{file=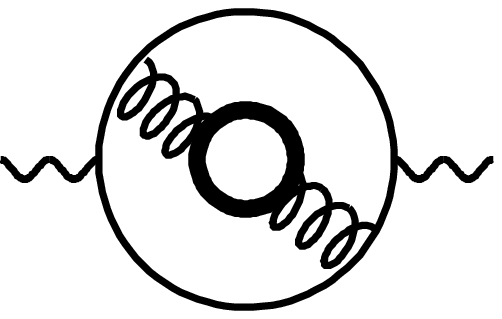,width=5.cm,height=5.cm}}
            }
\end{center}
\caption[]{\label{doubub}{{Double Bubble Diagram.}}}
\end{figure}
The leading term has also been determined
\cite{me93}
by employing the heavy mass expansion as
described in Section~\ref{decoupling}.
 In  the heavy top
limit the correction reads:
\beq
\EQN{mt2c}
r_{\rmfont{NS}}^{(0)} = \left[\apis\right]^2 \frac{s}{m_{\rmfont{t}}^2}
\left(\frac{44}{675}+ \frac{2}{135}\ln\,
\frac{m_{\rmfont{t}}^2}{s}
\right)
{}\, .
\eeq
As shown in Fig.~\ref{figadd},
the heavy mass expansion provides an excellent
approximation to the full answer from
$m_{\rmfont{t}} \gg s$,  even down to the threshold
$4m_{\rmfont{t}}^2=s$.
The result was
derived in the theory with $n_{\stand{f}} =6$, whence
$\as$ should be taken for the correction term
accordingly. However, since
$\as\Big|_{n_{\stand{f}} =5}=\as\Big|_{n_{\stand{f}} =6}+\ordas^2)$,
this distinction is irrelevant for the terms under
consideration.  Note that the diagrams
of Fig.~\ref{doubub} were
studied in Refs.~\cite{Bernreuther81,Bernreuther82},
where an exact double integral
representation was obtained.
The r.h.s. of (153)
was  numerically evaluated in
Ref.~\cite{Soper94}.

It seems appropriate at this point to already here
anticipate the mass corrections
arising from internal loops of quarks
with $m^2/s\ll 1 $. Also, these corrections are
universal. They will be derived in Sections
\ref{repnsm2} and \ref{repnsm4}.
The leading $\as^2m^2/s$ term is absent.
The first non-vanishing terms are of order
$\as^3m^2/s$ and $\as^2m^4/s^2$ and provide
a correction,
\beq
\ba{ll}\dsp
r^{(0)}_{\rmfont{NS}} =
& \dsp
\left[\apis\right]^3
\left[-15+\frac{2}{3}~n_{\stand{f}} \right]
\left[\frac{16}{3}-4\zeta(3)\right]
\sum_{\stand{f}} \frac{\ovl{m}_{\stand{f}}^2}{s}
\\ & \dsp
+\left[\apis\right]^2
\sum_{\stand{f}} \frac{\ovl{m}_{\stand{f}}^4}{s^2}
\left[
\frac{13}{3}-\ln\,\frac{\ovl{m}_{\stand{f}}^2}{s}
 -4\zeta(3)
\right]
{}\,  .
\ea
\EQN{mt2b}
\eeq
These corrections,
as well as those from a heavy top,
 apply equally well to
vector and axial correlators.
\begin{figure}
\begin{center}
\leavevmode
\mbox{}\epsffile[100 300 500 540]{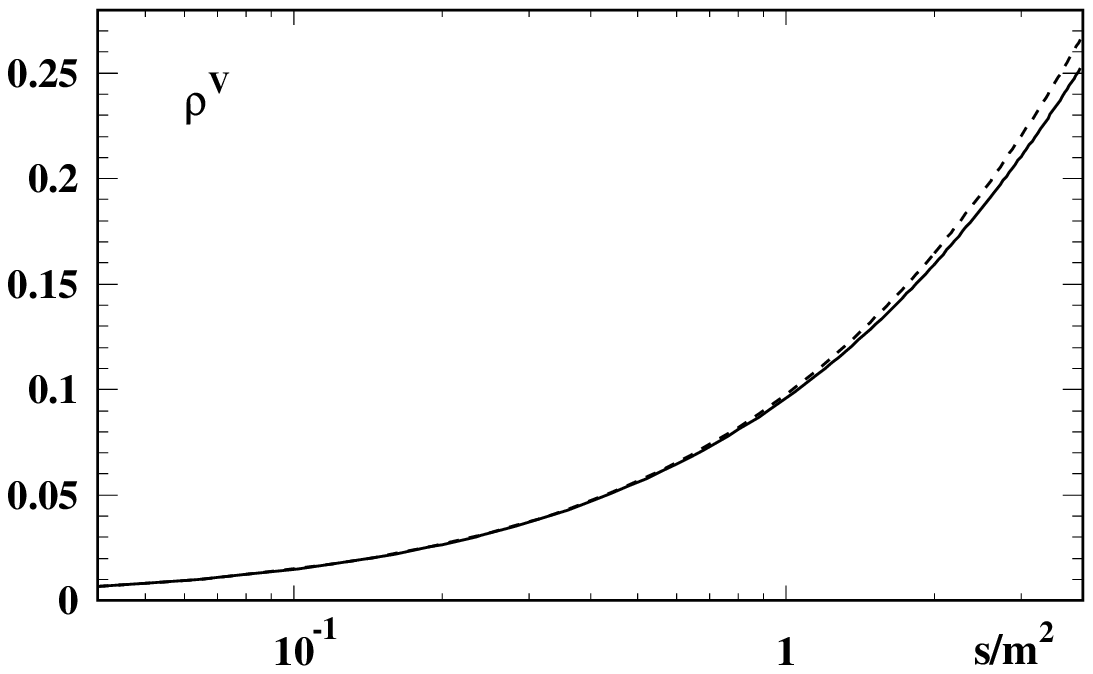}
\caption{\label{figadd}
The function
$\varrho^{\rmfont{V}}$ describing virtual corrections in the
range $s/m^2<4$ (solid curve) and
the approximation of the heavy mass expansion
(dashed curve).}
\end{center}
\end{figure}

\sect{Mass Corrections of Order $m^2/s$
\label{repnsm2}}
In view of the high precision reached in the
cross-section measurements the large size of the
first-order corrections made the knowledge
of higher order QCD corrections desirable.
Their exact computation for arbitrary
quark masses would be a tremendous task.
Fortunately
for many considerations and experimental
conditions, quark masses can be neglected
in comparison with   the characteristic energy of the
problem,  or are considered as small parameters.
This holds true for the light
$\rmfont{u,d}$ and $\rmfont{s}$ quarks, once the CMS energy
exceeds a few GeV, and is equally valid for charm and bottom quarks
at LEP energies of about $90$ GeV.
The problem may therefore be
simplified by performing an expansion in the
small parameter $m^2/s$,
which reduces the calculational effort
to massless propagator integrals.
The leading quadratic  terms $m^2/s$ and,
for the case of lower energies,    the
quartic mass terms $m^4/s^2$, represent
a very good approximation.

In first order this expansion is trivially
obtained in Eq.~(\ref{f1}) from the exact result.
We already noticed the large
logarithm $\ln\, m^2/s$ which makes the reliability
of perturbation theory questionable.
Its occurrence is connected to the use of  the pole
mass as an expansion parameter.
The problem may be overcome
by employing RG techniques and is
conveniently achieved in the $\msbar$-scheme.
In this calculational scheme leading logarithms
$\ln\, m^2/s$ are summed and absorbed in the
$\msbar$ mass $\ovl{m}(\mu^2)$ with $\mu$ being the
renormalization scale.
One can then write
\begin{eqnarray}
\dsp r^{\rmfont{V}}(f)& =&
 \dsp
r^{(0)} + \frac{\ovl{m}^2}{s}r^{V(1)}
+  \frac{\ovl{m}^4}{s^2}r^{V(2)}
+ {\cal O}(m^6/s^3)
{}\, ,
\nonumber\\ \dsp
\dsp r^{\rmfont{A}}(f) &=&
 \dsp
r^{(0)} + \frac{\ovl{m}^2}{s}r^{A(1)}
 + \frac{\ovl{m}^4}{s^2}r^{A(2)}
+ {\cal O}(m^6/s^3)
\, .
\end{eqnarray}
The massless results $r^{(0)}$ are identical
for the vector and the axial vector
correlators, whereas the mass corrections
$r^{V(n)}$ differ from $r^{A(n)}$ for $n\geq 1$.

It was found in Ref.~\cite{me82b}
(discussed  in some detail in Section \ref{large})
that the $\msbar$ scheme has the remarkable
property of all  coefficient functions for
QCD operators being  polynomial in masses and momenta.
{}From this,  and the fact that
no non-trivial operators of mass dimension two
exist in QCD,  follows that no logarithms
$\ln\, m^2/s$  appear in $r^{V(1)}$ and $r^{A(1)}$.
Therefore $r^{V/A(0)}$ and $r^{V/A(1)}$ can be
written as a perturbation expansion to all orders
in $\as$ with pure numerical coefficients.

Since to first order
$\as$ only trivial operators (unit operator
times a combination of quark masses) of mass
dimension four exist, logarithms
in $r^{(2)}$ are absent in order $\as$ and
show up for the first time in second order $\as^2$.
\subsect{Vector-Induced Corrections\label{vector}}
In this section we demonstrate that
the mass corrections of
 order ${\cal O}(\as^3)$ to the
flavour non-singlet contribution of the
vector-induced decay rate $\Gamma^{\rmfont{V}}$
can be obtained
from the three-loop vector current
correlator \cite{CheKue90}
without an explicit
four-loop calculation.
The argument
is based on the RG invariance
of the Adler function
{\renewcommand{\arraystretch}{2}
\bea
\dsp
D^{\rmfont{V}}(Q^2)  &=&
-12 \pi Q^2 \frac{d}{dQ^2}\left(
\frac{\Pi^{\rmfont{V}}_1}{Q^2}\right)
\nonumber
\\
&&=
3~\Bigg\{ \left[ 1 + \apimu +
\dots\right]
 - \frac{\ovl{m}^2(\mu)}{Q^2} \left[
b^{\rmfont{V}}_{00} + (b^{\rmfont{V}}_{01}+b^{\rmfont{V}}_{11}\ell)
\left(\apimu \right) \right.
\nonumber
\\
&&+
 ~(b^{\rmfont{V}}_{02}-a^{\rmfont{V}}_{02}+b^{\rmfont{V}}_{12}\ell
+b^{\rmfont{V}}_{22}\ell^2)\left(\apimu\right)^2
\EQN{nsmv1}
\\
&&+
\left.
(b^{\rmfont{V}}_{03}+b^{\rmfont{V}}_{13}\ell
+ b^{\rmfont{V}}_{23}\ell^2+b^{\rmfont{V}}_{33}
\ell^3)\left(\apimu \right)^3
\right] \Bigg\}
\nonumber
{}~,
\eea
}
\noindent
\hspace*{-0.1cm}where
$\ell\equiv \ln\,(\mu^2/Q^2)$.
The term $a^{\rmfont{V}}_{02}$ originates from the
$\rmfont{b}$ quark propagating in an inner
fermion loop.
Mass corrections  are  therefore also present for
the decay of the $\stand{Z}$-boson into massless quarks.
The coefficients up to and including
the second order $\ordas^2)$
were obtained in
\cite{GorKatLar86}
(the~$a^{\rmfont{V}}_{02}$~term~was~first~computed~in~
\cite{Bernreuther81}).~
They~read\footnote{
The  result
\re{V9} was  confirmed by
a  direct calculation
in Ref.~\protect\cite{Chetyrkin93}. Hence,
the correction of
the originally
published coefficient
$992$ of $\zeta(3)$ in $b^{\rmfont{V}}_{02}$
to $1008$ suggested in
Ref.~\protect\cite{Levan89}
and unfortunately used
in Ref.~[88]
turned out to be an error.
This fact has also been recently acknowledged by the very author
of Ref.~\cite{Levan89} in Ref.~\cite{Levan94}.
Numerically the use of the right result
leads to only a slight
increase (less than 0.6\%)
in the magnitude of $\lambda_6^{\rmfont{V}}$ in comparison
with that given in Ref.~\protect\cite{CheKue90}.}
\beq \ba{l}
b^{\rmfont{V}}_{00}  = 6\, , \\
b^{\rmfont{V}}_{01}  = 28\, , \;\;\;
b^{\rmfont{V}}_{11}  = 12\, , \\
b^{\rmfont{V}}_{02}  = [28799 + 992~\zeta(3)
- 8360~\zeta(5) - 882~n_f]/72\, , \\
b^{\rmfont{V}}_{12}  = (3303 - 114~n_f)/18\, , \;\;\;
b^{\rmfont{V}}_{22}  = (513 - 18~n_f)/18\, , \\
a^{\rmfont{V}}_{02}  = [32 - 24~\zeta(3)]/3
{}\, \,  .
\ea
\EQN{V9}
\eeq
The crucial point  for the  subsequent
calculation is the invariance of
$D^{\rmfont{V}}$ \re{nsmv1}
 under RG transformations
\beq
 \dmu D^{\rmfont{V}} = 0
{}\, ,
\eeq
combined with the absence of $\ln\,{\mu^2/m^2}$
terms in Eq.~(\ref{nsmv1}).
Recursion relations between the
coefficients of the Adler function
allow the calculation of
the  order $\ordas^3)$ coefficients $b^{\rmfont{V}}_{13},
b^{\rmfont{V}}_{23},b^{\rmfont{V}}_{33}$ from the
 lower order coefficients combined with
those of
anomalous mass dimension and the $\beta$-function:
\renewcommand{\arraystretch}{2}
\beq
\ba{l}
b^{\rmfont{V}}_{11}  \ds =  2b^{\rmfont{V}}_{00} \gm^0 \, ,\\
b^{\rmfont{V}}_{12}  \ds = (\beta_0 + 2\gm^0) b^{\rmfont{V}}_{01}
+ 2\gm^1 b^{\rmfont{V}}_{00}\, ,  \\
b^{\rmfont{V}}_{22}  \ds = \frac{1}{2} (\beta_0
                 + 2\gm^0) b^{\rmfont{V}}_{11}\, , \\
b^{\rmfont{V}}_{13}  \ds = 2(\beta_0 + \gm^0)(b^{\rmfont{V}}_{02}
-a^{\rmfont{V}}_{02})
                 +(\beta_1 + 2\gm^1) b^{\rmfont{V}}_{01}
+ 2\gm^2 b^{\rmfont{V}}_{00} \, ,\\
b^{\rmfont{V}}_{23}  \ds = (\beta_0 + \gm^0)b^{\rmfont{V}}_{12}
                 +\frac{1}{2}(\beta_1
+ 2\gm^1) b^{\rmfont{V}}_{11} \, ,\\
b^{\rmfont{V}}_{33}  \ds = \frac{2}{3}(\beta_0
+ \gm^0)b^{\rmfont{V}}_{22}~.

\EQN{V10}
\ea
\eeq
The coefficient $b^{\rmfont{V}}_{03}$
cannot be obtained via this recursion method.
However, the term proportional to this coefficient
does not contribute to $R^{\rmfont{V}}$.
The vector contribution
to the decay rate is then written in the
form:
\bea
\frac{\ovl{m}^2}{s} r_{\rmfont{V}}^{(1)}
   &=&\frac{\ovl{m}^2(\mu)}{s}
\Bigg\{
 \lambda_0^{V}
+\frac{\alpha_s (\mu)}{\pi}
\left[
     \lambda_1^{V}
   + \lambda_2^{V} \ln\,\frac{s}{\mu^2}
\right]
\label{V10b}
\\
&&
+
\left[ \frac{\alpha_s(\mu)}{\pi}\right]^2
\left[
  \lambda_3^{V} + \lambda_4^{V} \ln\,\frac{s}{\mu^2}
  + \lambda_5^{V} \ln^2\frac{s}{\mu^2}
\right]
 +
\left[ \frac{\alpha_s(\mu)}{\pi}\right]^3
 \left[\lambda_6^{V} + \dots
 \right]
 +
  \dots
\Bigg\}
{}\, .
\nonumber
\eea
If we set the normalization point $\mu^2=s$,
  the remaining logarithms of
$s/\mu^2$ are  absorbed in the running
 coupling constant and the running
mass. The  coefficients $\lambda$
can be obtained from the expansion
coefficients of the Adler function
by first integrating Eq.~(\ref{nsmv1})
to obtain $\Pi^{\rmfont{V}}/Q^2$ and subsequently taking
the imaginary part of $\Pi^{\rmfont{V}}/Q^2$ to arrive at
$r^{\rmfont{V}}$:
\beq
\ba{lll}
\lambda^{\rmfont{V}}_0=0\, ,
&
\lambda^{\rmfont{V}}_1=b^{\rmfont{V}}_{11}\, ,
&
\lambda^{\rmfont{V}}_2=0\, ,
\\
\lambda^{\rmfont{V}}_3=b^{\rmfont{V}}_{12}-2b^{\rmfont{V}}_{22}\, ,
&
\lambda^{\rmfont{V}}_4=-2 b^{\rmfont{V}}_{22}\, ,
&
\lambda^{\rmfont{V}}_5=0\, ,
\\
\lambda^{\rmfont{V}}_6=
        b^{\rmfont{V}}_{13}-2b^{\rmfont{V}}_{23}
         +(6-\pi^2)b^{\rmfont{V}}_{33}\, ,
&
\lambda^{\rmfont{V}}_7 = -2b^{\rmfont{V}}_{23}+6b^{\rmfont{V}}_{33}\, ,
&
\lambda^{\rmfont{V}}_8=3b^{\rmfont{V}}_{33}\, ,
\ \
\lambda^{\rmfont{V}}_9=0
{}\, .
\ea
\EQN{V11}
\eeq
The term  $\pi^2$ in $\lambda^{\rmfont{V}}_6$
is a consequence of the analytical
continuation
from space-like to time-like momenta
and arises from
the term $\ln^3\mu^2/Q^2 \rightarrow
(\ln\, \mu^2/|Q^2| \pm i\pi)^3$.
Explicitly, non-zero entries above read:
\beq
\ba{l}
\lambda^{\rmfont{V}}_1   =  12\, ,
\\ \dsp
\lambda^{\rmfont{V}}_3  =   -~\frac{13}{3}~n_{\stand{f}}
              +\frac{253}{2}\, ,
\\ \dsp
\lambda^{\rmfont{V}}_4  =  - 57  + 2~n_{\stand{f}}\, ,
\\ \dsp
\lambda^{\rmfont{V}}_6  =
-~\frac{1}{9}~n_{\stand{f}}^2 \pi^2
+\frac{125}{54}~n_{\stand{f}}^2
+\frac{17}{3}~n_{\stand{f}} \pi^2
-\frac{466}{27}~n_{\stand{f}} \zeta(3)
+\frac{1045}{27}~n_{\stand{f}} \zeta(5)
\\ \dsp
\phantom{ \lambda_6^{\rmfont{V}} = }
-
\frac{4846}{27}~n_f
-\frac{285}{4}~\pi^2
+\frac{490}{3}~\zeta(3)
-\frac{5225}{6}~\zeta(5)
+2442\, ,
\\ \dsp
\lambda^{\rmfont{V}}_7  =
-~\frac{13}{9}~n_{\stand{f}}^2
+\frac{175}{2}~n_f
-\frac{4505}{4}\, ,
\\ \dsp
\lambda^{\rmfont{V}}_8  =
\frac{1}{3}~n_{\stand{f}}^2
-17~n_f
+\frac{855}{4}
{}\,\, .
\ea
\EQN{V11B}
\eeq

The $a_{02}$ contribution originates from the $\rmfont{b}$
quark vacuum polarization graphs and is thus also
present for final states with massless quarks.
(More precisely, it originates in this case from
QCD corrections to ${\rm q\ovl{q}\rmfont{b}\ovl{\rmfont{b}}}$
configurations.) The same correction
would arise in $r^{\rmfont{A}}$.
This term has been anticipated in Eq.~(\ref{mt2b}).
The final answer  can still be interpreted
as an incoherent sum
of the contributions from different quark species.
In particular this implies that contributions
from three gluon intermediate final states
(singlet contributions) are absent in the
$\ordas^3)$ mass terms. This contrasts with the
corrections for $m=0$,  which receive
third-order contributions precisely
from this configuration --- see Eq.~\re{sm0v1} below.

Numerically,  one finds a quite decent decrease in
the terms of successively higher-orders, which
supports confidence in the applicability of
 these results for predictions of the rate. This
will be studied in more detail in
Part~\ref{numerical}.

\subsect{Axial Vector-Induced Corrections
\label{axial}}
The situation is  more involved if
  one wants to apply similar RG arguments
to the axial vector-induced rate
in order to again   compute the corresponding
 mass corrections
from nonsinglet diagrams.
The  comparison of
the expansion of  the Adler function
\beq \ba{ll}  \EQN{V12}    \dsp
 D^{\rmfont{A}} = & \dsp  -12\pi^2 Q^2
\frac{d}{dQ^2}\left(\frac{\Pi^{\rmfont{A}}_1}{Q^2}\right)
\\ \dsp
& = 3 \dsp
 \Bigg\{ \left( 1 + \apimu +
\dots\right)   - \frac{\ovl{m}^2(\mu)}{Q^2}
\left[
\sum_{\stackrel{\scriptsize i \leq j+1}
{i,j \geq0}} b^{\rmfont{A}}_{ij}\ell^{\stand{i}}
\left(\apimu\right)^{\stand{j}} \right]
 + {\cal O}(m^4) \Bigg\} {}\, ,
\ea\eeq
with Eq.~(\ref{nsmv1})  shows that in the
axial case the highest order
term of the power series in $\ell$
 within a given order $\ordas^{\stand{j}})$
is proportional to $\ell^{j+1}$,
 whereas in the vector case the
$\ell$-expansion terminated at $\ell^{\stand{j}}$.
 This structure is dictated by
 the anomalous dimension
\mbox{$\gaam$,}  which vanishes in
 the vector case.

\noindent
The expansion of $D^{\rmfont{V}}$ contained
the second order coefficient $a^{\rmfont{V}}_{02}$,
originating from an inner
 $\rmfont{b}$-quark loop.
 The same $\ordas^3)$
 term is also  present  in the axial case
for massless as well as massive
external quark lines. The mass correction
of $\ordas^3)$ from external quark loops
has not yet been calculated.

The coefficients of the expansion (\ref{V12}) read:
\beq \ba{lll}  \dsp
b^{\rmfont{A}}_{00} = -12\, ,& b^{\rmfont{A}}_{10} = -6\, , & \\   \dsp
  b^{\rmfont{A}}_{01} = -\frac{151}{2}
+ 24\zeta(3)\, ,\qquad&
b^{\rmfont{A}}_{11} = -34\, ,\qquad& b^{\rmfont{A}}_{21} = -6
{}\, .
\EQN{V16}
\ea
\eeq
The complication in deriving from
these  the second-order
coefficients of the logarithmic terms
arises from the fact that
the mass dependent part of  $D^{A}$
obeys the inhomogeneous
RG equation:
\beq
\dmu D^{\rmfont{A}} = \frac{\ovl{m}^2}{Q^2}\gaam \equiv
\frac{\ovl{m}^2(\mu)}{Q^2} \sum_{i\geq 0}
\left( \gaam\right)_{\stand{i}} \left[\apimu\right]^{\stand{i}}
{}\, .
\EQN{V14}
\eeq
Therefore, recursion relations can
be set up again, although in  this case
  order ${\cal O}(\as^{\stand{n}})$
  coefficients $b^{\rmfont{A}}_{kn}\;(k >  0)$
are not only expressed through the
$\{b_{ij}\}$ with $0 \leq i \leq j+1 \leq n$,
 but  also through the
expansion coefficients of the anomalous
dimension $(\gaam)_{\rmfont{r}}\;(r\leq n)$.
In fact, the second-order coefficients
 satisfy the  relations
{\renewcommand{\arraystretch}{2}
\beq \ba{ll}
b^{\rmfont{A}}_{10} & \ds = - \left( \gaam\right)_0 {}\, ,\\
b^{\rmfont{A}}_{11} & \ds = - \left( \gaam\right)_1
+ 2b^{\rmfont{A}}_{00} \gm^0 {}\, ,\\
b^{\rmfont{A}}_{21} & \ds =   b^{\rmfont{A}}_{10} \gm^0 {}\, ,\\
b^{\rmfont{A}}_{12} & \ds = - \left( \gaam\right)_2 +
  b^{\rmfont{A}}_{01} (\beta_0 + 2\gm^0)
                 + 2 b^{\rmfont{A}}_{00}\gm^1 {}\, , \\
b^{\rmfont{A}}_{22} & \ds = \frac{1}{2}~ b^{\rmfont{A}}_{11}
(\beta_0 + 2\gm^0)
                 +   b^{\rmfont{A}}_{10}\gm^1  {}\, ,\\
b^{\rmfont{A}}_{32} & \ds = \frac{1}{3}~
  b^{\rmfont{A}}_{21} (\beta_0+2\gm^0) {}\, .
\EQN{V15}\ea \eeq
[Note:  for the vanishing
anomalous dimension the Eqs.
(\ref{V10}) and (\ref{V15}) coincide.]
Therefore the anomalous dimension $\gaam$
 must be known to the same order
to which the decay rate is  computed.
The calculation of $\gaam$ is sketched
in Section~\ref{correlators}
and leads to the following
result \cite{CheKueKwi92}
\beq
\gaam = 6\left\{ 1+\frac{5}{3}~\api
+\left(\api\right)^2
\left[\frac{455}{72}-\frac{1}{3}~n_{\stand{f}}
      -\frac{1}{2}~\zeta(3)\right]
\right\}\, .
\EQN{V16b}
\eeq
As for the vector case
 we write a general expansion
for the axial vector-induced rate:
\begin{eqnarray}
\dsp
\frac{\ovl{m}^2}{s}r_{\rmfont{A}}^{(1)}\dsp
 & =&
\dsp
 \frac{\ovl{m}^2(\mu)}{s}
\Bigg\{ \lambda_0^{A}
+
 \frac{\alpha_s (\mu)}{\pi}
 \left[
\lambda_1^{A} + \lambda_2^{A} \ln\,\frac{s}{\mu^2}
\right]
\nonumber
\\
&&+
\left[ \frac{\alpha_s (\mu)}{\pi}\right]^2
\left[
\lambda_3^{A} + \lambda_4^{A} \ln\,\frac{s}{\mu^2}
 + \lambda_5^{A} \ln^2\frac{s}{\mu^2}
\right]
 \Bigg\}
{}\, .
\eea
The  coefficients
read as follows:
\beq\ba{lll}
 \lambda^{\rmfont{A}}_0=b^{\rmfont{A}}_{10}~,
&
\quad \lambda^{\rmfont{A}}_1=b^{\rmfont{A}}_{11}-2b^{\rmfont{A}}_{21}~,
&
\quad \lambda^{\rmfont{A}}_2=-2b^{\rmfont{A}}_{21}~,
\\
 \lambda^{\rmfont{A}}_3=
 b^{\rmfont{A}}_{12}-2b^{\rmfont{A}}_{22}
 +(6-\pi^2)b^{\rmfont{A}}_{32}{}\, ,
&\quad\lambda^{\rmfont{A}}_4=-2b^{\rmfont{A}}_{22}+6b^{\rmfont{A}}_{32}~,
&
 \quad\lambda^{\rmfont{A}}_5=3b^{\rmfont{A}}_{32}
{}\, .
\ea
\eeq
Or, explicitly,
\beq\ba{l}
\dsp
 \lambda^{\rmfont{A}}_0 =  -6~,
\\
\dsp
 \lambda^{\rmfont{A}}_1 =  -22~,
\\
\dsp
 \lambda^{\rmfont{A}}_2 =   12~,
\\
\dsp
 \lambda^{\rmfont{A}}_3=
 -~\frac{1}{3}~n_{\stand{f}} \pi^ 2
 - 4~n_{\stand{f}} \zeta(3) + \frac{151}{12}~n_{\stand{f}}
+ \frac{19}{2}~\pi^2 + 117~\zeta(3) - \frac{8221}{24}\, ,
\\
\dsp
 \lambda^{\rmfont{A}}_4 =
-~\frac{16}{3}~n_{\stand{f}} + 155~,
\\
\dsp
\lambda^{\rmfont{A}}_5=
n_{\stand{f}} - \frac{57}{2}
{}.
\ea
\eeq

The discussion in this and the previous
section is
tailored for an external current
 coupled to $\rmfont{b}\ovl{\rmfont{b}}$ and includes
 mass corrections
from internal
$\rmfont{b}$ quark loops as well as from the
loops coupled to the external current.
A slightly different situation occurs for a
nonsinglet correlator arising
 from massless quarks.
Internal bottom quarks as indicated in
the double bubble graph still induce
$m_{\rmfont{b}}^2/s$ corrections. However, a slight
generalization
 of the arguments presented above
demonstrates that these terms are
again absent in order
$\ordas^2)$. From corrections of the
diagrams in Fig.~\ref{doubub} one
obtains the terms of order
$\ordas^3)$\,,
which should for convenience
be incorporated into
$r_{\rmfont{NS}}^{(0)}$ and summed over all
massive quark species, adding the  term
\beq
\ba{rl}\dsp
r^{(0)}_{\rmfont{NS}}\longrightarrow
& \dsp
r^{(0)}_{\rmfont{NS}}+
\left(\api\right)^3
\sum_f
\frac{\ovl{m}_{\stand{f}}^2}{s}
(-2)(\beta_0+\gamma^0_{\stand{m}})a_{02}
\\ \dsp
=
& \dsp
r^{(0)}_{\rmfont{NS}}
-\left[\apis\right]^3
\left(15-\frac{2}{3}~n_{\stand{f}} \right)
\left[\frac{16}{3}-4\zeta(3)\right]
\sum_{\stand{f}}\frac{\ovl{m}_{\stand{f}}^2(s)}{s}
{}\, .
\ea
\EQN{V17}
\eeq\sect{Mass Corrections of Order $m^4/s^2$
\label{repnsm4}}
Terms of higher-order in $m^2/s$ are quite
unimportant as far as $\stand{Z}$ decays into b-quarks
are concerned. However, at lower energies
these should be taken into account in order to arrive
at an adequate description of the cross-section.
In fact, as shown in Fig.~\ref{kvexp},
the order $\as$ correction functions
$K_{\rmfont{V}}$ and $K_{\rmfont{A}}$ introduced in Eq.~(\ref{ap1})
are well described by the first few terms
of the expansion in $m^2/s$\,, not only at
high energies, but even fairly close to
threshold. Hence one should
arrive at a reliable result to
${\cal O}(\as^2)$ near the threshold
through the incorporation of the first
terms of the expansion in $\as^2(m^2/s)^{\stand{n}}$.
The second-order  calculation of quartic
mass corrections
presented below is based on
Ref.~\cite{TTP94-08}.
 The calculation was performed for
 vector and axial vector current
nonsinglet correlators. The first
is of course relevant for electron--positron
annihilation into heavy quarks
 at arbitrary energies, the second
for $\stand{Z}$ decays into $\rmfont{b}$ quarks and for
top production at a future linear
collider.

Quartic mass corrections were already
presented to
order $\as$ in Part~\ref{exact},  expressed
in terms of the pole mass. In this section
the result in the $\msbar$ scheme is given
and the second order
$\as^2$ contribution is discussed.
 The calculation is based on
 the operator product expansion of
the T-product of two vector currents,
$J_\mu=\ovl u\g_\mu d $ and
$J_\nu^+=\ovl d\g_\nu u $.
Here ${u}$ and ${d}$ are simply  two generically
 different quarks with
masses $m_{\rmfont{u}}$ and $m_{\rmfont{d}}$.
The operator product expansion includes power
law suppressed terms up to operators of
dimension four induced by
 non-vanishing quark masses.
Renormalization group arguments
 similar to those  already  employed   in
the previous section
allowed a deduction in  the $\as^2 m^4$ terms.
Quarks which are
 not coupled to the external
current will influence the result in order
 $\as^2$ through their
coupling to the gluon field.
The result may be immediately transformed
to the case of the
electromagnetic current of a heavy, say,
$\rmfont{t}$ (or $\rmfont{b}$) quark.

The asymptotic behaviour of the transverse part
of this
(operator valued) function for
  $Q^2 = -q^2\to\infty$  is given by
an OPE  of the following form
(different powers of $Q^2$ may be studied
separately and only operators of dimension
4 are displayed):
\beq
i\int T(J_\mu(x) J^+_\nu(0))\ex^{iqx} \dif x
\bbuildrel{=}_{q^2\to\infty}^{} \frac{1}{Q^4}
\sum_n
(q_\mu q_\nu -g_{\mu\nu} q^2)
\fos{\rmfont{T}}{C_{\stand{n}}}(Q^2,\mu^2,\alpha_s)
O_{\stand{n}}+\dots
\EQN{142}
\eeq
Only the gauge invariant operators
$G_{\mu\nu}^2, m_{\stand{i}}\ovl{q}_jq_{\stand{j}}$ and a polynomial
of fourth order in the masses contributes to
physical matrix elements.
Employing renormalization group arguments
the vacuum expectation value of
$\sum_{\stand{n}} C_n O_{\stand{n}}$
is under control up to terms of order $\alpha_s$
as far as the constant terms are concerned
and even up to $\alpha_s^2$ for the
logarithmic terms proportional to $\ln\, Q^2/\mu^2$.
Only these logarithmic terms  contribute to
the absorptive part. Hence one arrives at the
full answer for $\alpha_s^2m^4/s^2$ corrections.
Internal quark loops contribute in this order,
giving rise to the terms proportional
to $m^2 m_{\stand{i}}^2$ and $m_{\stand{i}}^4$ below.

 The result reads [below we set for brevity
the $\ovl{\rmfont{\rm MS}}$ normalization
scale $\mu = \sqrt{s}$
and $\ovl{m}_{\rmfont{u}}(s) = \ovl{m}_{\rmfont{d}}(s) = \ovl{m}$]:
\begin {figure}
\begin{center}
\mbox{\epsfig{file=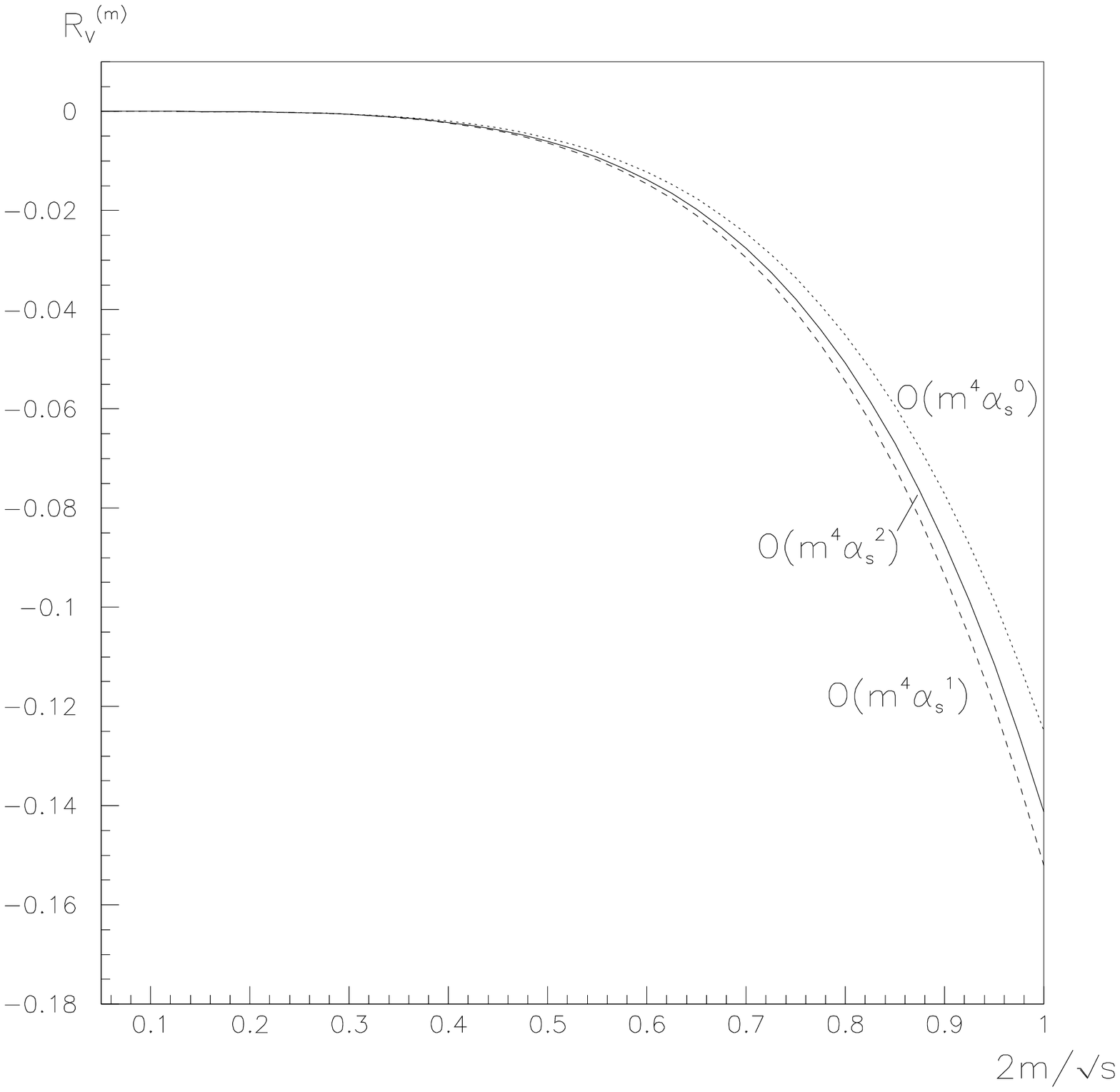,width=12.cm,height=9.cm,%
bbllx=0.cm,bblly=0.cm,bburx=18.cm,bbury=22.cm}
}
\end{center}
\caption{\label{alphasexp}{{Contributions
to $R^{\rmfont{V}}$ from $m^4$ terms
including successively higher
orders in $\alpha_s$ (order $\alpha_s^0$/
 $\alpha_s^1$/ $\alpha_s^2$
corresponding to dotted/ dashed/ solid lines)
as functions of
$2m_{\rm pole}/\protect\sqrt{s}$.}}}
\end{figure}
\ice{
 The result reads [below we set for brevity
the $\ovl{\rmfont{\rm MS}}$ normalization
scale $\mu = \sqrt{s}$
and $\ovl{m}_{\rmfont{u}}(s) = \ovl{m}_{\rmfont{d}}(s) = \ovl{m}$]:
}
\begin{eqnarray}
\dsp
\frac{\ovl{m}^4}{s^2}
r_{\rmfont{V}}^{(2)}
 \dsp &=&\frac{\ovl{m}^4}{s^2}\Bigg\{
 \dsp -6  -22 \apis
\nonumber
\\
&&
+
\left[\apis\right]^2 \left[
n_{\stand{f}} \left(
\frac{1}{3}  \ln\,\frac{\ovl{m}^2}{s}
-\frac{2}{3}\pi^2-\frac{8}{3}\zeta(3)
+\frac{143}{18}
  \right)
\right.
\nonumber
\\
&&
\left.
\dsp
-~\frac{11}{2} \ln\,\frac{\ovl{m}^2}{s}
+ 27\pi^2+112\zeta(3)-\frac{3173}{12}
+12  \sum_{\stand{i}} \frac{\ovl{m}_{\stand{i}}^2}{\ovl{m}^2}
\right.
\nonumber
\\
&&
\left.
\dsp
+\left(\frac{13}{3}-4\zeta(3)\right)
  \sum_{\stand{i}} \frac{\ovl{m}_{\stand{i}}^4}{\ovl{m}^4}
 -
\sum_{\stand{i}} \frac{\ovl{m}_{\stand{i}}^4}{\ovl{m}^4}
\ln\,\frac{\ovl{m_{\stand{i}}}^2}{s}
\right]\Bigg\}
\, ,
\EQN{VM4N}
\end{eqnarray}
\newpage
\begin{eqnarray}
\dsp
\frac{\ovl{m}^4}{s^2}
r_{\rmfont{A}}^{(2)}
 &=&\frac{\ovl{m}^4}{s^2}\Bigg\{
\dsp 6  +10 \apis
\nonumber
\\
&&
+
\left[\apis\right]^2 \left[
n_{\stand{f}} \left(
-~\frac{7}{3}  \ln\,\frac{\ovl{m}^2}{s}
+\frac{2}{3}\pi^2+\frac{16}{3}\zeta(3)
-\frac{41}{6}
  \right)
\right.
\nonumber
\\
&&
\left.
\dsp
+~\frac{77}{2} \ln\,\frac{\ovl{m}^2}{s}
- 27\pi^2-220\zeta(3)+\frac{3533}{12}
-12  \sum_{\stand{i}} \frac{\ovl{m}_{\stand{i}}^2}{\ovl{m}^2}
\right.
\nonumber
\\
&&
\left.
\dsp
+\left(\frac{13}{3}-4\zeta(3)\right)
  \sum_{\stand{i}} \frac{\ovl{m}_{\stand{i}}^4}{\ovl{m}^4}
 -
\sum_{\stand{i}} \frac{\ovl{m}_{\stand{i}}^4}{\ovl{m}^4}
\ln\,\frac{\ovl{m_{\stand{i}}}^2}{s}
\right]\Bigg\}~.
\EQN{VM4N1}
\end{eqnarray}

Note that the sum over $i$ includes also the quark
coupled to the external current and with
 mass denoted by $m$.
Hence in the case of one heavy quark $\rmfont{u}$  of
mass $m$ ($d \equiv u)$ one should set
$\sum_{\stand{i}} {\ovl{m}_{\stand{i}}^4}/{\ovl{m}^4} = 1$
and
$\sum_{\stand{i}}{\ovl{m}_{\stand{i}}^2}/{\ovl{m}^2} = 1$.
In the opposite case,  when one considers
the correlator of light
(massless) quarks the heavy quarks appear only
through their coupling to gluons. There one
 finds for the correction term:
\beq
\dsp
r_{\rmfont{V}}  = r_{\rmfont{A}} =
\left[\apis\right]^2
\sum_{\stand{f}} \frac{\ovl{m}_{\stand{f}}^4(s) }{s^2}
 \left[
\frac{13}{3}
-\ln\,\frac{\ovl{m}^2_{\stand{f}}(s)}{s}
-4 \zeta(3)
\right]
{}\, ,
\EQN{VAM4}
\eeq
as anticipated in Eq.~(\ref{mt2b}).

The $\stand{Z}$ decay rate is hardly affected by
the $m^4$ contributions. The
lowest order term in Eq.~\re{VM4N} evaluated with
$\ovl{m}=2.6  \ {\rm GeV}$
amounts to
$\pm 6 \ovl{m}/s^2=\pm 5\times 10^{-6}$
for the vector
(axial vector) current induced
 $Z\to  \rmfont{b}\bar{\rmfont{b}} $ rate.
Terms of increasing order in $\alpha_s$
 become successively smaller.
The $m_{\rmfont{b}}^4$ correction
to $\Gamma(Z\to \rmfont{q}\bar{\rmfont{q}})$\,,
which starts in
order $\alpha_s^2$\,,
is evidently even smaller.
It  is worth noting, however, that the
 corresponding series, evaluated in
the onshell scheme, leads to terms which
are larger by about one order
of magnitude and of oscillatory signs.
{}From these
considerations it is clear that $m^4$
corrections to the $\stand{Z}$ decay
rate are well under control --- despite
 the still missing singlet piece
--- and that they can be neglected for
 all practical purposes.

The situation is different in the low
energy region ---  say
several GeV above the charm or the bottom
threshold. For  definiteness
the second case will be considered and
for simplicity
all other masses will be put to
zero.
The contributions to $R^{\rmfont{V}}$ from $m^4$ terms are
presented  in Fig.~\ref{alphasexp}
as functions of $2m/\sqrt{s}$ in the range
from 0.05 to 1.
The input parameters
$M      _{\rm pole}=4.70$ GeV and
$\alpha_s(m_{\stand{Z}}^2)=0.12$ have
 been chosen.
Corrections of higher-orders are added
 successively. The prediction is
fairly stable with increasing order in
 $\alpha_s$ as a consequence of
the fact that most large logarithms were
absorbed in the running mass.
The relative magnitude of the sequence
 of terms from the $m^2$
expansion
is displayed in Fig.~\ref{massexp}. The
curves for $m^0$ and $m^2$  are based on
corrections up to third order in $\alpha_s$\,,
with the $m^2$ term starting
at first order. The $m^4$ curve receives
 corrections from order zero to
two.

\begin {figure}
\begin{center}
\mbox{\epsfig{file=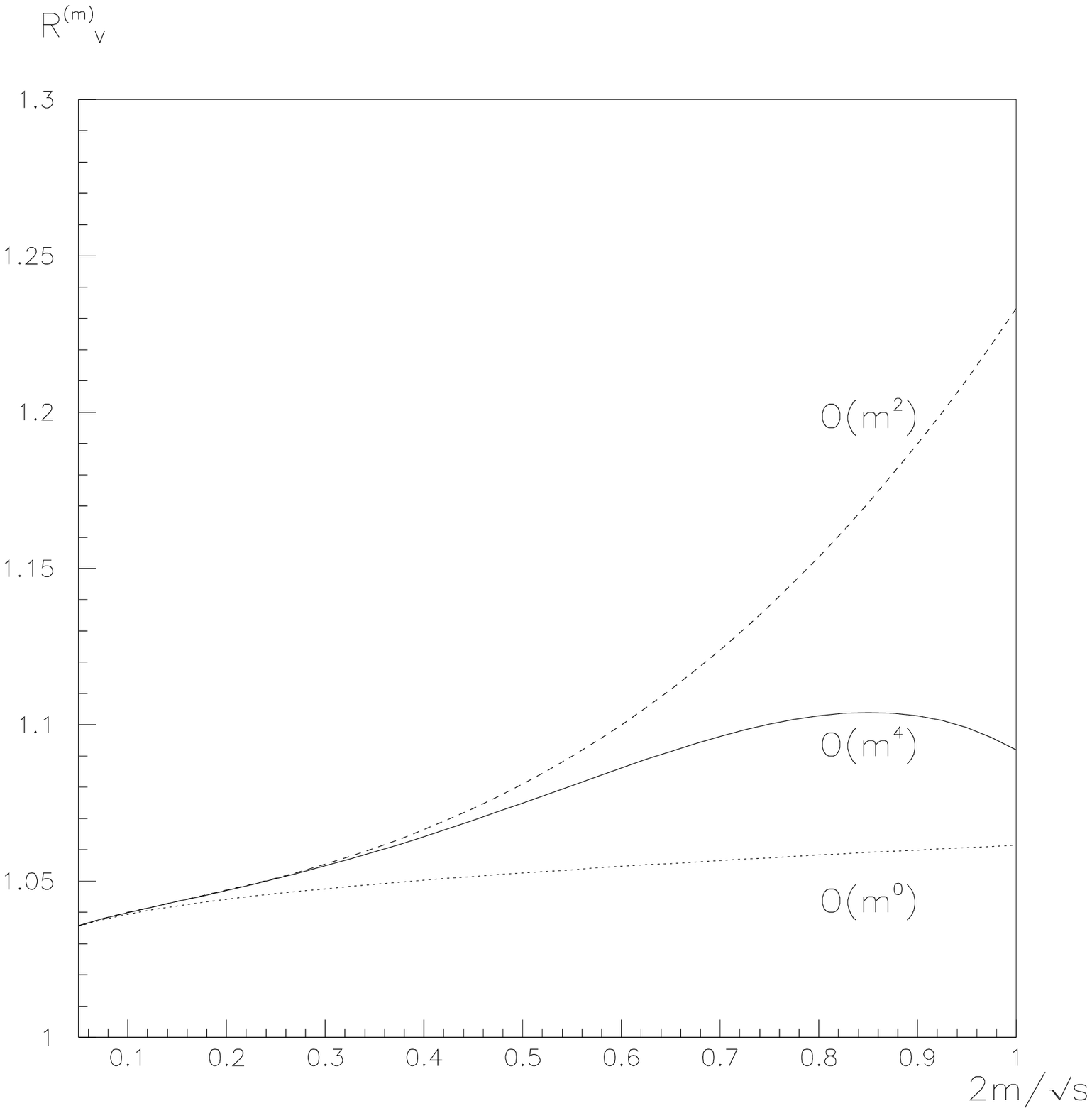,width=10.cm,height=8.cm,%
bbllx=0.cm,bblly=0.cm,bburx=17.cm,bbury=21.cm}
}
\end{center}
\caption{\label{massexp}{{Predictions for
 $R^{\rmfont{V}}$ including successively higher
orders in $m^2$.}}}
\end{figure}

Of course, very close to threshold --- say
above 0.75 (corresponding to
$\sqrt{s}$ below\break\hfill 13 GeV) --- the approximation
is expected to break down,
as indicated already in Fig.~\ref{kvexp}.
Below the $\rmfont{b}\bar{\rmfont{b}}$ threshold, however, one
may decouple the bottom quark and consider
 mass corrections from the
charmed quark within the same formalism.\sect{Partial Rates
\label{nspart}}
The formulae for the QCD corrections to the total
 rate $\Gamma_{{\rm had}}$
 have a simple, unambiguous meaning. The
theoretical predictions for individual
$\rmfont{q}\bar{\rmfont{q}}$
channels, however, require additional
interpretation. In fact, starting from order
${\cal O}(\alpha_s^2)$ it is no longer possible
to assign all hadronic final states to well
specified $\rmfont{q}\ovl{\rmfont{q}}$   channels in a unique manner.
The vector-  and axial vector-induced rates receive
(non-singlet) contributions from the diagrams,
 where the heavy quark pair is radiated off a
light $\rmfont{q}\ovl{\rmfont{q}}$
system (see Fig.~\ref{ffermi}a).
The analytical result for this contribution
for arbitrary $m^2/s$ can be found
in Ref.~\cite{Hoang94} and is reproduced in the
Appendix.
\begin {figure}
\begin{center}
\begin{tabular}{lll}
\parbox{3cm}{
\epsfig{file=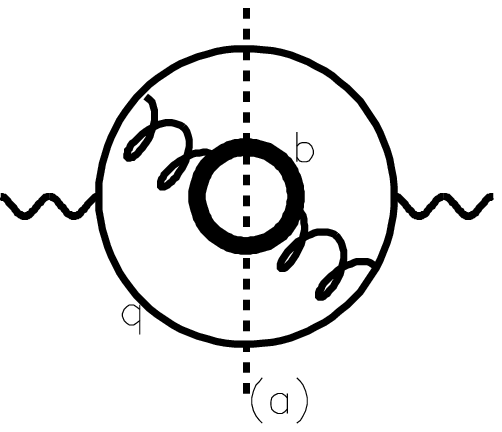,width=5.cm,height=5.cm}
            }
&
\hphantom{XXXXX}
&
\parbox{3cm}{
\epsfig{file=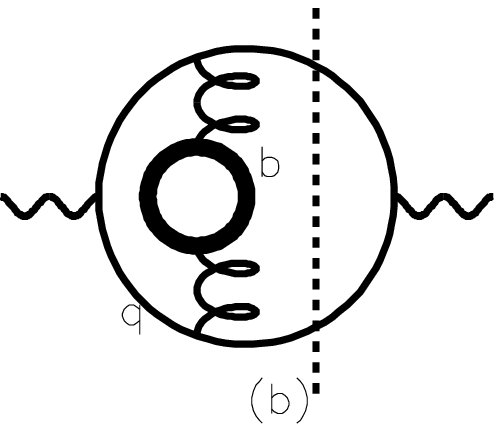,width=5.cm,height=5.cm}
            }
\end{tabular}
\end{center}
\caption[]{\label{ffermi}
{{Nonsinglet ${\cal O} (\alpha_s^2)$
four fermion (a) and virtual (b)
contributions.}}
}
\end{figure}
The rate for this
specific contribution to the
$\rmfont{q}\bar{\rmfont{q}}\rmfont{b}\ovl{\rmfont{b}}$
final state in the limit  of small $m^2/s$
is given by

\bea
\EQN{nspart1}
\dsp
R^{\rmfont{NS}}_{{\rm q\ovl{q}}\rmfont{b}\ovl{\rmfont{b}}}
=\frac{\Gamma^{\rmfont{NS}}
_{{\rm q\ovl{q}}\rmfont{b}\ovl{\rmfont{b}}}}%
{\Gamma^{{\rm Born}}_{{\rm {q\ovl{q}}}}}
 \dsp
&=&
\left(\frac{\alpha_s}{\pi}
\right)^2
\frac{1}{27}\left\{
\ln^3 \frac{s}{m_{\rmfont{b}}^2}
             \right.
-
 \frac{19}{2} \ln^2 \frac{s}{m_{\rmfont{b}}^2}
 + \left[\frac{146}{3}-12\zeta(2)
   \right] \ln\, \frac{s}{m_{\rmfont{b}}^2}
\\
\nonumber
\dsp
&&-
\left.
\frac{2123}{18}+38 \zeta(2) + 30\zeta(3)
\right\}
+ O(m_{\rmfont{b}}^2/s)
{}\, .
\eea
Numerically one obtains
\beq\EQN{nspart2}
R^{\rmfont{NS}}_%
{\rmfont{q}\ovl{\rmfont{q}}\rmfont{b}\ovl{\rmfont{b}}}=
\left(\frac{\alpha_s}{\pi}
 \right)^2
\{
0.922/0.987/1.059
\}~;\;\;\;\;\;
{\rm for}\; m_{\rmfont{b}}=4.9/4.7/4.5\;{\rm GeV}
\, .
\eeq
The contributions from this configuration
to the total rate (in particular the
logarithmic
mass singularities) are nearly
 cancelled by those from the corresponding
virtual corrections (see Fig. \ref{rhorv}b).

Despite the fact that $\rmfont{b}$ quarks are present
in the four-fermion final state,
the natural prescription is to assign
these events to the
${\rm q\ovl{q}}$ channel.
They must be subtracted experimentally
from the partial rate $\Gamma_{\rmfont{b}\ovl{\rmfont{b}}}$. This
should be possible, since their signature is
characterized by a large invariant mass of
the light quark pair and a small invariant mass
of the bottom system, which is emitted
collinear to the light quark momentum.
If this subtraction is not performed,
the inclusive bottom rate
is overestimated  (for $m_{\rmfont{b}}$ =\break\hfill 4.70 GeV
and $\as$ chosen between 0.12 and 0.18) by
\beq\EQN{nspart3}
\Delta \equiv
\frac{\dsp \sum_{\rmfont{q}=\rmfont{u,d,s,c,b}}
\Gamma^{\rmfont{NS}}%
_{\rmfont{q}\ovl{\rmfont{q}}\rmfont{b}\ovl{\rmfont{b}}}}
{\Gamma_{\rmfont{b}\ovl{\rmfont{b}}}}
\approx 0.007 \; .. \;  0.016~.
\eeq
As shown in Fig.~\ref{rhorv}a the
leading contribution well  matches the full
calculation (for bottom quarks it is about
10\% above the exact answer) leading  to
the analytic result presented in the
Appendix.
A slightly different approach to the evalutaion
of heavy quark multiplicities,  which attempts
the  resummation  of leading logarithms,  can be
found in Ref.~\cite{Seymor}.
\begin{figure}
\begin{center}
\leavevmode
\mbox{}\epsffile[100 300 500 540]{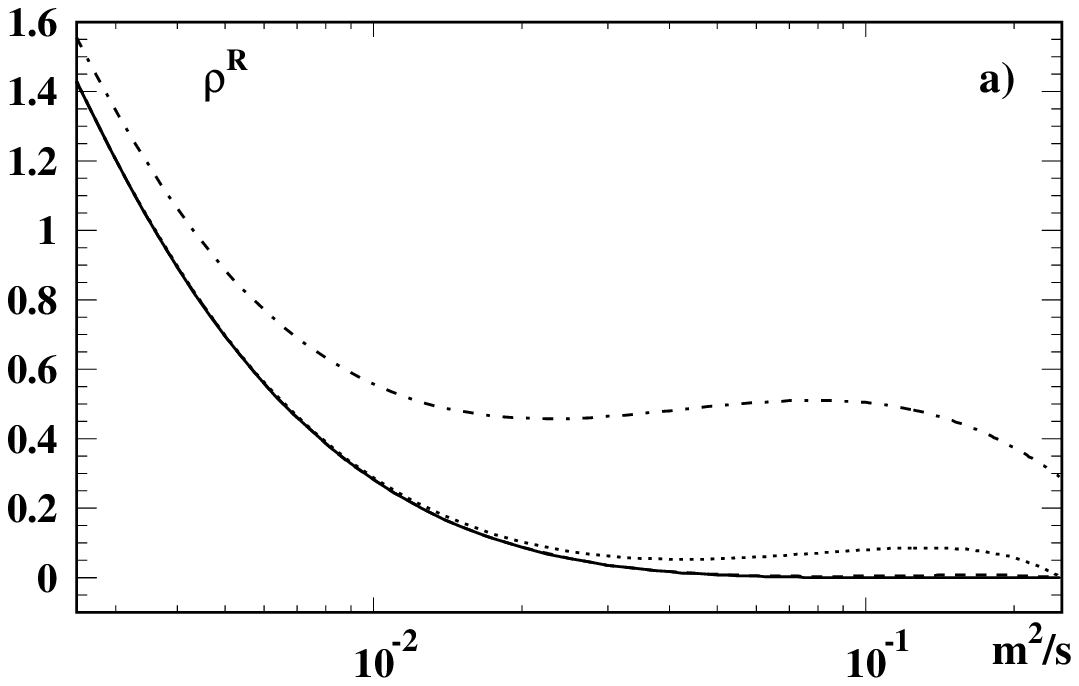} \\
\mbox{}\epsffile[100 300 500 540]{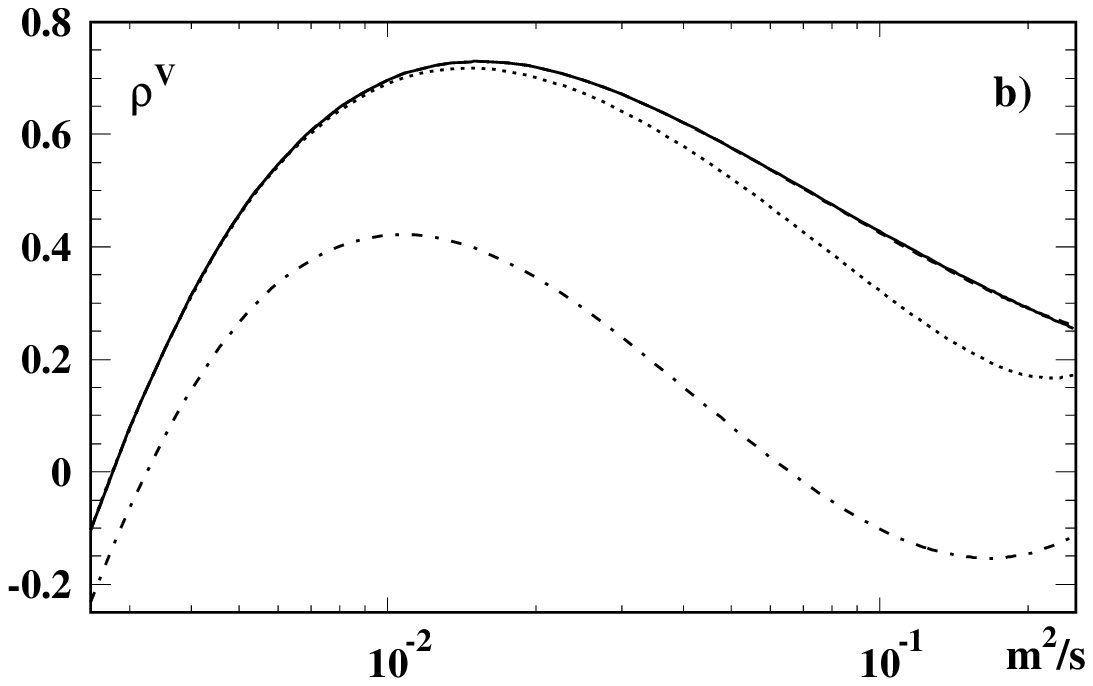}
\caption{\label{rhorv}
{{a)
The function $\varrho^{\rmfont{R}}$ describing the production of four
fermions in the region $0<x=m^2/s<1/4$. Solid line: exact result;
dashed-dotted line: logarithmic and constant terms only; dotted line:
including $m^2/s$ corrections; dashed line: including $m^2/s$ and
$m^4/s^2$ corrections.
b)~Corresponding curves for $\varrho^{\rmfont{V}}$ describing virtual
corrections. }}}
\end{center}
\end{figure}
\chap{Singlet Contribution\label{singlet}}
\sect{Massless Final State\label{singl-massless}}
\subsect{Vector Currents\label{singl-vector}}
Singlet contributions to the $\stand{Z}$ decay rate
or to the total cross-section originate from
 diagrams that can be split into two parts
by cutting gluon lines only. In the vector case
the first of these contributions arises in order
$\ordas^3)$ and is induced by  `light-by-light'
scattering diagrams (see Fig.~\ref{class}). The charge
structure of this contribution differs from the
non-singlet terms. Hence the lowest order singlet
contribution is UV finite. In the notation
introduced in Section~\ref{notations} one obtains:
\beq
\EQN{sm0v1}
\ba{ll}\dsp
r^{\rmfont{V}}_S
& \dsp
=\frac{1}{3}
\left(\frac{\alpha_s(s)}{4\pi}\right)^3
\left(
\frac{d_{abc}}{4}\right)^2
\left[\frac{176}{3}-128\zeta(3)\right]
\\ \dsp
& \dsp
=\frac{1}{3}
 \left(\apis\right)^3(- 1.240)
\, .
\ea
\eeq

At this point a brief comment concerning
mass-dependent terms is in order. As discussed in
Section~\ref{vector}, $m_{\rmfont{b}}^2/s$\,  terms from diagrams
depicted in Fig.~\ref{class}
are absent. This leaves
potential contributions with heavy top
quarks from the same diagram.
However, these are suppressed by a factor
$s/m_{\rmfont{t}}^2$ and asymptotically
decoupled. In Section~\ref{top}  the corrections
of order $\ordas^2s/m_{\rmfont{t}}^2)$ from non-singlet
diagrams were calculated and shown to be small.
Corrections of order $\ordas^3s/m_{\rmfont{t}}^2)$ will
therefore be ignored throughout%
\footnote{Recently these corrections
have  been evaluated in Ref.~\cite{timo2}
and proved  to be quite small.}.
Hence no mass corrections from singlet
diagrams will be considered in the
vector case.

It should be stressed again that the knowledge
of the two  functions $R^{\rmfont{V}}_{\rmfont{NS}}$
and $R^{\rmfont{V}}_{\rmfont{S}}$ (and hence of $r_{\rmfont{NS}}^{(0)},
r_{\rmfont{NS}}^{V(1)},
r_{\rmfont{S}}^{\rmfont{V}}$) is sufficient to evaluate
all possible vector current correlators
like $R^{{\rm em}},R^{\rmfont{V}}$ and $R^{{\rm int}}$.
\subsect{Axial Case\label{singl-axial}}
In the (fictitious) case of  mass degenerate
isospin doublets, singlet contributions
from up and down quarks
compensate exactly,
since $a_{\rmfont{u}}=-a_{\rmfont{d}}$. Interesting enough,
individual contributions to four-fermion final
states are nevertheless present. The individual
contributions from these cut double triangle
diagrams to the $\stand{Z}$ decay rate for a massless
($\rmfont{u}$ and $\rmfont{d}$)
doublet are given as follows:
\beq
\EQN{sm0a1}
\Gamma_{\rm u\ovl{\rm u} u\ovl{ \rm u}}
=\Gamma_{\rm d\ovl{\rm d}d\ovl{\rm d}}
=
-\frac{1}{2}\Gamma_{\rm u\ovl{\rm u} d\ovl{\rm d}}
{}\, .
\eeq
For the top and bottom quark this cancellation
is no longer operative as a consequence of the
large mass splitting within the multiplet.
For $m_{\rmfont{t}}\rightarrow \infty$ one recovers the
predictions based on an anomalous axial current.
The $\ordas^2)$ calculation has been performed
in Refs.~\cite{KniKue90b,KniKue90a} for arbitrary $m_{\rmfont{t}}^2/s$.
The additional term in the $\stand{Z}$ decay rate
can be decomposed into a term from   two-
($\rmfont{b}\ovl{\rmfont{b}}$),
three-  ($\rmfont{b}\ovl{\rmfont{bg}}$) and
four-parton ($\rmfont{b}\ovl{\rmfont{b}}\rmfont{b}\ovl{\rmfont{b}}$)
configurations.
The two gluon cut is forbidden according to the
Landau--Yang selection rules \cite{LanYan49}, which forbid
the decay of a parity odd particle with non-even
integer total angular momentum
into two massless vector bosons.

The respective diagrams contribute an additional
singlet piece,
\beq\EQN{sm0a2}
r^{\rmfont{A}}_{\rmfont{S}}=
d^2_{\rmfont{S}}\left(\api\right)^2
+  d^3_{\rmfont{S}}\left(\api\right)^3
{}\, .
\eeq
The first term can be decomposed into
contributions from two-, three- and
four-particle  intermediate states:
\beq\EQN{sm0a3}
d^2_{\rmfont{S}}=\frac{1}{3}(
{\rm Re}{\cal I}_2+\Delta {\cal I}_3
+{\cal I}_4 )
= \frac{1}{3}{\cal I}
{}\, .
\eeq
The functions ${\cal I}$ are well
 approximated by\footnote{The
exact formula in terms of
Clausens functions can be found in Ref.~\cite{KniKue90b}.}
\beq\EQN{sm0a4}
\ba{rl}\dsp
{\rm Re}{\cal I}_2
& \dsp
= -7.210+1.481\frac{s}{4m_{\rmfont{t}}^2}
 +1.363\left(\frac{s}{4m_{\rmfont{t}}^2}\right)^2
 +3\ln\,\frac{s}{m_{\rmfont{t}}^2}
{}\, ,
\\
\dsp
\Delta{\cal I}_3
& \dsp
= -1.580-0.444\frac{s}{4m_{\rmfont{t}}^2}
 -0.731\left(\frac{s}{4m_{\rmfont{t}}^2}\right)^2
{}\, ,
\\
\dsp
{\cal I}_4
& \dsp
= -0.460
{}\, ,
\\ \dsp
{\cal I}
& \dsp
= -9.250+1.037\frac{s}{4m_{\rmfont{t}}^2}
 +0.632\left(\frac{s}{4m_{\rmfont{t}}^2}\right)^2
 +3\ln\,\frac{s}{m_{\rmfont{t}}^2}
{}\, .
 \ea
\eeq
The asymptotic behaviour for large $m_{\rmfont{t}}^2$
is particularly simple\footnote{
Note that the above result for
${\cal I} $ at large $m_{\rmfont{t}}$
has been checked through completely
independent\break\hfill\hspace*{0.5cm} calculations in
Ref.~\cite{ChetKwiat93}
(up to terms of order $s/m_{\rmfont{t}}^2$)
and in Ref.~\cite{timo2}
(up to terms of order $s^3/m_{\rmfont{t}}^6$).
}:
\beq\EQN{sm0a5}
\ba{rl}\dsp
{\rm Re}{\cal I}_2
& \dsp
\Rightarrow 2\zeta(2)-\frac{21}{2}
+\frac{10}{27}~\frac{s}{m_{\rmfont{t}}^2}
 +3\ln\,\frac{s}{m_{\rmfont{t}}^2}
{}\, ,
\\
\dsp
\Delta{\cal I}_3
& \dsp
\Rightarrow -4\zeta(2)+5
-\frac{s}{9 m_{\rmfont{t}}^2}
{}\, ,
\\
\dsp
{\cal I}_4
& \dsp
\Rightarrow 2\zeta(2) - \frac{15}{4}
{}\, ,
\\ \dsp
{\cal I}
& \dsp
\Rightarrow -~\frac{37}{4}
+\frac{7}{27}~\frac{s}{m_{\rmfont{t}}^2}
  +3\ln\,\frac{s}{m_{\rmfont{t}}^2}
{}\, .
 \ea
\eeq
At this point the scale in $\as$ is still ambiguous,
as is the precise definition of $m_{\rmfont{t}}$.
In fact, since two
different mass scales,
$M_{\stand{Z}}$ and $m_{\rmfont{t}}$,  are present
in the problem, the asymptotic behaviour
for large $m_{\rmfont{t}}$ cannot be directly derived from this
result. The evaluation of the leading logarithms
in Ref.~\cite{CK4} allows the resolution of the
ambiguity between the choice of $\mu = m_{\rmfont{t}}$
and $\mu =  M_{\stand{Z}}$.
The remaining constant term has been evaluated
in Ref.~\cite{ChetTar93}. For the renormalization scale
$\mu^2=s$\,, and with
$m_{\rmfont{t}}^2=\ovl{m}_{\rmfont{t}}^2(s)$ \,,
one gets in the limit of large $m_{\rmfont{t}}$:
\begin{eqnarray}
\EQN{sm0a6}
d^2_{\rmfont{S}}&=&
 \dsp
\frac{1}{3}\Bigg\{
-\frac{37}{4}
+3 \ln\,\frac{s}{\ovl{m}_{\rmfont{t}}^2(s)}
\Bigg\}
{}\, ,
\nonumber\\ \dsp
d^3_{\rmfont{S}}&=&
 \dsp
\frac{1}{3}\Bigg\{
-\frac{5651}{72} + 3\zeta(3) +\frac{23}{12}\pi^2
+\frac{31}{6}
\ln\,\frac{s}{\ovl{m}_{\rmfont{t}}^2(s)}
+\frac{23}{4}\ln^2
\frac{s}{\ovl{m}_{\rmfont{t}}^2(s)}
\Bigg\}
{}\, .
\end{eqnarray}
For practical purposes it is more convenient
to employ the on-shell mass as the input parameter.
Relating the $\msbar$ mass at scale $\mu^2=s$
to the on-shell mass through Eq.~(\ref{m-from-M}),
one arrives at
\begin{eqnarray}
\EQN{sm0a7}
d^2_{\rmfont{S}}&=&
 \dsp
\frac{1}{3}\Bigg\{
-\frac{37}{4} + 3\ln\,\frac{s}{M_{t}^2}
\Bigg\}
{}\, ,
\nonumber\\ \dsp
d^3_{\rmfont{S}}&=&
\dsp
\frac{1}{3}\Bigg\{
-\frac{5075}{72}
+ 3\zeta(3) +\frac{23}{12}\pi^2
+\frac{67}{6}\ln\,\frac{s}{M_{\rmfont{t}}^2}
+\frac{23}{4}\ln^2\frac{s}{M_{\rmfont{t}}^2}
\Bigg\}
{}\, .
\end{eqnarray}
For $d^2_{\rmfont{S}}$ one should in fact include the
subleading terms $\sim 1/m_{\rmfont{t}}^2$ or, even more
appropriately, employ the complete answer,
or at least the approximation
Eq.~(\ref{sm0a4}).
\sect{Bottom Mass Corrections in the Singlet Term
\label{singl-bottom}}
Bottom quark mass effects have been neglected
in the previous section. Employing the techniques
of light/heavy mass expansions discussed in
Part~\ref{calc-tech},
 one may derive terms of order
$m_{\rmfont{b}}^2/s$ as well as
of order $m_{\rmfont{b}}^2/m_{\rmfont{t}}^2$.
Although the former are significantly more
important for realistic top masses than the latter,
we list  both contributions for completeness.
The corresponding results are \cite{ChetKwiat93,ChetKwiat92}:
\beq
\ba{ll}\dsp
r^{\rmfont{A}}_{\rmfont{S}}=
& \dsp
-6~\frac{\ovl{m}_{\rmfont{b}}^2}{s}\left(\api\right)^2
\left[-3+\ln\,\frac{s}{m_{\rmfont{t}}^2}\right]
\\ \dsp
& \dsp
-10~\frac{\ovl{m}_{\rmfont{b}}^2}{m_{\rmfont{t}}^2}\left(\api\right)^2
\left[\frac{8}{81}-\frac{1}{54}\ln\,\frac{s}{m_{\rmfont{t}}^2}\right]
{} \, .
\ea
\eeq

We would like to conclude this section with a brief
comment on $\ordas^3m_{\rmfont{b}}^2/s)$ singlet terms.
As stated in Section~\ref{vector}, these are absent
in the case of vector-current correlators.
Hence, for a complete treatment of order
$\ordas^3)$, including mass corrections,  only axial
singlet and  non-singlet contributions
of order $\ordas^3m_{\rmfont{b}}^2/s)$ are missing.
In fact,  only corrections of this order to the
currents $\ovl{b}\gamma_{\mu}\gamma_5 b$
and $\ovl{t}\gamma_{\mu}\gamma_5 t$ are not yet
available [see the discussion in Section~\ref{top},
Eq.~\re{mt2b}].
The Born contribution of these currents
amounts to 20\% of the $\stand{Z}$ decay rate only,
and the missing corrections to these
terms of $\ordas^3m_{\rmfont{b}}^2/s)$
can be safely neglected both for the
moment and the foreseeable future.
\sect{Partial Rates
\label{singl-partial}}
Assigning of singlet terms $\Gamma^{\rmfont{S}}$
to partial decay rates into individual
${\rm q\ovl{q}}$ final states $\Gamma_{{\rm q\ovl{q}}}$,
which in turn are attributed to individual
quark currents, is evidently not possible
in a straightforward manner. Singlet contributions
arise from the interference between different
 quark amplitudes. Nevertheless, at the
present level of precision where experiments are
only able to identify heavy flavour rates with
a relative precision of about 1\%, pragmatic
prescriptions for this separation can be found.
\subsect{Axial Rate\label{singl-partial-axial}}
Let us start with the ${\cal O}(\as^2)$ singlet
term induced by the interference of axial top
and bottom currents. It can be decomposed into
two-, three- and four-particle cuts, corresponding
to the interference terms in Figs.~\ref{sin23}
and \ref{sin4}. For final states with bottom quarks
only  one has \cite{KniKue90b}
[see also Eq.~(\ref{sm0a4})]:
\beq
\EQN{spart1}
\ba{ll}
R^{\rmfont{S}}_{\rmfont{b}\ovl{\rmfont{b}}}
& \dsp
=
\frac{1}{3}\left(\api\right)^2
\left\{
-7.210+1.481\frac{s}{4m_{\rmfont{t}}^2}
 +1.363\left(\frac{s}{4m_{\rmfont{t}}^2}\right)^2
 +3\ln\,\frac{s}{m_{\rmfont{t}}^2}
\right\}
\\
&
\dsp
=
-\left(\api\right)^2
(3.52\pm 0.25)
{}\, ,
\\
\dsp
R^{\rmfont{S}}_{\rmfont{b}\ovl{\rmfont{b}}\rmfont{g}}
& \dsp
=
\frac{1}{3}\left(\api\right)^2
\left\{
 -1.580-0.444\frac{s}{4m_{\rmfont{t}}^2}
 -0.731\left(\frac{s}{4m_{\rmfont{t}}^2}\right)^2
\right\}
\\
&
\dsp
=
-\left(\api\right)^2
(0.60\pm 0.01)
{}\, ,
\\
\dsp
R^{\rmfont{S}}_{\rmfont{b}\ovl{\rmfont{b}}\rmfont{b}\ovl{\rmfont{b}}}
& \dsp
=
-~
\frac{1}{3}\left(\api\right)^2
0.460
\\
&
\dsp
=
-\left(\api\right)^2
0.15
{}\, ,
\ea
\eeq
where $m_{\rmfont{t}}=174\pm 20$ GeV has been assumed in the
numerical evaluation. The first, logarithmically
enhanced term dominates and can reasonably be
assigned to $\Gamma_{\rmfont{b}\ovl{\rmfont{b}}}$. The same holds
true for the three particle contribution
from $\rmfont{b}\ovl{\rmfont{b}}g$. At this point it should be
stressed that   other ${\rm q\ovl{q}}$ final states
are affected by this mechanism:
\beq \EQN{spart2}
\ba{lllll}\dsp
R^{\rmfont{S}}_{\rmfont{d}\ovl{\rmfont{d}}}
& \dsp
=
-R^{\rmfont{S}}_{\rmfont{u}\ovl{\rmfont{u}}}
& \dsp
=
R^{\rmfont{S}}_{\rmfont{s}\ovl{\rmfont{s}}}
& \dsp
=
-R^{\rmfont{S}}_{\rmfont{c}\ovl{\rmfont{c}}}
& \dsp
=
R^{\rmfont{S}}_{\rmfont{b}\ovl{\rmfont{b}}}
{}\,~,
\\
\dsp
R^{\rmfont{S}}_{\rmfont{d}\ovl{\rmfont{d}}\rmfont{g}}
& \dsp
=
-R^{\rmfont{S}}_{\rmfont{u}\ovl{\rmfont{u}}\rmfont{g}}
& \dsp
=
R^{\rmfont{S}}_{\rmfont{s}\ovl{\rmfont{s}}\rmfont{g}}
& \dsp
=
-R^{\rmfont{S}}_{\rmfont{c}\ovl{\rmfont{c}}\rmfont{g}}
& \dsp
=
R^{\rmfont{S}}_{\rmfont{b}\ovl{\rmfont{b}}\rmfont{g}}
{}\,~.
\ea
\eeq
These terms are proportional to the
isospin of the quark, whence contributions
from the massless doublet cancel.
The assignment of singlet ${\rmfont{q}\ovl{\rmfont{q}}}$ and
${\rmfont{q}\ovl{\rmfont{q}}\rmfont{g}}$
terms can therefore be performed
in a convincing manner.

The situation is more intricate for the four-fermion
final states,
say $\rmfont{q}\ovl{\rmfont{q}}\rmfont{b}\ovl{\rmfont{b}}$,
which originate
from the interference between ${\rm q\ovl{q}}$
and $\rmfont{b}\ovl{\rmfont{b}}$ induced amplitudes
(Fig.~\ref{sin4}). Neglecting masses one
obtains
\beq\EQN{spart3}
R^{\rmfont{S}}_{\rmfont{b}\ovl{\rmfont{b}}\rmfont{b}\ovl{\rmfont{b}}}
= -~\frac{1}{2}R^{\rmfont{S}}_{\rmfont{b}\ovl{\rmfont{b}}
                    \rmfont{u}\ovl{\rmfont{u}}}
=  \frac{1}{2}R^{\rmfont{S}}_{\rmfont{b}\ovl{\rmfont{b}}
                    \rmfont{d}\ovl{\rmfont{d}}}
= -~\frac{1}{2}R^{\rmfont{S}}_{\rmfont{b}\ovl{\rmfont{b}}
                    \rmfont{c}\ovl{\rmfont{c}}}
=  \frac{1}{2}R^{\rmfont{S}}_{\rmfont{b}\ovl{\rmfont{b}}
                    \rmfont{s}\ovl{\rmfont{s}}}
{}\,~.
\eeq
Among the final states from these
diagrams with at least one $\rmfont{b}\ovl{\rmfont{b}}$ pair,
only the
$\rmfont{b}\ovl{\rmfont{b}}\rmfont{b}\ovl{\rmfont{b}}$
term remains after all
compensations have been taken into account.
Numerically it is tiny, about a factor 25
below the
$\rmfont{b}\ovl{\rmfont{b}}+\rmfont{b}\ovl{\rmfont{b}}
\rmfont{g}
$ rate,
and can therefore
be ignored at present.
The four-fermion contribution is
present also for other light quarks. Within
one complete doublet the cancellation
occurs between mixed and pure configurations:
\beq\EQN{spart4}
R^{\rmfont{S}}_{\rmfont{u}\ovl{\rmfont{u}}\rmfont{d}\ovl{\rmfont{d}}}
=
R^{\rmfont{S}}_{\rmfont{d}\ovl{\rmfont{d}}\rmfont{d}\ovl{\rmfont{d}}}
= -
{}~\frac{1}{2}
R^{\rmfont{S}}_{\rmfont{u}\ovl{\rmfont{u}}\rmfont{d}\ovl{\rmfont{d}}}
=
R^{\rmfont{S}}_{\rmfont{b}\ovl{\rmfont{b}}\rmfont{b}\ovl{\rmfont{b}}}
{}\,~.
\eeq
If one were to  insist on distributing
the four-fermion singlet part to specific partial rates,
the separation could only be performed in an
analysis tailored to the specific experimental
cuts. Mixed configurations like
$\rmfont{b}\ovl{\rmfont{b}}
\rmfont{c}\ovl{\rmfont{c}}
$ could either be assigned to
$\Gamma_{\rmfont{b}\ovl{\rmfont{b}}}$ (`hierarchical'    assignment)
or with equal weight to
$\Gamma_{\rmfont{b}\ovl{\rmfont{b}}}$ and
$\Gamma_{\rmfont{c}\ovl{\rmfont{c}}}
$
(`democratic'    assignment).
Furthermore, four fermion events with
secondary radiation (which exhibit mass
singularities --- see\break\hfill Section~\ref{nspart})
lead to similar
signatures and must be subtracted from the
singlet parts with the help of Monte Carlo
simulations. However, as stated above,
the issue can be ignored at present and the
assignment of the axial singlet terms to
partial rates according to the quark
isospin seems adequate.

\begin {figure}
\begin{center}
\vspace*{-1cm}
\begin {tabular}{lll}
\parbox{3cm}{
\epsfig{file=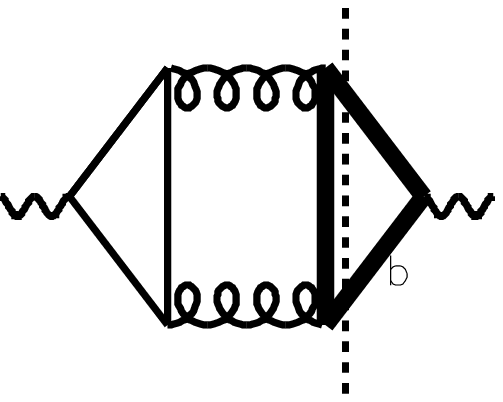,width=5.cm,height=5.cm}
            }
&
\hphantom{XXXXXX}
&
\parbox{3cm}{
\epsfig{file=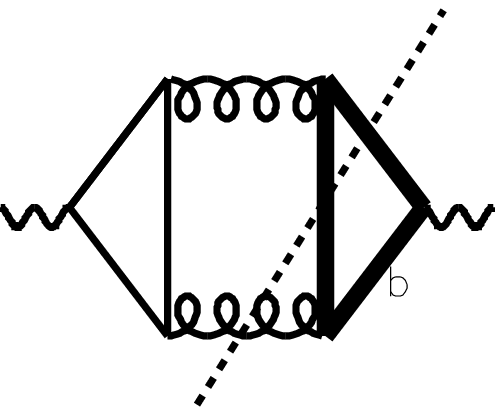,width=5.cm,height=5.cm}
            }
\end {tabular}
\caption {\label{sin23}
{{Singlet contributions with final states
$\rmfont{b}\ovl{\rmfont{b}}$ and $\rmfont{b}\ovl{\rmfont{b}}g$.
}}}
\end{center}
\end{figure}
\begin {figure}
\begin{center}
\vspace*{-1cm}
\begin {tabular}{lll}
\parbox{4cm}{
\epsfig{file=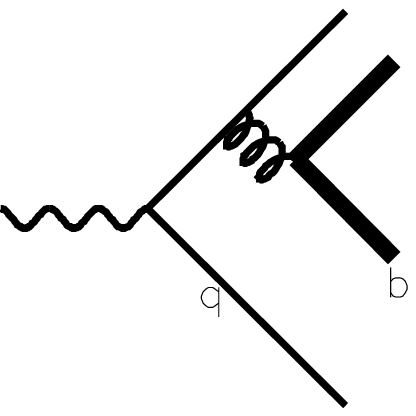,width=4.cm,height=5.cm}
            }
$*$
&
\parbox{4cm}{
\epsfig{file=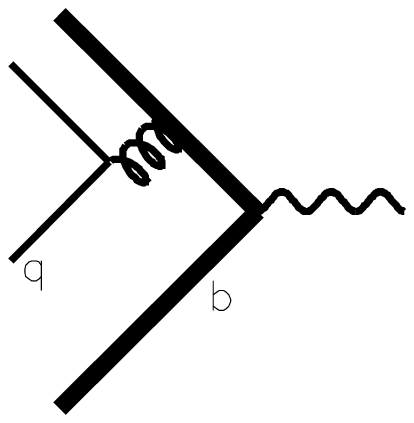,width=4.cm,height=5.cm}
            }
$=$
&
\parbox{4cm}{
\epsfig{file=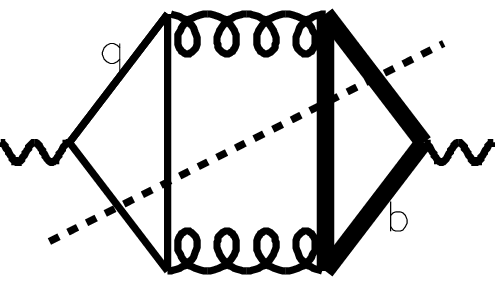,width=4.cm,height=5.cm}
            }
\end {tabular}
\end{center}
\caption {\label{sin4}
{{Singlet contribution
 with four-fermion
final state ${\rm q\ovl{q}}\rmfont{b}\ovl{\rmfont{b}}$.
}}}
\end{figure}

The $\as^3$ term has not been decomposed
into individual cuts, except for  the three gluon
final state discussed below. Nevertheless it
is evident from the structure of the
calculation that the leading logarithmically
enhanced terms can be interpreted as  a  correction
to the $\rmfont{b}\ovl{\rmfont{b}}$ configuration and hence are
again proportional to the weak isospin.
Therefore the complete axial
singlet rate  equations ~(\ref{sm0a4})  and (\ref{sm0a7})
can be assigned to $\Gamma_{\rmfont{q}\ovl{\rmfont{q}}}$
with a weight proportional to the weak isospin
$I_3^{\rmfont{q}}$.

As noted above, the three-gluon rate induced
 by the axial current has been calculated
with the truly tiny result
\cite{HopBij94},
\beq\EQN{spart5}
\ba{ll}\dsp
\Gamma_{\rmfont{ggg}}^{\rmfont{A}}
& \dsp
=\frac{G_FM_{\stand{Z}}^3}{24 \sqrt{2}\pi}
\left(\frac{\alpha_s}{\pi}\right)^3
\frac{1}{16}
\left[ \frac{2981}{3}-58\pi^2-\frac{21}{5}\pi^4
-8\zeta(3)\right]
= 0.00072\;\; {\rm MeV}
{}\, .
\ea
\eeq

\subsect{Vector Rate\label{singl-partial-vector}}
In order $\as^2$ there are no singlet
terms as a consequence of charge conjugation
(Furry's Theorem). The $\as^3$ term given in
Section~\ref{singl-vector}  receives contributions from
two  to five parton configurations. Again,
only the three-gluon contribution has been
calculated separately \cite{HopBij94}:
\begin{eqnarray}
\Gamma_{\rmfont{ggg}}^{\rmfont{V}}
&=&\frac{G_FM_{\stand{Z}}^3}{24 \sqrt{2}\pi}
\left(\frac{\alpha_s}{\pi}\right)^3
\left(\sum_{\stand{f}} v_{\stand{f}} \right)^2
\frac{5}{144}
\Biggl[ -124+\frac{41}{3}\pi^2+\frac{7}{15}\pi^4\nonumber\\
&&-128\zeta(3)+200\zeta(5)-8\pi^2
\zeta(3)\Biggl]
=0.0041 \;\;{\rm MeV}
{}\, .
\end{eqnarray}
The prediction is again
far below the level  of detectability.
\chap{Numerical Discussion\label{numerical}}
\sect{$\stand{Z}$ Decays \label{Zdecays}}
One of the central tasks at LEP is the
extraction of a precise value for $\as$
from the hadronic $\stand{Z}$ decay rate
(or from derived quantities such as  $R_{\rmfont{had}}$
or $\sigma$). Another quantity of interest
is the ration
$\Gamma_{\rmfont{b}\ovl{\rmfont{b}}}/\Gamma_{\rmfont{had}}$\,,
which provides important limits on the
mass of the top quark and, indirectly, on
new physics. It is therefore mandatory
to explore the sensitivity of these predictions
with respect to uncertainties in the input
parameters, such as quark masses or $\as$\,,
and to deduce estimates on the uncertainties
from as yet uncalculated higher-orders.

For the convenience of the reader we shall now
extract from the previous parts
a summary of the main results
combined with a numerical
evaluation that is particularly tailored for
the energy regime around the $\stand{Z}$. If not stated
otherwise, $\as$ will denote the QCD
coupling $\as(s)$ in the $\msbar$-scheme
 evaluated for five flavours at the scale $s$.
As input for
$\Lambda_{\ovl{\rmfont{MS}}}$
 we shall use the value
$\Lambda_{\ovl{\rmfont{MS}}}=233$ MeV
 corresponding to
$\as(M_{\stand{Z}}^2)=0.1200$.
The b-mass $\ovl{m}_{\rmfont{b}}(s)$ will be
taken as $\msbar$-mass in a five-flavour
theory at the mass scale ${s}$.
For the bottom pole mass we shall use the value of
$M=M_{\rmfont{b}}=(4.7\pm 0.2)$ GeV.
It is related to
$\ovl{m_{\rmfont{b}}}({M_{\rmfont{b}}})$ through Eq.~(\ref{m-from-M}),
evaluated at $\mu=M_{\rmfont{b}}$.
The running mass is therefore dependent on
$\alpha_s$.
 A few typical values are given in
Table 2, where we  also  anticipate the
values relevant for the subsequent discussion
at lower energies. Our default value
corresponds to a running mass at the
$\stand{Z}$ peak of
$m \equiv \ovl{m}_{\rmfont{b}}(M_{\stand{Z}}^2)=2.77$ GeV.
The running charm mass is about a factor of five
smaller than $\ovl{m}_{\rmfont{b}}$. Corrections from
$\ovl{m}_{\rmfont{c}}^2/s$ terms are hence entirely
negligible for $\stand{Z}$ decays.
For $\sin^2\theta_{\rmfont{W}}$ the value 0.2321 is adopted.
We also use
the value
$1/{\alpha}(M_{\stand{Z}}) = 127.9 \pm 0.1  $
\cite{Sirlin93}
for the running fine
structure constant at the scale of $M_{\stand{Z}}$.
For the top (pole) mass we choose $M_{\rmfont{t}}=174 \pm 20$ GeV.

\begin{table}
\hspace*{-0.5cm}
\begin{center}
{\bf Table 2}\\
\vskip0.2cm
{\small{Table of bottom masses.}}
\vskip0.2cm
\begin{tabular}{|c|c|c|c|c|c|c|}
\hline
$\Lambda_{\ovl{\rmfont{MS}}}$
&
$\as(M_{\stand{Z}})$
&
$\as(10~{\rm GeV} )$
&
$M$
&
$\ovl{m}(M)$
&
$\ovl{m}(M_{\stand{Z}})$
&
$\ovl{m}( 10~{\rm GeV} )$
\\ \hline\hline
0.150 & 0.112 & 0.165 & 4.70 & 4.10 & 2.95 & 3.69
 \\ \hline
0.200 & 0.117 & 0.176 & 4.70 & 4.04 & 2.84 & 3.59
 \\ \hline
0.233 & 0.120 & 0.183 & 4.70 & 3.99 & 2.77 & 3.54
 \\ \hline
0.300 & 0.125 & 0.195 & 4.70 & 3.91 & 2.64 & 3.43
 \\ \hline
0.400 & 0.131 & 0.210 & 4.70 & 3.80 & 2.48 & 3.28
 \\ \hline
\end{tabular}
\end{center}

\end{table}

\subsect{The Total Hadronic Decay Rate
            $\Gamma_{\rmfont{had}}$ \label{total}}
The hadronic decay rate  can be cast into following
form:
\beq\EQN{num1}
\Gamma_{\rmfont{had}} = \sum_{i=1}^{12} \Gamma_i
= \sum_{i=1}^{12} \Gamma_0 R_i
{}\, ,
\eeq
with
$\Gamma_0=G_{\rmfont{F}} M_{\stand{Z}}^3/24 \sqrt{2} \pi=82.94$ MeV,
$v_{\stand{f}} = 2 I^{\stand{f}}_3 - 4Q_{\stand{f}}
\sin^2 \theta_{\rmfont{w}}, a^{\stand{f}}_3 = 2 I_{\stand{f}}$,
and the following separate contributions:
\\
\vskip0.2cm
\noindent
Massless non-singlet corrections:
\beq\EQN{num2}
\ba{ll}\dsp
R_1
& \dsp
= 3\sum_{\stand{f}} (v_{\stand{f}}^2+a_{\stand{f}}^2)r_1
\\ &  \dsp
= 3\sum_{\stand{f}} (v_{\stand{f}}^2+a_{\stand{f}}^2)
\left[
1+\api+1.40932\left(\api\right)^2
-12.76706 \left(\api\right)^3
\right]
{}\, ;
\ea
\eeq
\noindent
Massive universal corrections
 (`double bubble'):
\beq\EQN{num3}
\ba{ll}\dsp
R_2
& \dsp
= 3\sum_{\stand{f}} (v_{\stand{f}}^2+a_{\stand{f}}^2)r_2
\\ &  \dsp
= 3\sum_{\stand{f}} (v_{\stand{f}}^2+a_{\stand{f}}^2)
(-6.126)\left(\api\right)^3
\sum_{f'=b}\frac{\ovl{m}_{f'}^2}{s}
\\ \dsp
R_3
& \dsp
= 3\sum_{\stand{f}} (v_{\stand{f}}^2+a_{\stand{f}}^2)r_3
\\ &  \dsp
= 3\sum_{\stand{f}} (v_{\stand{f}}^2+a_{\stand{f}}^2)
\left(\api\right)^2
\sum_{f'=b}\frac{\ovl{m}_{f'}^4}{s^2}
\left[
-0.4749 -\ln\,\frac{\ovl{m}_{f'}^2}{s}
\right]
\\ \dsp
R_4
& \dsp
= 3\sum_{\stand{f}} (v_{\stand{f}}^2+a_{\stand{f}}^2)r_4
\\ &  \dsp
= 3\sum_{\stand{f}} (v_{\stand{f}}^2+a_{\stand{f}}^2)
\left(\api\right)^2
\frac{s}{M_{\rmfont{t}}^2}
\left[
0.0652 +0.0148\ln\,\frac{M_{\rmfont{t}}^2}{s}
\right]
{}\, ;
\ea
\eeq

Massive non-singlet corrections (vector):
\beq\EQN{num4}
\ba{ll}\dsp
R_5
& \dsp
= 3v_{\rmfont{b}}^2 r_5
\\ &  \dsp
= 3v_{\rmfont{b}}^2 ~\frac{\ovl{m}_{\rmfont{b}}^2}{s}
12
\left[
\api+8.736\left(\api\right)^2
+45.657 \left(\api\right)^3
\right]
\\ \dsp
R_6
& \dsp
= 3v_{\rmfont{b}}^2 r_6
\\ & \dsp
= 3v_{\rmfont{b}}^2~\frac{\ovl{m}_{\rmfont{b}}^4}{s^2}
\left[
-6-22~\api+\left(\api\right)^2
\left(
139.489-3.833\ln\,\frac{\ovl{m}_{\rmfont{b}}^2}{s}
\right)
\right]
{}\, ;
\ea
\eeq

Massive non-singlet corrections (axial):
\beq\EQN{num5}
\ba{ll}\dsp
R_7
& \dsp
= 3r_7
\\ \dsp
& \dsp
= 3 ~\frac{\ovl{m}_{\rmfont{b}}^2}{s}(-6)
\left[
1+3.667~\api+14.286\left(\api\right)^2
\right]
\\ \dsp
R_8
& \dsp
= 3r_8
\\ \dsp
 & \dsp
= 3~\frac{\ovl{m}_{\rmfont{b}}^4}{s^2}
\left[
6+10~\api+\left(\api\right)^2
\left(
-217.728+26.833\ln\,\frac{\ovl{m}_{\rmfont{b}}^2}{s}
\right)
\right]
{}\, ;
\ea
\eeq

Singlet corrections (axial):
\begin{eqnarray}
\EQN{num6}
R_9
&=& 3r_9
\nonumber\\
&=&3~\Bigg\{
\left(\api\right)^2\frac{1}{3}
\left[
-9.250+1.037 \frac{s}{4M_{\rmfont{t}}^2}
+0.632~\Biggl(\frac{s}{4M_{\rmfont{t}}^2}\Biggl)^2
+3\ln\,\frac{s}{M_{\rmfont{t}}^2}
\right]
\nonumber\\
&&+\left(\api\right)^3\frac{1}{3}
\left[
-47.963+11.167\ln\,\frac{s}{M_{\rmfont{t}}^2}
+5.75\ln^2\frac{s}{M_{\rmfont{t}}^2}
\right]\Bigg\}~,
\\
R_{10}
& =& 3r_{10}
\nonumber\\
&=&3\left(\api\right)^2
\left\{
-6\frac{\ovl{m}_{\rmfont{b}}^2}{s}
\left[
-3+\ln\,\frac{s}{M_{\rmfont{t}}^2}
\right]
\right.
\left.
-~10~\frac{\ovl{m}_{\rmfont{b}}^2}{M_{\rmfont{t}}^2}
\left[
0.0988-0.0185\ln\,\frac{s}{M_{\rmfont{t}}^2}
\right]
\right\}\nonumber
{}\, ;
\end{eqnarray}

Singlet corrections (vector):
\beq\EQN{num7}
\ba{ll}\dsp
R_{11}
& \dsp
= 3\Biggl(\sum_fv_{\stand{f}}\Biggl)^2 r_{11}
\\ & \dsp
= \Biggl(\sum_fv_{\stand{f}}\Biggl)^2
\left(\api\right)^3 (-1.2395)
{}\, ;
\ea
\eeq

${\cal O}(\alpha\alpha_s)$ corrections:
\beq\EQN{num8}
\ba{ll}\dsp
R_{12}
& \dsp
= 3\sum_{\stand{f}}(v_{\stand{f}}^2
 +a_{\stand{f}}^2)Q_{\stand{f}}^2 r_{12}
\\ & \dsp
= 3\sum_{\stand{f}}(v_{\stand{f}}^2
+a_{\stand{f}}^2)Q_{\stand{f}}^2~\frac{3}{4}~
\frac{\alpha}{\pi}
\left[
1-\frac{1}{3}~\api
\right]
{}\, .
\ea
\eeq
The various contributions are listed
in Table 3, where the terms of order
$\as,\as^2$ and $\as^3$ are also
separately   displayed.

Electroweak corrections are not incorporated.
To predict  precise numerical results for the
width,  the formulas should be used only in
conjunction with electroweak corrections.
\begin{table}
\begin{center}
{\bf Table 3}\\
\vskip0.2cm
{\bf Numerical Values}
\end{center}
\begin{tabular}{|l||r|r|r|r|r|r|r|}
\hline
 & \multicolumn{1}{|c|}{$\as^0$} & \multicolumn{1}{|c|}{$\as^1$} &
\multicolumn{1}{|c|}{$\as^2$} &
\multicolumn{1}{|c|}{$\as^3$} & \multicolumn{1}{|c|}{$\sum$}
 & \multicolumn{1}{|c|}{$R_{\stand{i}}$} &
\multicolumn{1}{|c|}{$\Gamma_{\stand{i}}$[MeV]}
  \\
  \hline
  \hline
$r_1$ & \multicolumn{1}{|c|}{$\scriptstyle 1$} & $ \scriptstyle 3.821\times
10^{-2}$
    & $\scriptstyle 2.057\times 10^{-3}$
    & $\scriptstyle -7.120\times 10^{-4}$
    & $ \scriptstyle 1.0396 $
    & $\scriptstyle 20.96025$
    & $\scriptstyle 1738.443$
\\ \hline
$r_2$ & \multicolumn{1}{|c|}{--} & \multicolumn{1}{|c|}{--} &
\multicolumn{1}{|c|}{--}
    & $\scriptstyle -3.153\times 10^{-7}$
    & $\scriptstyle -3.153\times 10^{-7}$
    & $\scriptstyle -0.0000064$
    & $\scriptstyle -0.000527$
\\ \hline
$r_3$ & \multicolumn{1}{|c|}{--} & \multicolumn{1}{|c|}{--}
    & $\scriptstyle 8.095\times 10^{-9}$
    & \multicolumn{1}{|c|}{--} & $\scriptstyle 8.095\times 10^{-9}$
    &$\scriptstyle  0.0000002$ & $\scriptstyle 0.000014$
\\ \hline
$r_4$ & \multicolumn{1}{|c|}{--} & \multicolumn{1}{|c|}{--}
    & $\scriptstyle 3.381\times10^{-5}$
    & \multicolumn{1}{|c|}{--} & $\scriptstyle 3.381\times 10^{-5}$
    & $\scriptstyle 0.0006817$ & $\scriptstyle 0.056537$
\\ \hline
$r_5$ &\multicolumn{1}{|c|}{--} & $\scriptstyle 4.231\times 10^{-4}$
      & $\scriptstyle 1.412\times 10^{-4}$
      & $\scriptstyle 2.819\times 10^{-5}$
      & $\scriptstyle 5.924\times 10^{-4}$
      & $\scriptstyle 0.0008475$ & $\scriptstyle 0.070292 $
\\ \hline
$r_6$ & $\scriptstyle -5.109\times 10^{-6}$
      & $\scriptstyle -7.157\times 10^{-7}$
      & $\scriptstyle 2.067\times 10^{-7}$
      & \multicolumn{1}{|c|}{--} & $\scriptstyle -5.618\times 10^{-6}$
      & $\scriptstyle -0.0000080$ & $\scriptstyle -0.000667$
\\ \hline
$r_7$ & $\scriptstyle -5.537\times 10^{-3}$
  & $\scriptstyle -7.756\times10^{-4}$
  & $\scriptstyle -1.154\times 10^{-4}$
  & \multicolumn{1}{|c|}{--} & $\scriptstyle -6.428\times 10^{-3}$
  & $\scriptstyle -0.0192826$ &$\scriptstyle  -1.599301$
\\ \hline
$r_8$ & $\scriptstyle 5.109\times 10^{-6}$
      & $\scriptstyle 3.253\times 10^{-7}$
      & $\scriptstyle -5.037\times 10^{-7}$
      &\multicolumn{1}{|c|}{ --} & $\scriptstyle 4.930\times 10^{-6}$
      & $\scriptstyle 0.0000148$ & $\scriptstyle 0.001227$
\\ \hline
$r_9$ & \multicolumn{1}{|c|}{--} & \multicolumn{1}{|c|}{--}
      & $\scriptstyle -6.351\times10^{-3}$
      & $\scriptstyle -9.814\times 10^{-4}$
      & $\scriptstyle -7.332\times 10^{-3}$
      & $\scriptstyle -0.0219966$ & $\scriptstyle -1.824400$
\\ \hline
$r_{10}$ & \multicolumn{1}{|c|}{--} & \multicolumn{1}{|c|}{--}
      & $\scriptstyle 3.432\times 10^{-5}$
      & \multicolumn{1}{|c|}{--} & $\scriptstyle 3.423\times 10^{-5}$
      & $\scriptstyle -0.0001027$ & $\scriptstyle 0.008518 $
\\ \hline
$r_{11}$ & \multicolumn{1}{|c|}{--} & \multicolumn{1}{|c|}{--} &
\multicolumn{1}{|c|}{--}
      & $\scriptstyle -2.304\times 10^{-5}$
      & $\scriptstyle -2.304\times 10^{-5}$
      & $\scriptstyle -0.0001185$ &$\scriptstyle  -0.009831$
\\ \hline
$r_{12}$ & $\scriptstyle 1.867\times 10^{-3}$
   & $\scriptstyle -2.378\times 10^{-5}$
   & \multicolumn{1}{|c|}{--} & \multicolumn{1}{|c|}{--} & $\scriptstyle
1.843\times10^{-5}$
   & $\scriptstyle 0.0083492$ & $\scriptstyle 0.692480$
\\ \hline
\end{tabular}
\end{table}
Estimates of the theoretical error from as
yet uncalculated higher-orders are to some
extent subjective. They are frequently based
 on the stability of the prediction against
a variation of the renormalization scale or
the comparison of predictions in different
 schemes. Alternatively,  one may simply
identify the last calculated term with
an upper limit on the uncertainty.

Let us start with the discussion of the mass
corrections. In Fig.~\ref{scheme} mass
corrections of successively higher-orders
in the $\msbar$-scheme are compared with  those
in the on-shell scheme. The poor convergence of
 the OS prediction, which results from the
large logarithms in the coefficients, is
evident. The $\msbar$ prediction, however,
is fairly stable. $R^{\rmfont{V}}$ and
$R^{\rmfont{A}}$ are
 calculated to order ${\cal O}(\as^3)$
and  ${\cal O}(\as^2)$ respectively, whence
all error estimates can be limited to the axial
contribution. The size of the $\as^2$ term
in the sum $R_7 + R_8$\,,
resulting from the non-singlet contribution,
amounts
to $0.03$ MeV, which can be taken as a safe
error estimate.
(Including the corresponding singlet term
would reduce it  to\break\hfill $0.02$ MeV.)
\begin{figure}[H]
\begin{center}
\mbox{
\epsfig{file=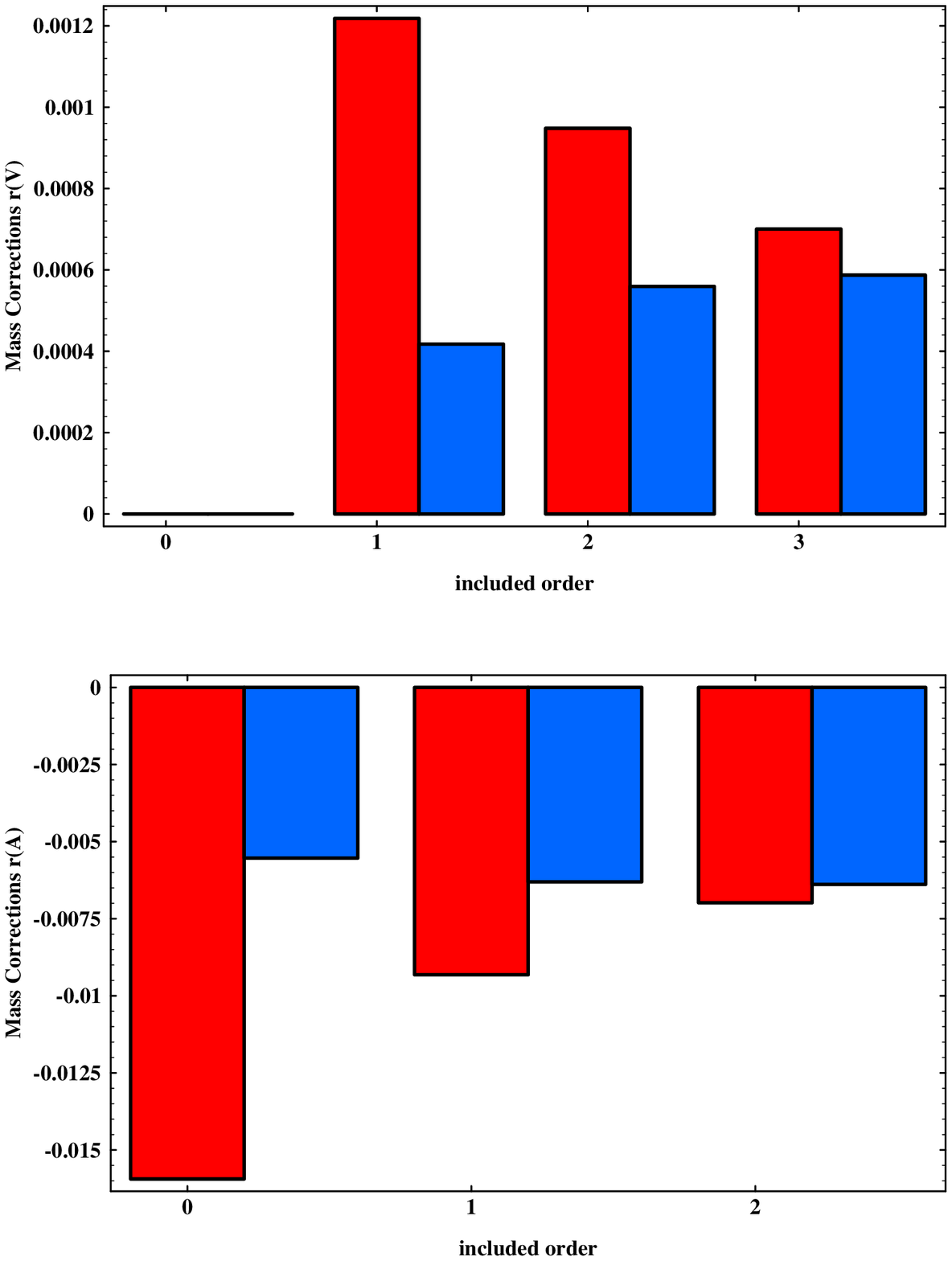,width=15.cm,height=20.cm}
 }
\end{center}
\caption{\label{scheme}
Mass Corrections from  $r^{(1)}_{\rmfont{V}}$ (upper graph)
and $r^{(1)}_{\rmfont{A}}$ (lower graph).
The left-hand bars represent the result in the on-shell
scheme, the  right-hand  ones are obtained in the
$\msbar$-scheme.
}
\end{figure}
{}From the study of the stability of the prediction
with respect to a variation of the renormalization
scale  an alternative error estimate can be deduced.
It is obtained by using an equation equivalent
to Eq.~(\ref{num5}), but with arbitrary scale $\mu^2$
(see Appendix).
In Fig.~\ref{mscale}
the scale is varied between $\mu^2=s/4$ and
$\mu^2=4s$. The corresponding error in the
prediction for the decay rate amounts to
$\delta \Gamma^{(m)}={ +0.022 \choose -0.006}$ MeV.

The  uncertainty in the prediction from the input mass
is essentially proportional to the relative
error in $m^2$. Adopting
$M_{\rmfont{b}}= 4.7 \pm 0.2\; {\rm GeV}$,
one is lead to
$\delta m^2/m^2\approx \pm 0.11$
and hence to a variation of the $m^2$ terms
by $\pm 11 \% $.
It is clear that this contribution
leads to the dominant error in the corrections
corresponding to  $\pm 0.17 $ MeV.
The combined uncertainty from mass term
is  therefore below
$\delta \Gamma^{(m)} =  \pm 0.21 $ MeV.

\begin{figure}[H]
\begin{center}
\mbox{\epsfig{file=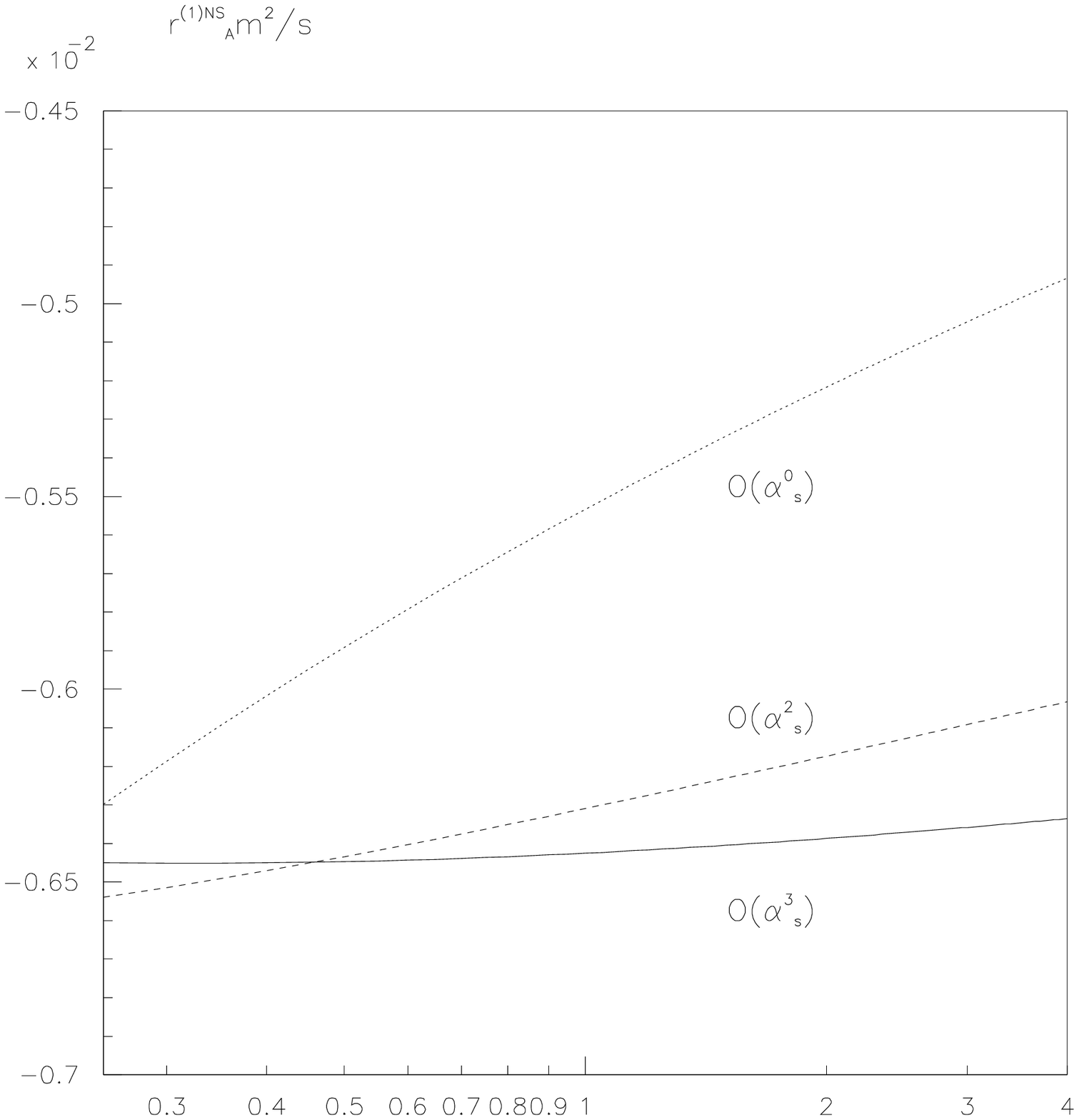,width=12.cm,height=9.cm,%
bbllx=0.cm,bblly=0.cm,bburx=18.cm,bbury=22.cm}
}
\end{center}
\caption[]{\label{mscale}
Renormalization scale dependence of the massive axial
QCD corrections;
$[\alpha_s(M_{\stand{Z}})=0.12]~.$}
\end{figure}

We now move to the massless limit. As discussed in
Section~\ref{singl-massless} the reliability of the singlet terms
is significantly improved through
the inclusion of the $\as^3$ corrections. This is
illustrated in Fig.~\ref{sinscale}, where the
${\cal O}(\as^2)$ and  ${\cal O}(\as^3)$
predictions as functions of the renormalization
scale are compared. From the variation of the
full prediction an uncertainty
$\delta \Gamma_{\rmfont{S}}={+0.12 \choose -0.05}$ MeV
can be deduced.
(Taking the last calculated term of order $\alpha_s^3$
as an error estimate would lead to
$\delta \Gamma_{\rmfont{S}}=\pm 0.25$ MeV.
)
The singlet terms  furthermore depend on the
top mass. A change of our default value
$M_{\rmfont{t}}=174$ GeV by $\pm 20$ GeV leads to a variation
of $\delta\Gamma_{\rmfont{S}}$ by ${+0.10 \choose -0.08}$ MeV.
\begin{figure}[H]
\vspace*{-1cm}

\begin{center}
\mbox{\epsfig{file=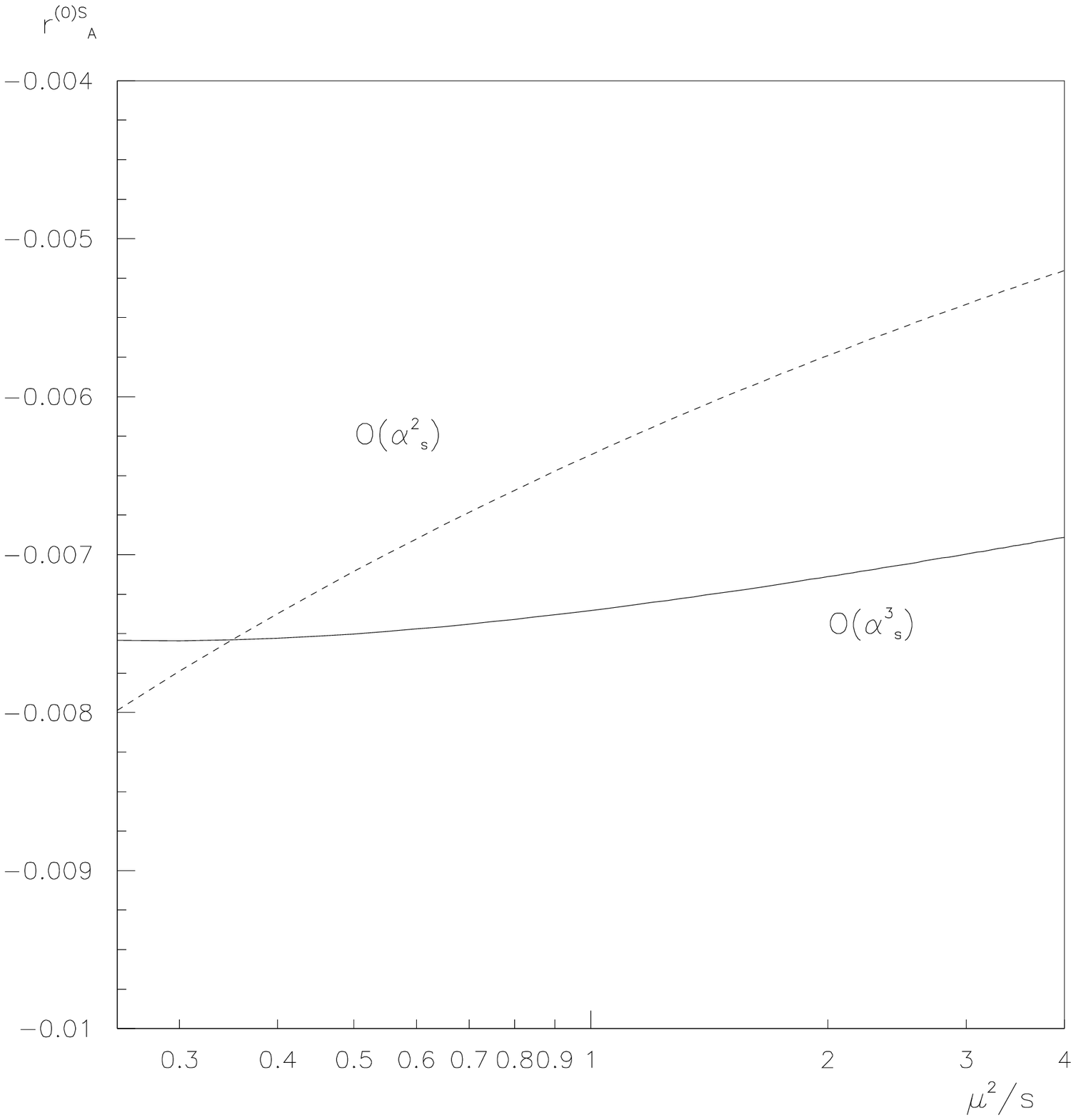,width=12.cm,height=9.cm,%
bbllx=0.cm,bblly=0.cm,bburx=18.cm,bbury=22.cm}
}
\end{center}
\caption[]{\label{sinscale}
{{Renormalization scale dependence of the axial singlet
massless QCD corrections;
$[\alpha_s(M_{\stand{Z}})=0.12]$.}}
}.
\end{figure}
\begin{figure}[H]
\begin{center}
\mbox{\epsfig{file=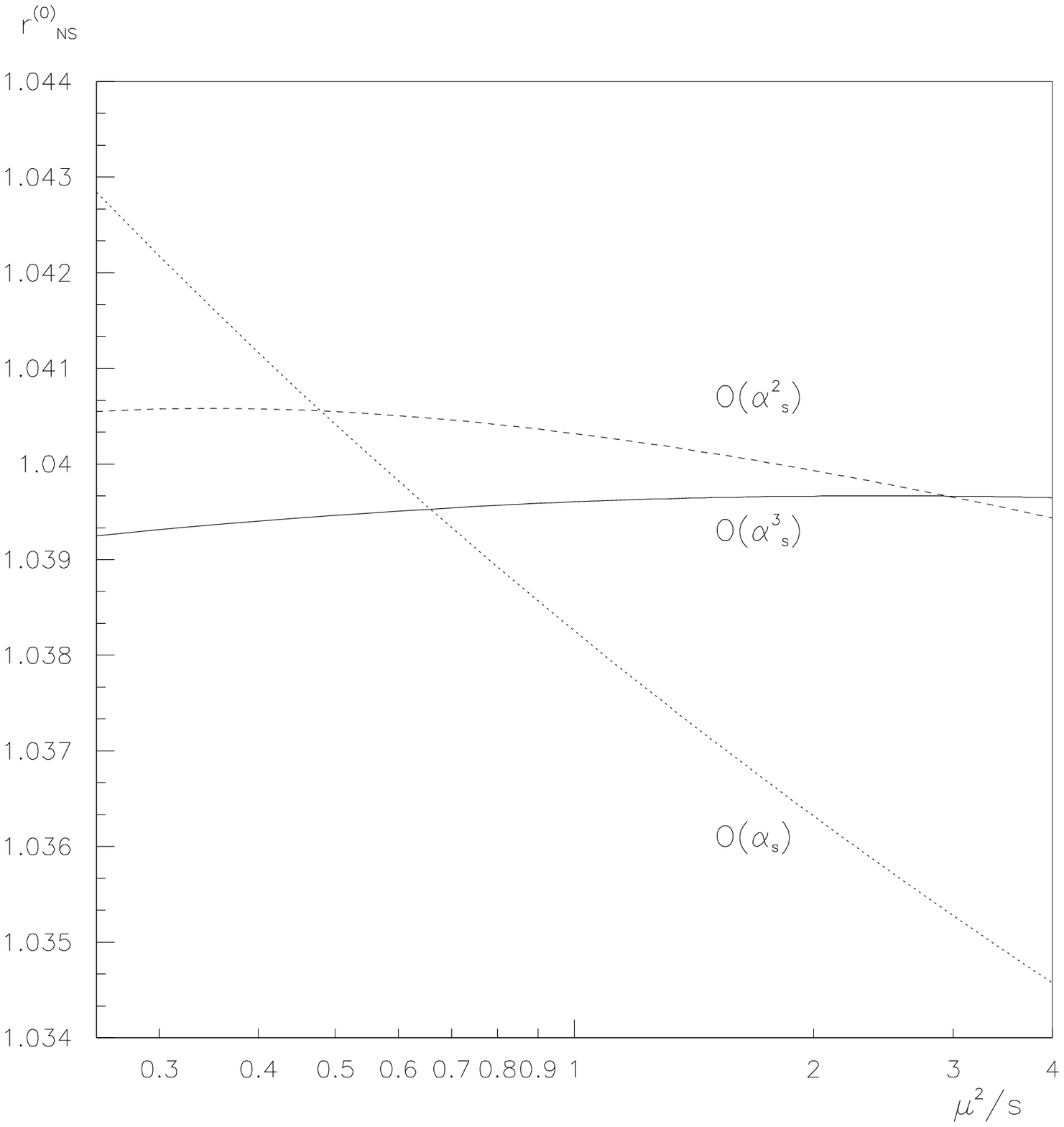,width=12.cm,height=9.cm,%
bbllx=0.cm,bblly=0.cm,bburx=18.cm,bbury=22.cm}
}
\end{center}
\caption[]{\label{nonsinscale}
{{Renormalization scale dependence of the non-singlet
massless QCD corrections.
$[\alpha_s(M_{\stand{Z}})=0.12]$
}}}
\end{figure}
Clearly, even the combining linearly of the modulus of
the three
 dominant errors (from the bottom mass, the
 singlet contribution and $M_{\rmfont{t}}$) leads
 to an uncertainty in the nonuniversal
corrections of only
 ${+0.43\choose -0.33}$ MeV.
This corresponds to
an uncertainty in the value of $\as$ extracted
from $\Gamma_{\rmfont{had}}$ of
${7.8 \choose -6.2} \times 10^{-4}$ and a relative error in
$\Gamma_{\rmfont{b}\ovl{\rmfont{b}}}/\Gamma_{\rmfont{had}}$ of
${11 \choose -9}\times 10^{-4}$,
significantly below the anticipated
experimental precision.

The remaining uncertainty in the
$\as$ determination
results from
the unknown terms of
 ${\cal O}(\as^4)$ in the non-singlet correction
$r^{(0)}_{\rmfont{NS}}$. In Fig.~\ref{nonsinscale}
the variation of $r^{(0)}_{\rmfont{NS}}\equiv r_1$ with $\mu^2$
is displayed. Evidently one arrives at a fairly
stable answer in order  ${\cal O}(\as^3)$.
The variation of
$ \delta r^{(0)}_{\rmfont{NS}}=
{+0.4\choose -3.5}\times 10^{-4}$
 may be interpreted
as an error estimate. It translates into
$ \delta \Gamma^{(0)}=
{+0.07\choose -0.6}$ MeV
and corresponds to an
uncertainty in $\as$ of
$ \delta \as=
{+0.11\choose -1.05}\times 10^{-3}$.

This result is of course  dependent on the
input for $\as$. For  central values of
$\as=0.115$
 and $\as=0.125$ one obtains
 $\delta\as={+0.11\choose -0.87}\times 10^{-3}$
and $\delta\as={+0.16\choose -1.26}\times 10^{-3}$
 respectively.
An alternative approach has been advocated
in Ref.~\cite{katai} (see also Ref.~\cite{Bjorken}), where
an attempt is made to actually arrive
at an estimate for the $(\as/\pi)^4$ term.
Adopting their value for the coefficient
of $-97$ one obtains a shift of
$\delta r^{(0)}_{\rmfont{NS}}=-2.1\times 10^{-4}$
and hence of $\as$ by $-6.3\times 10^{-4}$,
quite comparable to the error estimate
presented above.
As  a third option one may again take the last
calculated term in
$\delta r^{(0)}_{\rmfont{NS}}$ for an error estimate, resulting in
$\delta r^{(0)}_{\rmfont{NS}} = \pm 7.1 \times 10^{-4}$ and
$\delta \alpha_s = \pm 21.7 \times 10^{-4}$.  The choice is left
to the reader.
\subsect{The Partial Rate $\Gamma_{\rmfont{b}\ovl{\rmfont{b}}}$
\label{num-partial}}
Another quantity of interest is the ratio
$\Gamma_{\rmfont{b}\ovl{\rmfont{b}}}/\Gamma_{\rmfont{had}}$. Assuming
that ${\rm q \ovl{q}}$ events with secondary
radiation of $\rmfont{b}$ quarks (see Section~\ref{nspart})
are assigned to $\Gamma_{{\rm q \ovl{q}}}$, the
universal QCD corrections
 $\Gamma_{\stand{i}},\;i=1\dots 4$ cancel to a large extent
and  the
nonuniversal parts dominate. Their
 contribution
is small
and the resulting uncertainty
hence even smaller. The ratio can thus be
predicted quite unambiguously. The near
independence of the prediction on $\as$
is shown in Fig.~\ref{gbb}.
The flatness of the solid curve
is the result of two (accidental) cancellations.
With increasing  $\alpha_s(M_{\stand{Z}})$
the mass correction is lowered through the
 reduction  of the  running $\rmfont{b}$ mass
(for fixed pole mass) the singlet correction,
however, essentially increases proportional to $\alpha_s^2$.
This is illustrated by the dashed curve, where the running mass
has been kept fixed to the default  value.

The deviation
from the parton model prediction
(with $m_{\rmfont{b}}=0$) amounts to 0.7\% with a negligible
error from higher-orders in $\alpha_s$.
The uncertainties from the input values for $M_{\rmfont{b}}$
and $M_{\rmfont{t}}$ amount to $\pm 3\times 10^{-4}$
and $\pm 2\times 10^{-4}$  respectively.

\begin{figure}[H]
\vspace*{-1cm}
\epsfig{file=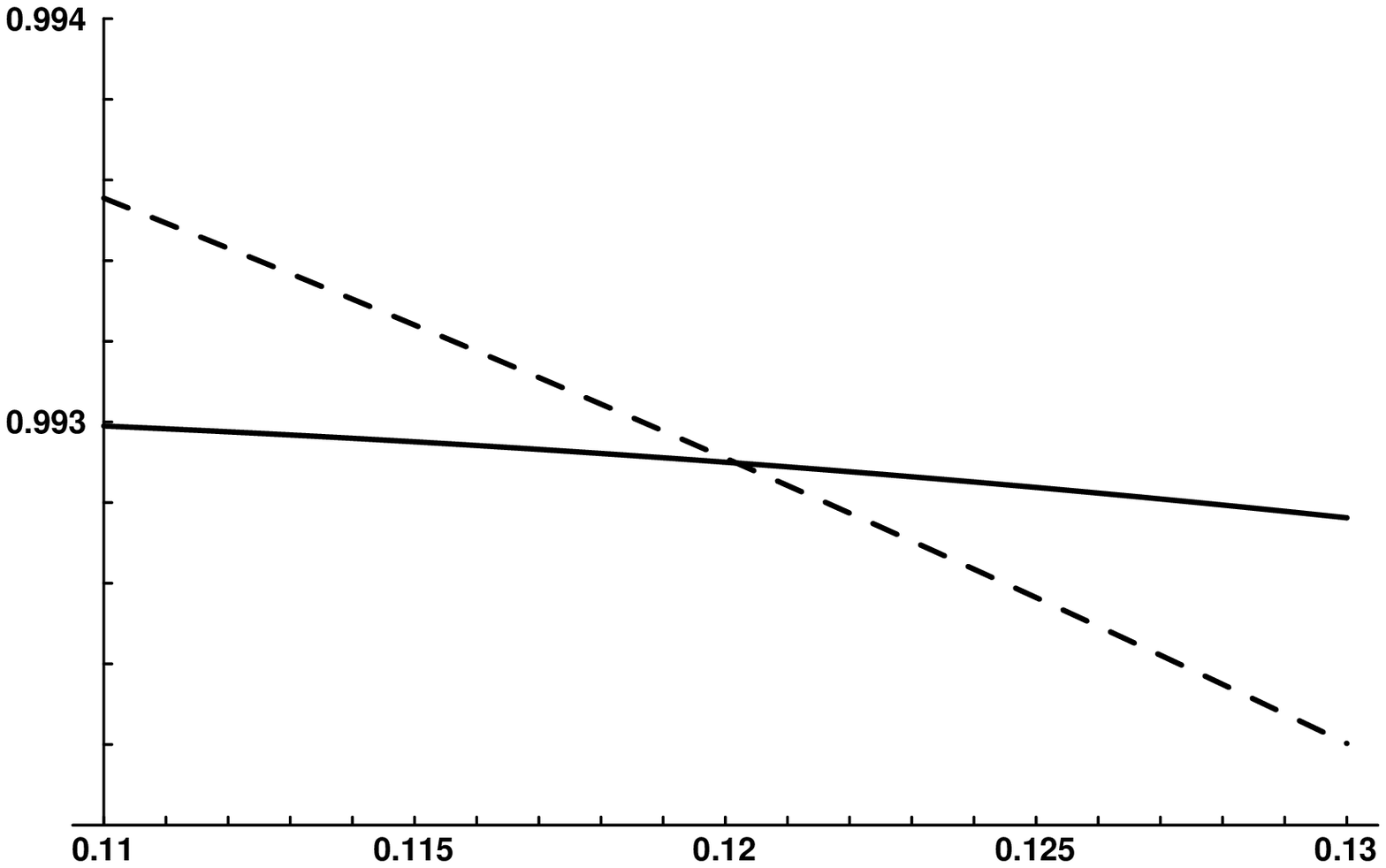,width=12.cm,height=9.cm,%
bbllx=0.cm,bblly=0.cm,bburx=18.cm,bbury=22.cm}
\vspace*{-3cm}
\caption[] {\label{gbb}
{{The ratio
$
\dsp
\Biggl\{\left(
{\Gamma_{\rmfont{b}\ovl{\rmfont{b}}}}%
/{v_{\rmfont{b}}^2 + a_{\rmfont{b}}^2}\right)/
\left[{\Gamma_{\rmfont{had}}}/{\sum(v_{\stand{f}}^2 + a_{\stand{f}}^2)}
\right]\Biggl\}
$
versus   \protect $\alpha_s$.
The dashed curve corresponds to a case of a fixed value
of $m_{\rmfont{b}}(M_{\stand{Z}}) = 2.77$ GeV, while
the continuous one takes into account  the implicit $\alpha_s$ dependence
of $m_{\rmfont{b}}(M_{\stand{Z}})$.}}
}
\end{figure}

\subsect{Quick Estimates}
The previous formulae display the full
dependence of the QCD corrections on the
input parameters $m_{\rmfont{b}}$ and $M_{\rmfont{t}}$ as well as
their effect on the vector and axial vector rate
of the various quark species separately.
These are the formulae most suited as building
blocks for detailed fitting programs.
It seems, however, also useful to provide
a short numerical formula suited for quick
estimates. For this purpose we set
$m= \ovl{m_{\rmfont{b}}}(M_{\stand{Z}}) = 2.77~\pm$\break\hfill 0.15  GeV
(corresponding approximately  to
$M= {m_{\rmfont{b}}|_{pole}} = 4.7 \pm 0.20 $ GeV)
and
$M_{\rmfont{t}}|_{pole}=174~\pm$ 20 GeV.
One obtains:
\begin{eqnarray}
\EQN{num9}
\frac{\Gamma_{\rmfont{had}}}{\Gamma_0}
&=&
3\sum_f
\dsp
(v_{\stand{f}}^2+a_{\stand{f}}^2)
\Bigg\{
1+\api+1.409\left(\api\right)^2
-12.767 \left(\api\right)^3
\nonumber\\
&&
+(0.023 \pm 0.005)\left(\api\right)^2
+
(-0.006 \pm 0.001)\left(\api\right)^3
\Bigg\}
\nonumber\\
&&
+~3v_{\rmfont{b}}^2
\Bigg\{
\dsp
(
-5\times 10^{-6} \pm 1 \times 10^{-6}
)
+
(
0.011 \pm 0.001
)
\api
\nonumber\\
&&
\dsp
+~(
0.097 \pm 0.01
)
\left(\api\right)^2
+
(
0.51  \pm 0.05
)
\left(\api\right)^3
\Bigg\}
\\
&&
\dsp
+~
3\Bigg\{
\dsp
(
-0.0055 \pm 0.0006
)
+
(
-0.020 \pm  0.002
)
\api
\nonumber\\
&&
\dsp
+~(-4.41 \pm 0.26)
\left(\api\right)^2
+
(-17.60 \pm 0.30)
\left(\api\right)^3
\Bigg\}
\nonumber\\ \dsp
&& \dsp
+~3\left(\sum_{\stand{f}} v_{\stand{f}}\right)^2
\Bigg\{
-0.413\left(\api\right)^3
\Bigg\}
\nonumber\\
&&
\dsp
+~3\sum_{\stand{f}}
(v_{\stand{f}}^2+a_{\stand{f}}^2)Q_{\stand{f}}^2
\Bigg\{
 0.001867
-0.000622
\api
\Bigg\}
{}\, .\nonumber
\end{eqnarray}
The origin of the terms is still evident
from the structure of the couplings.
For a simple
evaluation of QCD corrections to the total rate
one may now  combine the correction
coefficients with the numerically evaluated
weights $v_{\rmfont{b}}^2/\sum_{\stand{f}}
(v_{\stand{f}}^2+a_{\stand{f}}^2)$ etc.\
and arrive at
\begin{eqnarray}
\EQN{num10}
\Gamma_{\rmfont{had}}
 \dsp
&=&
\Gamma_{\rmfont{had}}
\Bigg|_{{\scriptstyle m_{\rmfont{b}}=0\atop\scriptstyle\as=0}}
\Bigg\{
1+\api+1.40932\left(\api\right)^2
-12.76706 \left(\api\right)^3
\nonumber\\ \dsp
 \dsp
%
%
&&-~(0.00040\pm 0.00008)
-(0.0023\pm 0.0002)\api
\nonumber\\ \dsp
 \dsp
&&-~(0.63\pm 0.04)\left(\api\right)^2\nonumber\\
&&-~(2.69 \pm +0.06 )\left(\api\right)^3
\Bigg\}
{}\, .
\end{eqnarray}
A similar treatment of $\Gamma_{\rmfont{b}\ovl{\rmfont{b}}}$
implies
\begin{eqnarray}
\EQN{num10b}
\Gamma_{\rmfont{b}\ovl{\rmfont{b}}}
& =&
\Gamma_{\rmfont{b}\ovl{\rmfont{b}}}
\Bigg|_{{\scriptstyle m_{\rmfont{b}}=0\atop\scriptstyle\as=0}}
\Bigg\{
1+\api+1.40932\left(\api\right)^2
-12.76706 \left(\api\right)^3
\nonumber\\ \dsp
 \dsp
%
%
&&-~(0.0035\pm 0.0004)
-(0.010 \pm 0.001)
\api
\\ \dsp
 \dsp
&&-~(2.95\pm 0.17)\left(\api\right)^2
-(11.9 \pm 0.3 )\left(\api\right)^3
\Bigg\}\nonumber
\end{eqnarray}
and
\begin{eqnarray}
\EQN{num11}
\frac{\Gamma_{\rmfont{b}\ovl{\rmfont{b}}}}{\Gamma_{\rmfont{had}}}
& =&
\left(\frac{\Gamma_{\rmfont{b}\ovl{\rmfont{b}}}}%
{\Gamma_{\rmfont{had}}}\right)
\Bigg|_{{\scriptstyle m_{\rmfont{b}}=0\atop\scriptstyle\as=0}}
\Bigg\{
1-(0.0031\pm 0.0003)
-(0.005\pm 0.0005)\api
\nonumber\\
 \dsp
&&-~(2.3\pm 0.1)\left(\api\right)^2
-(6.9\pm 0.3)\left(\api\right)^3
\Bigg\}
{}\, .
\end{eqnarray}
{}From these formulae it is evident that the
coefficients of the $\as^2$ and the $\as^3$
terms are entirely different from the massless
non-singlet case as a consequence of the
bottom mass effects and virtual top loops
discussed in this work.
The deviation of this  from 1  can be
traced to two sources: the bottom mass term,
responsible for the term independent of $\alpha_s$,
and the singlet term mainly responsible
for the $\alpha_s^2$ (and $\alpha_s^3$)
contribution.

\sect{The Low-Energy Region \protect\\
Near the Bottom Threshold\label{low}}
The previous discussion has dealt mainly  with the
applications  of the theoretical results to the high
energy region. However, as indicated already  in
Section~4.4 the results are also
applicable  for energies relatively  close to the threshold
of heavy  quarks, if $m^2/s, m^4/s^2$ and $m^6/s^3$
terms are included.  This was demonstrated for the Born
and the $O(\alpha_s)$ formulae in Section~4.4, where
it was shown that these leading terms  provide an excellent
approximation even if the ratio  $4 m^2/s$
approaches  $0.8$.
With this justification a detailed analysis  of $R_{\rmfont{had}}$
can be performed for the region above
the charmonium resonances
and below the bottom threshold (excluding,
of course,  the narrow $\Upsilon$
resonances). Furthermore  the region above the
$\ovl{\rmfont{b}}\rmfont{b}$
resonances --- say, above 11.5  or 12 GeV --- can be treated in
the same approximation.

With increasing statistics  and precision   at LEP  the uncertainty
in $\alpha_s$  from the\break\hfill measurement of the hadronic decay rate of
 the $\stand{Z}$   can be reduced  to
$\pm 0.002$. It would be highly desirable  to test the evolution
of the strong coupling as predicted by the beta function
through a determination of $\alpha_s$  from essentially the same
observable ---    at lower energy, however. The region from several
GeV above charm threshold (corresponding to  the maximal energy of
BEPC around $5.0$ GeV) to just below the $B$ meson
threshold  at around  $10.5$ GeV (corresponding to the
`off  resonance' measurements  of CESR) seems particularly suited for
this  purpose.  As a  consequence of  the favourable error propagation,
the accuracy in the  measurement (compared to $91$ GeV)
may decrease by  a factor of about three or even  four
at 10 and 5.6 GeV
respectively, to achieve comparable precision
in $\Lambda_{\ovl{\rmfont{MS}}}$:
\[
\delta \alpha_s(s)
= \frac{\alpha_s^2(s)}{\alpha_s^2(M_{\stand{Z}}^2)}
\delta \alpha_s(M_{\stand{Z}}^2)~.
\]
Most of the results discussed above
for massless quarks are applicable also for the case under
consideration. However, two additional complications arise:
\\
\begin{itemize}
\item[i)] Charm quark effects cannot be ignored  completely
and  should be taken into consideration  through an expansion in the
ratio $m_{\rmfont{c}}^2/s$, employing the results of
Sections~\ref{repnsm2} and \ref{repnsm4}
for terms of order $m_{\rmfont{c}}^2/s$  and $m_{\rmfont{c}}^4/s^2$.
\item[ii)] Contributions involving virtual bottom quarks are present,
starting from order $\alpha_s^2$. Their contribution depends
in a nontrivial manner on $m_{\rmfont{b}}^2/s$.
In order $\alpha_s^2$  these
are discussed in Section~\ref{top} and are shown to be small.
Estimates for the corresponding contributions of order
$\alpha_s^3$   indicate that they  are under control and can be
safely neglected, provided that one works within the correctly
defined effective four-quark theory.
 \end{itemize}

The results presented below are formulated for
a theory with $n_{\stand{f}} =4$ effective flavours and with the
corresponding definitions of the coupling constant and  the
quark mass.  The relation to a formulation with
$n_f = 5$  appropriate for  the measurements  above the
$\rmfont{b}\bar{\rmfont{b}}$  threshold
was discussed in Section~\ref{decoupling}  and
will be given at the end
of the paper.

We shall now list the  independent contributions
and their relative importance.  Neglecting for the moment
the masses of the charmed quark and {\em a forteriori}
of the $\rmfont{u}$, $\rmfont{d}$ and
$\rmfont{s}$ quarks one predicts in order $\alpha_s^3$,
\begin{equation}
R_{\rmfont{NS}} = \sum_{f = \rmfont{u,d,s,c}}  3 Q_{\stand{f}}^2
\left[
1 +  \frac{\alpha_s}{\pi}
+ 1.5245
\left(\frac{\alpha_s}{\pi}\right)^2
-11.52033
\left(\frac{\alpha_s}{\pi}\right)^3
\right]
\label{R3ns}
\end{equation}
for the non-singlet contribution.
The second and the third order
coefficients are
evaluated with $n_{\stand{f}}=4$,  which means that the bottom
quark loops are absent. In order $\alpha_s^2$ the bottom quark
loops can be taken into consideration with their
full mass dependence given in Section~\ref{top}.
However, the leading term of order
$\alpha_s^2/m_{\rmfont{b}}^2$
provides
a fairly accurate description even up to  the very  threshold
$s = 4 m_{\rmfont{b}}^2$.
Hence one has to add a correction
\begin{equation}
\delta R_{m_{\rmfont{b}}} = \sum_{f = \rmfont{u,d,s,c}}
3 Q_{\stand{f}}^2
\left(\frac{\alpha_s}{\pi}\right)^2 \frac{s}{\bar m_{\rmfont{b}}^2}
\left[
\frac{44}{675}   +  \frac{2}{135} \ln\, \frac{\bar m_{\rmfont{b}}^2}{s}
\right] .
\label{Rmb}
\end{equation}
For the singlet term one obtains
\begin{equation}
R_{\rmfont{S}}=
-\left(\frac{\alpha_s}{\pi}\right)^3
\left(\sum_{\rmfont{u,d,s,c}} Q_{\stand{f}}\right)^2 \, 1.239 =
-0.55091
\left(\frac{\alpha_s}{\pi}\right)^3
{}.
\label{Rs}
\end{equation}
The bottom quark is absent in this sum.
In view of the smallness of the
$\alpha_s^2 s/m^2_{\rmfont{b}}$ correction
(even close to the $\rmfont{b}\bar{\rmfont{b}} $ threshold!)
all other terms of $ {\cal O}(\alpha_s^3)$  from virtual
$\rmfont{b}$ quarks are also neglected.
This can be justified  with the results of Ref.~\cite{timo2}
where $s/m^2$ terms  are evaluated.
In the same spirit
it is legitimate to use  the scale invariant value of
the $\rmfont{b}$  quark mass $\bar m_{\rmfont{b}}
= \bar m_{\rmfont{b}}(\bar m^2_{\rmfont{b}})$ as defined in
the five-quark theory.

In contrast to the bottom mass
the effects of the charmed  mass can be incorporated
through an expansion in $m_{\rmfont{c}}^2/s$.
Quadratic mass corrections are included
up to order $\alpha_s^3$, quartic
mass terms  up to order $\alpha_s^2$.
Since $m_{\rmfont{c}}^2/s$ is in itself a small expansion parameter,
the order $\alpha_s^2 m_{\rmfont{c}}^4/s^2$ terms should be sufficient
for the present purpose.

The charmed mass corrections are thefore given by
\begin{eqnarray}
\delta R_{m_{\rmfont{c}}} &=&
3\,Q_{\rmfont{c}}^2 \,12\,
\frac{m_{\rmfont{c}}^2}{s}~\frac{\alpha_s}{\pi}
\left[
1
+
9.097
{}~\frac{\alpha_s}{\pi}
+
53.453
\left(\frac{\alpha_s}{\pi}\right)^2
\right]\nonumber\\
&&-~3 \sum_{f=\rmfont{u,d,s,c}}Q_{\stand{f}}^2~\frac{m_{\rmfont{c}}^2}{s}
 \left(\frac{\alpha_s}{\pi}\right)^3
   6.476
\nonumber\\
\displaystyle
&&+
{}~3\,Q_{\rmfont{c}}^2~\frac{m_{\rmfont{c}}^4}{s^2}
\left\{
-6
-22~\frac{\alpha_s}{\pi}
+
\left[
141.329 - \frac{25}{6}\ln\,\left(\frac{m_{\rmfont{c}}^2}{s}\right)
\right]
\left(\frac{\alpha_s}{\pi}\right)^2
\right\}
\nonumber\\
\displaystyle
&&+~3 \sum_{f=\rmfont{u,d,s,c}} Q_{\stand{f}}^2
{}~\frac{m_{\rmfont{c}}^4}{s^2}
\left(\frac{\alpha_s}{\pi}\right)^2
\left[
-0.4749
- \ln\, \left(\frac{m_{\rmfont{c}}^2}{s}\right)
\right]
\nonumber\\
\displaystyle
&&-~3\,Q_{\rmfont{c}}^2~\frac{m_{\rmfont{c}}^6}{s^3}
\left\{
8
+\frac{16}{27}
{}~\frac{\alpha_s}{\pi}
\left[
6\ln\,\left(\frac{m_{\rmfont{c}}^2}{s}\right) + 155
\right]
\right\}\, ,
\end{eqnarray}
\label{Rmc}
where $n_{\stand{f}}=4$ has been adopted everywhere.  Note that
terms of order  $\alpha_s^3 m^2_{\rmfont{c}}/s$ are more important than
those of  order  $\alpha_s^2 m^4_{\rmfont{c}}/s^2$
in the whole energy region under consideration.
For completeness,
  $m_{\rmfont{c}}^6/s^3$ and   $\alpha_s m_{\rmfont{c}}^6/s^3$
terms are also listed, which, however,  are insignificant and
will be ignored in the numerical analysis.

The charm quark mass is to be taken
as $m_{\rmfont{c}} = \bar m_{\rmfont{c}}^{(4)}(s)$
and is to be evaluated in the four-flavour theory
via the standard RG equation with the initial
value $\bar m_{\rmfont{c}}(\bar m_{\rmfont{c}}) = 1.12 \ \rmfont{GeV}$,
corresponding to a pole mass of $1.6$  GeV
in the case of  $\alpha_s^{(5)}(M_{\stand{Z}})= 0.120$
(see Section~\ref{CandB}).
A  similar line of reasoning  can be
pursued for   bottom mass terms in the
region several GeV above the $B$ meson
threshold.  The formula given below is
expected to provide a reliable answer
for $\sqrt{s}$ around  15 GeV and
perhaps even down to 13 GeV:
\newpage
\begin{eqnarray}
\delta R_{m} &=&
3~\left(Q_{\rmfont{c}}^2~\frac{m_{\rmfont{c}}^2}{s}
+ Q_{\rmfont{b}}^2~\frac{m_{\rmfont{b}}^2}{s}\right)
{}~12~\frac{\alpha_s^{(5)}}{\pi}
\left[
1
+
8.736~\frac{\alpha_s^{(5)}}{\pi}
+
45.657
\left(\frac{\alpha_s^{(5)}}{\pi}\right)^2
\right]
\nonumber\\
&&-~3\sum_{f=\rmfont{u,d,s,c,b}}Q_{\stand{f}}^2
\left(\frac{m_{\rmfont{c}}^2}{s} +\frac{m_{\rmfont{b}}^2}{s}\right)
 \left(\frac{\alpha_s^{(5)}}{\pi}\right)^3
6.126
\nonumber\\
&&+~3\,Q_{\rmfont{c}}^2~\frac{m_{\rmfont{c}}^4}{s^2}
\left\{
-6
-22~\frac{\alpha_s^{(5)}}{\pi}
+
\left[
139.489 - \frac{23}{6} \ln\,\left(\frac{m_{\rmfont{c}}^2}{s}\right)
+ 12~\frac{m_{\rmfont{b}}^2}{m_{\rmfont{c}}^2}
\right]
\left(\frac{\alpha_s^{(5)}}{\pi}\right)^2
\right\}
\nonumber\\
&&+ ~3\,Q_{\rmfont{b}}^2~\frac{m_{\rmfont{b}}^4}{s^2}
\left\{
-6
-22~\frac{\alpha_s^{(5)}}{\pi}
+
\left[
139.489 - \frac{23}{6} \ln\,\left(\frac{m_{\rmfont{b}}^2}{s}\right)
+ 12~\frac{m_{\rmfont{c}}^2}{m_{\rmfont{b}}^2}
\right]
\left(\frac{\alpha_s^{(5)}}{\pi}\right)^2
\right\}
\nonumber\\
&&+~3\sum_{f=\rmfont{u,d,s,c,b}} Q_{\stand{f}}^2~
\frac{m_{\rmfont{c}}^4}{s^2}
\left(\frac{\alpha_s^{(5)}}{\pi}\right)^2
\left[
-0.4749
- \ln\, \left(\frac{m_{\rmfont{c}}^2}{s}\right)
\right]
\nonumber\\
&&+~3\sum_{f=\rmfont{u,d,s,c,b}} Q_{\stand{f}}^2~
\frac{m_{\rmfont{b}}^4}{s^2}
\left(\frac{\alpha_s^{(5)}}{\pi}\right)^2
\left[
-0.4749
- \ln\, \left(\frac{m_{\rmfont{b}}^2}{s}\right)
\right]
\nonumber\\
&&-~3\,Q_{\rmfont{b}}^2~ \frac{m_{\rmfont{b}}^6}{s^3}
\left\{
8
+\frac{16}{27}
{}~\frac{\alpha_s^{(5)}}{\pi}
\left[
6\ln\,\left(\frac{m_{\rmfont{b}}^2}{s}\right) + 155
\right]
\right\}\,.
\end{eqnarray}
\label{Rm}

Above the $B$ meson threshold it is more convenient to express all
quantities for  $n_{\stand{f}} = 5 $ theory and thus in  (\ref{Rm})
all the coupling constant and quark masses  are evaluated
in the five-flavour theory at the scale  $\mu = \sqrt{s}$.

The transition from four- to
five-flavour theory is performed as follows:
The  charm mass is naturally defined
in the  $n_{\stand{f}} =4 $ theory. In order to obtain  the value of
$m_{\rmfont{c}} =\bar m_{\rmfont{c}}^{(5)}(s)$  the initial value
$\bar m_{\rmfont{c}}^{(4)}(1~\rmfont{GeV})$ is evolved   via
the $n_{\stand{f}}=4$ RG equation
to the point $\mu^2 = M_{\rmfont{b}}^2$
and from there up to $\mu^2 = s$, now,  however,
 with the $n_{\stand{f}}=5$ RG equation.
The bottom mass, on the other
hand,  is naturally  defined in the $n_{\stand{f}} = 5$
theory irrespective of the characteristic momentum scale of
the problem under consideration.  Hence  we take
 $\ovl{m}_{\rmfont{b}}(s)$ obtained from the scale invariant mass
$\ovl{m}_{\rmfont{b}}(\ovl{m}_{\rmfont{b}})$ after running the latter
with the help of the $n_{\stand{f}}=5$ RG equation.
Finally,
$\alpha_s^{(4)}$ and $\alpha_s^{(5)}$
are related through the matching Eq.~\re{match2aPole}.


In Tables~4--7  the predictions for $R$
at $10.5$  and $13$ GeV
are listed  for different
values  of $\alpha_s^{(5)}(M_{\stand{Z}})$ together with the values of
$\alpha_s(s)$ and the running masses\footnote{
The results for   the  5 GeV region   and for  a larger
variety  of values for $\alpha_s$ are given in
Ref.~\cite{TTP94-12}.
Some
slight differences between
the numbers in Tables~4--7
and those in  Ref.~\cite{TTP94-12}  stem from
different
input values for $M_{\rmfont{c}}$.}. Note
that our predictions are presented without QED corrections from the running of
$\alpha$ and from initial state radiation. Figure~\ref{last}  shows the
behaviour of the  ratio $R(s)$
as a function of energy
below and above the bottom threshold, for
$\alpha_s(M_{\stand{Z}}) = 0.120, 0.125$ and $0.130$.
The light quark $\rmfont{u,d,s,c}$)  contribution is  also
displayed separately above $10.5$ GeV.
It is evident that the predictions from the four- and five-flavour
theories join smoothly.  The additional contribution from the
$\rmfont{b} \ovl{\rmfont{b}}$
channel  is presented  down to $11.5$  GeV,
where resonances start to contribute and the perturbative
treatment necessarily ceases to apply.
\ice{
 Evidently the
 $b \ovl{b}$
channel is present with full strength  down to the resonance
region --- an important consequence of QCD corrections.
{}From this  discussion it should be  evident  that the theoretical
prediction is well under control. Mass effects are small below
the $\rmfont{b} \ovl{\rmfont{b}}$
threshold as well as a  few GeV above.
An experimental test is of prime
importance.
}
\newpage

\begin{table}[H]
\begin{center}
{\bf Table 4}
\end{center}
\vskip0.2cm
{\small{Values of
$  \protect \Lambda^{(5)}_{\overline{{\scriptstyle MS}}}, \
\Lambda^{(4)}_{\overline{{\scriptstyle MS}}}, \
\alpha_s^{(4)}(s),\     m_{\rmfont{c}}^{(4)}(s)  \
\rmfont{and} \  \ovl{m}_{\rmfont{b}}(\ovl{m}_{\rmfont{b}})
$
at $\protect\sqrt{s}= 10.5$ GeV
for different values of
$\alpha_s^{(5)}(M_{\stand{Z}}^2)$.}}

\begin{center}
\begin{tabular}{|r r r r r r|}
\hline
$\alpha_s^{(5)}(M_{\stand{Z}}^2)$              &
$\Lambda^{(5)}_{\overline{{\rm MS}}}$       &
$\Lambda^{(4)}_{\overline{{\rm MS}}}$      &
$\alpha_s^{(4)}(s)$                  &
$m_{\rmfont{c}}^{(4)}(s)$                  &
$\ovl{m}_{\rmfont{b}}(\ovl{m}_{\rmfont{b}})$          \\
0.1200   &
233 MeV &
320 MeV  &
0.177        &
0.751 GeV    &
4.09 GeV  \\
0.1250   &
302 MeV &
403 MeV  &
0.188        &
0.637 GeV    &
4.03 GeV  \\
0.1300   &
383 MeV &
498 MeV  &
0.199        &
0.500 GeV    &
3.96 GeV  \\
\hline
\end{tabular}
\end{center}
\end{table}
\begin{table}
\begin{center}
{\bf Table 5}
\end{center}
{\small{ Predictions for $R(s)$
at $ \protect \sqrt{s}= 10.5$ GeV;
the contributions to $\delta R_{m_{\rmfont{c}}}  $ are shown
separately for every  power of the quark mass.}}
\vspace{6mm}
\begin{center}
\begin{tabular}{|r|r|r|r|r|r|r|}
\hline
$\alpha_s^{(5)}(M_{\stand{Z}}^2)$              &
\multicolumn{1}{|c|}{$ R_{\rmfont{NS}}$}                       &
\multicolumn{1}{|c|}{$R_{\rmfont{S}}$}                           &
\multicolumn{1}{|c|}{$\delta R_{m^2_{\rmfont{c}}}$}              &
\multicolumn{1}{|c|}{$\delta R_{m^4_{\rmfont{c}}}$}             &
\multicolumn{1}{|c|}{$\delta R_{m_{\rmfont{b}}} $}               &
\multicolumn{1}{|c|}{$       R       $ }              \\
\hline
\hline
0.1200   &
3.530   &
--0.000098   &
0.0077   &
--0.00023   &
0.0026   &
3.540   \\
0.1250   &
3.543   &
--0.00012   &
0.0061   &
--0.000121   &
0.0030   &
3.551   \\
0.1300   &
3.556   &
--0.00014   &
0.0041   &
--0.000046   &
0.0034   &
3.563   \\
\hline
\end{tabular}
\end{center}
\end{table}

\begin{table}
\begin{center}
{\bf Table 6}
\end{center}
{\small{Values of
$  \protect \Lambda^{(5)}_{\overline{{\rm MS}}}, \
\alpha_s^{(5)}(s),
\ m_{\rmfont{c}}^{(5)}(s) \  \rmfont{and} \ m_{\rmfont{b}}^{(5)}(s) $
at $\protect\sqrt{s}= 13$ GeV
for different values of
$\alpha_s^{(5)}(M_{\stand{Z}}^2)$\,.}}
\vspace{6mm}
\begin{center}
\begin{tabular}{|r|r|r|r|r|}
\hline
$\alpha_s^{(5)}(M_{\stand{Z}})$                &
\multicolumn{1}{|c|}{$\protect \Lambda^{(5)}_{\overline{{\rm MS}}}$} &
\multicolumn{1}{|c|}{$\alpha_s^{(5)}(s)$}                  &
\multicolumn{1}{|c|}{$m_{\rmfont{c}}^{(5)}(s)$}                  &
\multicolumn{1}{|c|}{$m_{\rmfont{b}}^{(5)}(s)$}                  \\
\hline
\hline
0.1200   &
233 MeV &
0.172    &
0.729 GeV       &
3.41 GeV  \\
0.1250   &
302 MeV &
0.183    &
0.617 GeV       &
3.29 GeV  \\
0.1300   &
383 MeV &
0.194    &
0.483 GeV       &
3.17 GeV  \\
\hline
\end{tabular}
\end{center}
\end{table}
\begin{table}
\begin{center}
{\bf Table 7}
\end{center}
{\small{ Predictions for $R(s)$
at $ \protect \sqrt{s}= 13$ GeV;
the contributions to $\delta R_{m}  $ are shown
separately for every  power of the quark masses.}}
\vspace{6mm}
 \begin{center}
\begin{tabular}{|r|r|r|r|r|r|r|}
\hline
\multicolumn{1}{|c|}{$\alpha_s^{(5)}(M_{\stand{Z}}^2)$}              &
\multicolumn{1}{|c|}{$ R_{\rmfont{NS}}$}                       &
\multicolumn{1}{|c|}{$R_{\rmfont{S}}$}                           &
\multicolumn{1}{|c|}{$\delta R_{m^2}$}                &
\multicolumn{1}{|c|}{$\delta R_{m^4}$}                &
\multicolumn{1}{|c|}{$\delta R_{m^6}$}                &
\multicolumn{1}{|c|}{$R            $}                 \\
\hline\hline
0.1200   &
3.875   &
--0.000023   &
0.023   &
--0.011   &
--0.0014   &
3.887   \\
0.1250   &
3.888   &
--0.000027   &
0.023   &
--0.0092   &
--0.0011   &
3.901   \\
0.1300   &
3.901   &
--0.000032   &
0.022   &
--0.0079   &
--0.00091   &
3.914   \\
\hline
\end{tabular}
\end{center}
\end{table}


\begin{figure}[H]
\begin{center}
\vspace*{-3cm}
\mbox{\epsfig{file=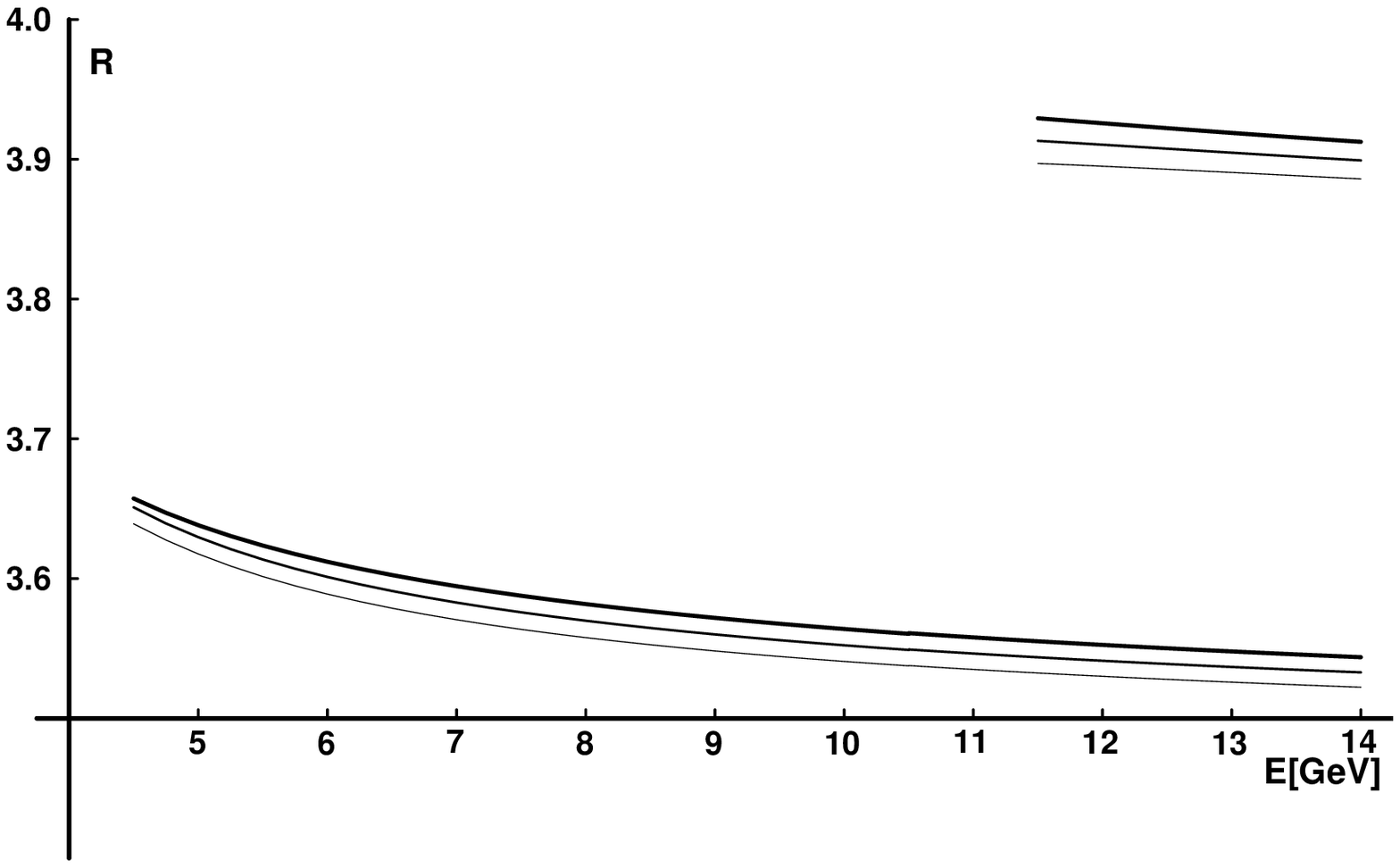,width=10.5cm}}
\end{center}
\vspace*{-5cm}
\caption{\label{last}
The ratio $R(s)$  below and above
the $\rmfont{b}$ quark production threshold at $10.5$ GeV
for  $\alpha_s(M_{\stand{Z}}) = 0.120, \, 0.125$ and $0.130$.
The contributions from  light quarks are displayed separately.}
\end{figure}
%
\noindent
Evidently the
 $b \ovl{b}$
channel is present with full strength  down to the resonance
region --- an important consequence of QCD corrections.
{}From this  discussion it should be  evident  that the theoretical
prediction is well under control. Mass effects are small below
the $\rmfont{b} \ovl{\rmfont{b}}$
threshold as well as a  few GeV above.
An experimental test is of prime
importance.

\sect{Conclusions}
In this report we have tried to present, in a comprehensive form,
the theoretical framework  and all formulae presently available
that are required to predict the QCD corrected  total
cross-section of  ${\rm e}^+  {\rm e}^-$ annihilation and
the $\stand{Z}$ decay rate
into hadrons, with optimal accuracy. The presentation is supposed to
be  self-contained --- and hopefully self-consistent ---
such that all formulae relevant for the prediction of  experimental
quantities can be  deduced  from this work without the need
to combine results from different publications. Particular emphasis
has been put on the influence of the non-vanishing bottom quark
mass and on contributions from virtual top quarks, which are
of particular importance for the so-called singlet contributions.
Much of the discussion has been tailored for the\break\hfill 90 GeV region,
wher
experiments at LEP  provide highly accurate data, but,
applications to `low energies',  around 10 GeV or even lower,
have been  mentioned whenever appropriate.

The topic is approached from three different viewpoints:
from the purely theoretical, laying the ground for the
discussion,  from the calculational, providing  the formulae,
and from the practical viewpoint, discussing
the relative importance of the various
contributions and  associated
uncertainties. In the first two parts the  basis of the
subsequent calculations is presented. They contain,  essentially,
a brief review of the field theoretical ingredients:
$\beta$ function and anomalous dimensions, the relations
between various definitions of the mass,  the decoupling of heavy
quarks, and the corresponding transitions between different
`effective theories'. In many circumstances quarks are either
light ($m^2 \ll s$) or heavy ($m^2 \gg s$). The corresponding
expansions provide powerful tools.
They are described in Part~2.

The results of higher-order calculations are scattered in
numerous original publications, often with conflicting
conventions and notations.
In Parts~\ref{calc-tech},  \ref{nonsinglet}  and \ref{singlet}   the
formulae are collected and presented in a uniform way. The
brief presentation of exact results of $O(\alpha_s)$ (i.e.
with arbitrary $m^2/s$)  in Part~\ref{exact} leads quickly to
Parts~\ref{nonsinglet}  and \ref{singlet}
where  $m^2/s$ or $s/m^2$ are treated as
expansion parameters, and results of up to order $\alpha_s^3$
are collected.  An important classification of amplitudes
originates from the distinction between singlet and
non-singlet diagrams, with markedly different behaviour in
the limit $m^2/s \gg 1$, and  our presentation follows this
classification. Whilst most of the discussion is concerned
with predictions for the total cross-section or the decay
rate, occasionally also results for partial rates, for example into
four fermion states or for the inclusive rate into
$\rmfont{b}\bar{\rmfont{b}}$  quarks, are presented.
The formulae
are displayed in two different forms:  first in analytical form
with the relevant coefficients  given by
fractions and Riemann's Zeta function, as functions of
$n_f$, and then entirely numerically (Part~\ref{numerical}
and the  Appendices) in terms of decimal
fractions.

The final Part is most relevant to practical
applications. The numerical relevance (or irrelevance) of
the various contributions is clarified. Their stability
with respect to variations of the renormalization scale
$\mu$ is studied. Estimates are presented for the errors
from the truncation of the perturbation series, the
`theoretical error', and for the error induced by  the
uncertainty in the input parameters $M_{\rmfont{b}}$
and $M_{\rmfont{t}}$. A safe
upper limit on the combined uncertainty of
$\Gamma(Z \to   \rmfont{ hadrons})$
from quark mass ($M_{\rmfont{b}}$
and $M_{\rmfont{t}}$) dependent
corrections amounts  to about $0.4$ MeV. The uncertainty
from the  truncation of the perturbative expansion in the
massless limit in $O(\alpha_s^3)$  is highly  subjective,
with estimates ranging from $0.6$ MeV up to  $1.2$ MeV.
Fortunately enough,  in the foreseeable
future, both sources of errors will not effect
the precision of  $\alpha_s$ as determined  at LEP through
total cross-section measurements.

For the convenience  of the reader, simple numerical
formulae are presented which allow a quick estimate
of $\Gamma_{\rmfont{had}}$ ,
$\Gamma_{\rmfont{b}\bar{\rmfont{b}}}$ and their
ratio.  A similar discussion for the energy region
around the bottom threshold concludes this Part.
For easy access the formulas
most frequently used in practical applications
 are collected  and rewritten in the Appendices.
\vskip0.5cm
 \noindent
{\Large{\bf Acknowledgements}}
\vskip0.3cm

We would like to thank our collaborators
A.~Hoang,
M.~Je\.{z}abek,
B.~Kniehl,
D.~Pirjol,
K.~Schilcher,
O.~Tarasov,
T.~Teubner
and
P.~Zerwas for
providing important input and
essential ingredients to this report.
We have benefitted from  helpful and
clarifying discussions with
G.~Passarino.  We are grateful to
M.~Steinhauser  for  reading  the
manuscript and  help with drawing some
figures.  The review would not have
been completed without the
encouragement and initiative provided
by the LEP study group, in particular
by Dima Bardin.

We are  indebted to W.~Bernreuther, T.~Hebbeker,
W.~Hollik,
A.~Kataev,
B.~Kniehl and  P.~Zerwas for
reading  the manuscript and
providing us with their comments and
corrections.

Over the past years this work has been supported
by
the  Deutsche   Forschungsgemeinschaft
(grants   Ku 502/3-1, Ku 502/6-1),
by
the
Bundesministerium f\"ur Forschung und Technologie
(contracts 005KA94P1, 056KA93P),
and by  the  HERAEUS-Stiftung.
K.~Chetyrkin
thanks  the Universit\"at Karlsruhe for a  guest
professorship and the   Institute of Theoretical Particle
Physics for the warm hospitality extended
to him during recent years.  His work
was partially  supported by the Russian
Fund of  the Fundamental Research
(grant 93-02-14428).
A. Kwiatkowski  thanks
the  Deutsche   Forschungsgemeinschaft
for financial
support (grant no. Kw 8/1-1). Partial
support by US DoE under Contract
DE-AC03-76SF00098 is gratefully
acknowledged.
\vskip0.8cm

\noindent
{\Large{\bf Appendixes}}

\vspace{.1cm}

\setcounter{section}{0}
\setcounter{equation}{0}
\chap{Some Useful Formulae
\label{useful}}



\vspace*{2ex}
{\bf \large Zeta function}

\vspace*{1ex}
The Riemann Zeta function is defined by
\beq\EQN{apxa1}
\zeta(s) = \sum_{k=1}^{\infty}k^{-s}
{}\, .
\eeq
Some particular values are:
\beq\EQN{apxa2}
\ba{ll}\dsp
\zeta(0)=-~\frac{1}{2}~,
& \dsp
\quad\zeta(2)=\frac{\pi^2}{6}
        = 1.6449341~,
\\ \dsp
\zeta(3)=1.2020569~,
&\quad
\zeta(4)=\frac{\pi^4}{90} = 1.0823232~,
\\
\zeta(5)=1.0369278~.

&
\ea
.\eeq


{}
\vspace*{2ex}
\noindent
{\bf \large Dilogarithm}

\vspace*{1ex}
\beq\EQN{dilog1}
\mbox{Li}_2(x)= \sum_{n=1}^{\infty}
\frac{x^2}{n^2}~, \ \ \ \ \ \ (|x| < 1 )
{}
\eeq
or
\beq\EQN{dilog2}
\mbox{Li}_2(x)= -\int_{0}^{x}
\frac{\ln\,(1-t)}{t}~{\rm d t}
{}\, .
\eeq
\newpage

\chap{{ Renormalization Group Functions}}

\setcounter{equation}{0}
The RG equation for the quark mass reads:
\beq    \EQN{a1}
\dmu \bar{m}(\mu) =
 \bar{m}(\mu )\gm \equiv
-\bar{m}(\mu )\sum_{i\geq 0}
\g_m^i\left(\api\right)^{i+1}.
\eeq
It is solved by
\renewcommand{\arraystretch}{3}
\beq
\EQN{a2:app}
\ba{rl} \dsp
\bar{m}(\mu ) =
&
\dsp
\bar{m}(\mu_0 ) \exp \,\left\{
\frac{1}{\pi}
\int^{\as(\mu)}_
{\as(\mu_0)}
     \rmfont{d} x \frac{\g_m(x)}{\beta(x)}\right\}
\\
 =
& \dsp
\bar{m}(\mu_0 )\left[
\frac{\as(\mu )}{\as(\mu_0 )}\right]
^{{\g_m^0}/{\beta_0}}
\left\{ 1+  \left(\frac{\g_m^1}{\beta_0}
-\frac{\beta_1\g_m^0}{\beta_0^2}\right)
\left[\frac{\as(\mu )}{\pi}
-\frac{\as(\mu_0 )}{\pi}\right]
\right.
\\ & \dsp
  +~\frac{1}{2}
  \left(\frac{\g_m^1}{\beta_0}
-\frac{\beta_1\g_m^0}{\beta_0^2}\right)^2
\left[\frac{\as(\mu )}{\pi}
-\frac{\as(\mu_0 )}{\pi}\right]^2
\\ & \dsp
 \left.
  +~\frac{1}{2}
\left( \frac{\g_m^2}{\beta_0}
-\frac{\beta_1\g_m^1}{\beta_0^2}
   -\frac{\beta_2\g_m^0}{\beta_0^2}
+\frac{\beta_1^2\g_m^0}{\beta_0^3}\right)
\left[\left(\frac{\as(\mu )}{\pi}\right)^2
-\left(\frac{\as(\mu_0 )}{\pi}\right)^2\right]
\right\}.
\ea\eeq
\renewcommand{\arraystretch}{2}
with
\beq \EQN{a3}
\ba{ll}\dsp
\g_m^0 &=\dsp 1, \  \ \ \g_m^1=
\left(\frac{202}{3}-\frac{20}{9}~n_f\right)/16,
\\[3ex] \dsp \g_m^2
&=\dsp \left\{1249 -
\left[\frac{2216}{27}
+\frac{160}{3}\zeta(3)\right]
n_f-\frac{140}{81}~n_f^2\right\}/64
{}\, .
\ea
\eeq
Similarly,  one has for the strong coupling
constant
($L\equiv \ln\, \mu^2/\Lambda^2_{\ovl{\rmfont{MS}}}$):
\beq  \EQN{a4}
\dmu \left[ \frac{\as(\mu )}{\pi} \right] =
 \beta \equiv
-\sum_{i\geq0}\beta_i\left(\api\right)^{i+2},
 \eeq
Integration gives
\beq \EQN{a5}
\frac{\as(\mu )}{\pi} = \frac{1}{\beta_0 L}
\left\{
1 - \frac{1}{\beta_0L}\frac{\beta_1\ln\, L}
{\beta_0}
 + \frac{1}{\beta_0^2L^2}
\left[\frac{\beta_1^2}{\beta_0^2}
(\ln^2L -
\ln\, L - 1) + \frac{\beta_2}{\beta_0}
 \right] \right\},
\eeq
with
\beq\EQN{a6}
\ba{ll}\dsp
\beta_0&=\dsp\left(11-\frac{2}{3}~n_f\right)/4,
  \;\;\;\;
\beta_1=\left(102-\frac{38}{3}~n_f\right)/16~,
\\[3ex] \dsp
\beta_2&=\dsp\left(\frac{2857}{2}
-\frac{5033}{18}~n_f+
\frac{325}{54}~n_f^2\right)/64
{}\,~.
\ea
\eeq

\chap{{ List of Radiative Corrections}}

\setcounter{equation}{0}
In this section all contributions
to the total hadronic $\stand{Z}$
decay rate are collected.
  Complete order $\as$ prediction with
full mass dependence
($v^2=1-4m^2/s$):
\begin{eqnarray}
\Gamma_{\rmfont{had}}
& =&
\Gamma_0
3\Bigg\{
\sum_fv_f^2
  v\frac{3-v^2}{2}\left[1+\frac{4}{3}\apis K_{\rmfont{V}}\right]
\nonumber\\ & &
+\sum_fa_f^2
   v^3\left[1+\frac{4}{3}\apis K_{\rmfont{A}}\right]
\Bigg\}
{}\, ,
\end{eqnarray}
with

\beq
 \ba{ll}\EQN{apxc01}
K_{\rmfont{V}} = & \dsp
\frac{1}{v}\left[ A(v) +
 \frac{P_{\rmfont{V}}(v)}{(1-v^2/3)}\ln\,\frac{1+v}{1-v}
     + \frac{Q_{\rmfont{V}}(v)}{(1-v^2/3)}\right],
\\ \dsp
K_{\rmfont{A}} = & \dsp
\frac{1}{v}\left[ A(v)
+ \frac{P_{\rmfont{A}}(v)}{v^2}\ln\,\frac{1+v}{1-v}
                       + \frac{Q_{\rmfont{A}}(v)}{v^2}\right]
{}\, ,
\ea
\eeq
and
\beq
\ba{ll}\EQN{apxc02}   \dsp
A(v) =& \dsp (1+v^2)
\left\{
{\rm Li}_2\left[
      \left(\frac{1-v}{1+v}\right)^2
          \right]
+2{\rm Li}_2\left(\frac{1-v}{1+v}\right)
+\ln\,\frac{1+v}{1-v}\ln\,\frac{(1+v)^3}{8v^2}
 \right\}
\\ & \dsp
          +~3v\ln\,\frac{1-v^2}{4v}-v\ln\, v
{}\, ,
\ea
\eeq
\beq
\ba{ll}\EQN{apxc03}   \dsp
P_{\rmfont{V}}(v) = \frac{33}{24}+\frac{22}{24}v^2
-\frac{7}{24}v^4
{}\, ,
& \dsp
\quad Q_{\rmfont{V}}(v) = \frac{5}{4}v-\frac{3}{4}v^3
{}\, ,
\\ \dsp
P_{\rmfont{A}}(v) = \frac{21}{32}+\frac{59}{32}v^2
-\frac{19}{32}v^4-\frac{3}{32}v^6~,
&\;\;\;\;
\dsp
Q_{\rmfont{A}}(v) = -~\frac{21}{16}v+\frac{30}{16}v^3
+\frac{3}{16}v^5
{}\, ,
\ea
\eeq

\noindent
where
$\Gamma_0=G_F M_Z^3/24 \sqrt{2} \pi=82.94$ MeV.
\vspace{4mm}

Including higher-order corrections the decay
rate can be written in the following form:

\beq\EQN{apxc1}
\Gamma_{\rmfont{had}} = \sum_{i=1}^{12} \Gamma_i
= \sum_{i=1}^{12} \Gamma_0 R_i
{}\,~.
\eeq
The separate contributions are given by the
following expressions. [Below in Eqs. (6--15)
$\alpha_s$,  $\ovl{m}_f$ and $\ovl{m}_b$ stand for
$\alpha^{(5)}_s(\mu)$,
$\ovl{m}_f^{(5)}(\mu)$
and
$\ovl{m}_b^{(5)}(\mu)$
respectively.]
\vskip0.2cm
\noindent
{\em
Massless non-singlet corrections:
}
\begin{eqnarray}
R_1& =
&
 3\sum_f (v_f^2+a_f^2)
\Bigg\{
1+\api
\nonumber\\
& &+\left(\api\right)^2
\left[
\frac{365}{24}-11\zeta(3)
+~n_f\left(
          -~\frac{11}{12}+\frac{2}{3}\zeta(3)
    \right)
+\left(
-~\frac{11}{4}+\frac{1}{6}n_f
\right)\ln\,\frac{s}{\mu^2}
\right]
\nonumber\\
&&+\left(\api\right)^3
\Bigg[
\frac{87029}{288}-\frac{121}{48}\pi^2
   -\frac{1103}{4}\zeta(3) +\frac{275}{6}\zeta(5)
\nonumber\\
&&+~n_f\left(
  -~\frac{7847}{216}+\frac{11}{36}\pi^2
  +\frac{262}{9}\zeta(3)-\frac{25}{9}\zeta(5)
    \right)
+n_f^2\left(
  \frac{151}{162}-\frac{1}{108}\pi^2
  -\frac{19}{27}\zeta(3)
     \right)
\nonumber\\
&&+\left(
  -~\frac{4321}{48}+\frac{121}{2}\zeta(3)
  +~n_f\left[
       \frac{785}{72}-\frac{22}{3}\zeta(3)
      \right]
  +~n_f^2\left[
       -\frac{11}{36}+\frac{22}{3}\zeta(3)
        \right]
\right)\ln\,\frac{s}{\mu^2}
\nonumber\\
&&+\left(
  \frac{121}{16}-\frac{11}{12}n_f
   +\frac{1}{36}n_f^2
\right)\ln^2\frac{s}{\mu^2}
\Bigg]
\Bigg\}\nonumber\\
&=&
 3\sum_f (v_f^2+a_f^2)
\Bigg\{
1+\api
\nonumber\\
&&+\left(\api\right)^2
\left[
1.9857-0.1152~n_f
+\left(
-2.75+0.167~n_f
\right)\ln\,\frac{s}{\mu^2}
\right]
\nonumber\\
&&+\left(\api\right)^3
\Bigg[
-6.6369-1.2001~n_f-0.0052~n_f^2
\nonumber\\
&&
+\left(
-17.2964+2.0877~n_f-0.0384~n_f^2
\right)\ln\,\frac{s}{\mu^2}
\nonumber\\
&&+\left(
7.5625-0.9167~n_f+0.0278~n_f^2
\right)\ln^2\frac{s}{\mu^2}
\Bigg]
\Bigg\}
{}\, ;
\end{eqnarray}

\noindent
{\em
Massive universal corrections
 (`double bubble'):
}
\begin{eqnarray}
R_2
& =& 3\sum_f (v_f^2+a_f^2)
\left(\api\right)^3
\sum_{f'=b}\frac{\ovl{m}_{f'}^2}{s}
\Bigg\{
-80+60\zeta(3)
+n_f\left[
  \frac{32}{9}-\frac{8}{3}\zeta(3)
    \right]
\Bigg\}
\nonumber\\
 & =& 3\sum_f (v_f^2+a_f^2)
\left(\api\right)^3
\sum_{f'=b}\frac{\ovl{m}_{f'}^2}{s}
\Bigg\{
   -7.8766+0.3501~n_f
\Bigg\}
\nonumber\\
R_3
& =& 3\sum_f (v_f^2+a_f^2)
\left(\api\right)^2
\sum_{f'=b}\frac{\ovl{m}_{f'}^4}{s^2}
\left\{
\frac{13}{3}-4\zeta(3)-\ln\,\frac{\ovl{m}_{f'}^2}{s}
\right\}
\nonumber\\
 & =& 3\sum_f (v_f^2+a_f^2)
\left(\api\right)^2
\sum_{f'=b}\frac{\ovl{m}_{f'}^4}{s^2}
\left\{
-0.4749 -\ln\,\frac{\ovl{m}_{f'}^2}{s}
\right\}
\nonumber\\
R_4&=& 3\sum_f (v_f^2+a_f^2)
\left(\api\right)^2
\frac{s}{M_{\rmfont{t}}^2}
\left\{
\frac{44}{675} +\frac{2}{135}\ln\,\frac{M_{\rmfont{t}}^2}{s}
\right\}
\nonumber\\
& =& 3\sum_f (v_f^2+a_f^2)
\left(\api\right)^2
\frac{s}{M_{\rmfont{t}}^2}
\left\{
0.0652 +0.0148\ln\,\frac{M_{\rmfont{t}}^2}{s}
\right\}
{}\, ;
\end{eqnarray}

\noindent
{\em
Massive non-singlet corrections (vector):
}
\begin{eqnarray}
R_5
&=&3 v_{\rmfont{b}}^2~\frac{\ovl{m}_{\rmfont{b}}^2}{s}
\Bigg\{
12\api
+\left(\api\right)^2
\left[\frac{253}{2} - \frac{13}{3}~n_f
+(-57+2~n_f)\ln\,\frac{s}{\mu^2}
\right]
\nonumber\\
&&+\left(\api\right)^3
\Bigg[
2522-\frac{285}{4}~\pi^2+\frac{310}{3}~\zeta(3)
-\frac{5225}{6}~\zeta(5)
\nonumber\\
&&+~n_f\left(-~\frac{4942}{27}+\frac{17}{3}~\pi^2
   -\frac{394}{27}~\zeta(3)+\frac{1045}{27}~\zeta(5)
\right)
+n_f^2\left(
   \frac{125}{54}-\frac{1}{9}~\pi^2
\right)
\nonumber\\
&&+\left( -~\frac{4505}{4}+\frac{175}{2}~n_f
   -\frac{13}{9}~n_f^2
\right)\ln\,\frac{s}{\mu^2}
+\left(
    \frac{855}{4}-17~n_f
   +\frac{1}{3}~n_f^2
\right)\ln^2\frac{s}{\mu^2}
\Bigg]
\Bigg\}
\nonumber\\
&=& 3v_{\rmfont{b}}^2~\frac{\ovl{m}_{\rmfont{b}}^2}{s}
\Bigg\{
12~\api
+\left(\api\right)^2
\left[126.5-4.3333~n_f
+(-57+2~n_f)\ln\,\frac{s}{\mu^2}
\right]
\nonumber\\
&&+\left(\api\right)^3
\Bigg[
1040.01 -104.517~n_f + 1.2182~n_f^2
\nonumber\\
&&+\left(
   -1126.25+87.5~n_f
   -1.4444~n_f^2
\right)\ln\,\frac{s}{\mu^2}
\nonumber\\
&&+\left(
    213.75-17~n_f
   +0.3333~n_f^2
\right)\ln^2\frac{s}{\mu^2}
\Bigg]
\Bigg\}
\end{eqnarray}
\begin{eqnarray}
\EQN{apxc4p}
\dsp
R_6
 \dsp
&=&3 v_{\rmfont{b}}^2~\frac{\ovl{m}_{\rmfont{b}}^4}{s^2}
\Bigg\{
-6
+\api
\left[
-22 + 24 \ln \frac{s}{\mu^2}
\right]
\nonumber\\  \dsp
&&+\left(\api\right)^2
\Bigg[
-\frac{3029}{12}+27~\pi^2+112~\zeta(3)
     -\frac{11}{2}\ln\,\frac{\ovl{m}_{\rmfont{b}}^2}{s}
+225 \ln \frac{s}{\mu^2}  - 81 \ln^2 \frac{s}{\mu^2}
\nonumber\\  \dsp
&&+~n_f\left(
    \frac{143}{18}-\frac{2}{3}~\pi^2
    -\frac{8}{3}\zeta(3)
     +\frac{1}{3}\ln\,\frac{\ovl{m}_{\rmfont{b}}^2}{s}
-\frac{22}{3} \ln \frac{s}{\mu^2}  + 2 \ln^2 \frac{s}{\mu^2}
   \right)
\Bigg]
\Bigg\}
\nonumber\\  \dsp
&=&3 v_{\rmfont{b}}^2 ~\frac{\ovl{m}_{\rmfont{b}}^4}{s^2}
\Bigg\{
-6
+\api
\left[
-22 +24 \ln \frac{s}{\mu^2}
\right]
\nonumber\\
&&+\left(\api\right)^2
\Bigg[
148.693 -5.5\ln\,\frac{\ovl{m}_{\rmfont{b}}^2}{s}
+255 \ln \frac{s}{\mu^2}  - 81\ln^2 \frac{s}{\mu^2}
\nonumber\\
&&+
{}~n_f\left(
     -1.8408+0.3333\ln\,\frac{\ovl{m}_{\rmfont{b}}^2}{s}
     -7.3333 \ln \frac{s}{\mu^2}
     +2 \ln^2 \frac{s}{\mu^2}
     \right)
\Bigg]
\Bigg\}
{}\,  ;
\end{eqnarray}

\noindent
{\em
Massive non-singlet corrections (axial):
}
\begin{eqnarray}
\EQN{apxc5}
R_7
 \dsp
&=& 3~\frac{\ovl{m}_{\rmfont{b}}^2}{s}
\Bigg\{
-6+\api\left[
-22+12\ln\,\frac{s}{\mu^2}
\right]
\nonumber\\  \dsp
&&+\left(\api\right)^2
\Bigg[
-\frac{8221}{24}+\frac{19}{2}~\pi^2+117~\zeta(3)
+n_f\left(
\frac{151}{12}-\frac{1}{3}~\pi^2-4~\zeta(3)
\right)
\nonumber\\  \dsp
&&+\left(
   155-\frac{16}{3}~n_f
\right)\ln\,\frac{s}{\mu^2}
+\left(
   -~\frac{57}{2}+n_f
\right)\ln^2\frac{s}{\mu^2}
\Bigg]
\Bigg\}
\nonumber\\  \dsp
&=& 3~\frac{\ovl{m}_{\rmfont{b}}^2}{s}
\Bigg\{
-6+\api\left[
-22+12\ln\,\frac{s}{\mu^2}
\right]
\\  \dsp
&&+\left(\api\right)^2
\Bigg[
-108.14+4.4852~n_f
+\left(
   155-5.3333~n_f
\right)\ln\,\frac{s}{\mu^2}
+\left(
   -28.5+n_f
\right)\ln^2\frac{s}{\mu^2}
\Bigg]
\Bigg\}\nonumber
\end{eqnarray}

\begin{eqnarray}
\EQN{apxc5p}
\dsp
R_8
 \dsp
&=& 3~\frac{\ovl{m}_{\rmfont{b}}^4}{s^2}
\Bigg\{
6+
\api
\left[
10 - 24  \ln \frac{s}{\mu^2}
\right]
\nonumber\\  \dsp
&&+\left(\api\right)^2
\Bigg[
\frac{3389}{12}-27~\pi^2-220~\zeta(3)
     +\frac{77}{2}\ln\,\frac{\ovl{m}_{\rmfont{b}}^2}{s}
-207 \ln \frac{s}{\mu^2}  + 81 \ln^2 \frac{s}{\mu^2}
\nonumber\\  \dsp
&&+~n_f\left(
-~\frac{41}{6}+\frac{2}{3}~\pi^2+\frac{16}{3}~\zeta(3)
     -\frac{7}{3}\ln\,\frac{\ovl{m}_{\rmfont{b}}^2}{s}
+\frac{22}{3} \ln \frac{s}{\mu^2}  - 2\ln^2 \frac{s}{\mu^2}
   \right)
\Bigg]
\Bigg\}
\nonumber\\  \dsp
&=& 3~\frac{\ovl{m}_{\rmfont{b}}^4}{s^2}
\Bigg\{
6~
\api
\left[
10 - 24  \ln \frac{s}{\mu^2}
\right]
\nonumber\\  \dsp
&&+\left(\api\right)^2
\Bigg[
-248.515 +38.5\ln\,\frac{\ovl{m}_{\rmfont{b}}^2}{s}
-207 \ln \frac{s}{\mu^2}    + 81 \ln^2 \frac{s}{\mu^2}
\nonumber\\  \dsp
&&+
{}~n_f\left(
    6.1574-2.3333\ln\,\frac{\ovl{m}_{\rmfont{b}}^2}{s}
    +7.3333 \ln \frac{s}{\mu^2}    -2 \ln^2 \frac{s}{\mu^2}
    \right)
\Bigg]
\Bigg\}
{}\, ;
\end{eqnarray}

\noindent
{\em
Singlet corrections (axial):
}
\begin{eqnarray}
\EQN{apxc6}
R_9
  \dsp
&=&\left(\api\right)^2
\left\{
-9.250+1.037 \frac{s}{4M_{\rmfont{t}}^2}
+0.632 \left(\frac{s}{4M_{\rmfont{t}}^2}\right)^2
+3\ln\,\frac{s}{M_{\rmfont{t}}^2}
\right\}
\nonumber\\  \dsp
&&+\left(\api\right)^3
\Bigg\{
-\frac{5075}{72}+\frac{23}{12}~\pi^2+3~\zeta(3)
+\frac{67}{6}\ln\,\frac{\mu^2}{M_{\rmfont{t}}^2}
\nonumber\\
  &&+~\frac{23}{4}\ln^2\frac{\mu^2}{M_{\rmfont{t}}^2}
  +\frac{373}{8}\ln\,\frac{s}{\mu^2}
  -\frac{23}{4}\ln^2\frac{s}{\mu^2}
\Bigg\}
\nonumber\\  \dsp
&=&\left(\api\right)^2
\left\{
-9.250+1.037~\frac{s}{4M_{\rmfont{t}}^2}
+0.632 \left(\frac{s}{4M_{\rmfont{t}}^2}\right)^2
+3\ln\,\frac{s}{M_{\rmfont{t}}^2}
\right\}
\nonumber\\\dsp
&&+\left(\api\right)^3
\Bigg\{
-47.963
+11.167\ln\,\frac{\mu^2}{M_{\rmfont{t}}^2}
\nonumber\\  \dsp
 && +~5.75\ln^2\frac{\mu^2}{M_{\rmfont{t}}^2}
  +46.625\ln\,\frac{s}{\mu^2}
  -5.75\ln^2\frac{s}{\mu^2}
\Bigg\}
\end{eqnarray}
\begin{eqnarray}
\EQN{apxc6p}
R_{10}
  \dsp
&=&3\left(\api\right)^2
\left\{
-6\frac{\ovl{m}^2_{\rmfont{b}}}{s}
\left[
-3+\ln\,\frac{s}{M_{\rmfont{t}}^2}
\right]
-~10\frac{\ovl{m}^2_{\rmfont{b}}}{M_{\rmfont{t}}^2}
\left[
\frac{8}{81}-\frac{1}{54}\ln\,\frac{s}{M_{\rmfont{t}}^2}
\right]
\right\}
{}\, ;
\end{eqnarray}

\noindent
{\em
Singlet corrections (vector):
}
\beq\EQN{apxc7}
R_{11}
= \left(\sum_fv_f\right)^2
\left(\api\right)^3 (-1.2395)
{}\, ;
\eeq

\noindent
{\em
${\cal O}(\alpha\alpha_s)$ corrections:
}
\beq\EQN{apxc8}
R_{12}
= 3\sum_f(v_f^2+a_f^2)Q_f^2~\frac{3}{4}
{}~\frac{\alpha}{\pi}
\left[
1-\frac{1}{3}~\api
\right]
{}\, ;
\eeq

\noindent
{\em
$R_{\rmfont{had}}$ from virtual photons:
}
\vskip0.2cm
\noindent
The prediction for the `classical' production through a
value
$R^{\rmfont{em}} = \sigma(\rmfont{e}^+ \rmfont{e}^-
\bbuildrel{\to}_{\gamma}^{} \mbox{hadrons})
/ \sigma_{\rmfont{point}}$
is obtained
from the above equations  by
 setting $a_f=0$ and
$
R^{\rmfont{em}} = \sum_1^6 R_i
$
after setting $v_f \to Q_f$ and $a_f \to 0$;
\newpage

\vskip0.2cm
\noindent
{\em
Secondary radiation of heavy quarks:
}
\vskip0.2cm
\noindent
The rate for secondary radiation of a pair
of quarks with mass $m$ is given by
\beq\EQN{apxc9}
\frac{\Gamma_{{ \rmfont{Q}\ovl{\rmfont{Q}} \rmfont{q}
 \ovl{\rmfont{q}}}}}{\Gamma_{\rmfont{q}\ovl{\rmfont{q}}}}
=\frac{2}{3}\left(\api\right)^2
\varrho^R\left(\frac{m^2}{s}\right)
{}\, ,
\eeq
where
\begin{eqnarray}\EQN{apxc10}
\varrho^R(x) &=  &
 \frac{4}{3}\,\left( 1 - 6 x^2 \right) \,
   \left\{\frac{1}{2}\,{\rm Li}_3\left({{1 - w}\over 2}\right) -
         \frac{1}{2}\,{\rm Li}_3\left({{1 + w}\over 2}\right)
 \right.  \nonumber \\ & & \left.
+\, {\rm Li}_3\left({{1 + w}\over {1 + a}}\right) -
      {\rm Li}_3\left({{1 - w}\over {1 - a}}\right) +
      {\rm Li}_3\left({{1 + w}\over {1 - a}}\right) -
      {\rm Li}_3\left({{1 - w}\over {1 + a}}\right)
 \right.  \nonumber \\ & &
 +\,\frac{1}{2}\,\ln\, \left({{1 + w}\over {1 - w}}\right)
  \left[
   \zeta(2) -
   \frac{1}{12}\,\ln^2 \left(\frac{1+w}{1-w}\right)\right.\nonumber\\&& +
   \frac{1}{2}\,\ln^2\left(\frac{a-1}{a+1}\right) -\left.\left.
   \frac{1}{2}\,\ln\, \left(\frac{1+w}{2}\right) \ln\,
\left(\frac{1-w}{2}\right)
  \right]
    \right\}
    \nonumber\\ & &
 +\,\frac{1}{9}\,a\,\left( 19 + 46 x \right) \,
  \left[
   {\rm Li}_2\left({{1 + w}\over {1 + a}}\right) +
        {\rm Li}_2\left({{1 - w}\over {1 - a}}\right) -
        {\rm Li}_2\left({{1 + w}\over {1 - a}}\right) -
        {\rm Li}_2\left({{1 - w}\over {1 + a}}\right) \right.
    \nonumber\\ & &
  \left.
+\,\ln\, \left({{a - 1}\over {a + 1}}\right)\,\ln\,
\left({{1 + w}\over {1 - w}}\right)
    \right]
  \nonumber \\ & &
 +\,4 \left( {{19}\over {72}} + x + x^2  \right) \,
   \left[
    {\rm Li}_2\left(-~\frac{1+w}{1-w}\right) - {\rm
Li}_2\left(-~\frac{1-w}{1+w}\right)
    -\ln\, x\, \ln\, \left(\frac{1+w}{1-w}\right) \right]
  \nonumber\\ & &
 +\,7 \left( {{73}\over {189}} + {{74}\over {63}} x +
   x^2\right) \,\ln\, \left({{1 + w}\over {1 - w}}\right)
-\,\frac{1}{3}\left( {{2123}\over {108}}
+ \frac{2489}{54} x \right) \,w
\,,
\end{eqnarray}
with
\begin{equation}
\EQN{apxc11}
a=\sqrt{1+4 x}\,, \quad w=\sqrt{1-4 x}~.
\nonumber
\end{equation}

\end{document}